\documentclass[a4paper,11pt]{article}

\usepackage[colorlinks=true,linkcolor=red,citecolor=blue]{hyperref}
\usepackage{natbib}
\usepackage[latin1]{inputenc}
\usepackage[T1]{fontenc}
\usepackage{amsmath,amsfonts,amssymb}
\usepackage[francais]{babel}
\usepackage{theorem}
\usepackage{graphicx}

\newtheorem{Theorem}{Theorem}[section]
\newtheorem{Proposition}{Proposition}[section]
\newtheorem{lemma}{Lemma}[section]

\newtheorem{Remark}{Remark}[section]
\newtheorem{Conclusion}{Conclusion}[section]
\numberwithin{equation}{section}

\title{\bfseries Local asymptotically optimal test  in ARCH model}
\author{\bfseries Lounis tewfik\\Laboratoire de
mathématiques Nicolas Oresme,\\ CNRS UMR  6139 Université de Caen
}

\begin{document}

\maketitle
 \vspace{3mm} \hrule \vspace{3mm} {\small \noindent{\bf Abstract.}
 This work is an extension in Arch models of the
theorem of S.Y. Hwang and I.V. Basawa \cite{HB} which was used
before in nonlinear time series contiguous to $AR(1)$ processes.
Our results are established under some general assumptions and
stationarity and ergodicity conditions. Local asymptotic normality
(LAN) for the log likelihood ratio was established.An optimal test
was constructed when the parameter is assumed known. Also the
optimality of our test was proved when the parameter is
unspecified.
The method is based on the introducing of a new estimator.\\

Le Cam's third lemma, discrete estimate, modified estimator, time series models, ARCH models.\\

---------------------------------------------------------------------------------------------------------------------

\section{Introduction}
\noindent The study  of the chronicles   emanating from  economic,
biological, financial, hydrological, biomedical data or others
make use of relevant  mathematical models, namely  the time series
models that allow to model  this type of problems provided that
this framework takes into account  several criteria, such as, for
instance, the dependance of the observations, or the mean and the
variance  which are functions that depend on time. This often
leads us to choose  a class of well adapted models to aggregate
these differences best. The chosen class will be that of
stochastic models which will be detailed in the following.
\noindent Let $\{(Y_{i},X_{i})\}$ be  a sequence  of stationary
and ergodic random vectors with finite second and third moment
such that for all $i \in \mathbb{Z}$, $Y_{i}$ a univariate random
variable and $X_{i}$ a \emph{d}-variate random vector. We consider
the class of stochastics models\begin{eqnarray}
    Y_i =T(Z_i) + V(Z_i)\,\epsilon_i, \quad
    i\in\mathbb{Z},\label{modelprincipal}
\end{eqnarray} where the random vectors $ Z_i = \Big(Y_{i-1}, Y_{i-2},\ldots, Y_{i-s},X_i,X_{i-1},\ldots,X_{i-q}\Big)$,
for given non negative  integers $q$  and $s$, the
$\epsilon_{i}$'s are centred iid random variables with unit
variance and density function $f$, such that for all $i \in
\mathbb{Z}$, $\epsilon_{i}$ is independent of $\mathcal{F}_{i} =
\sigma(Z_j , j \leq i),$  the real-valued functions $T (\cdot)$
and $V (\cdot)$ are unknown.\\
 In this paper we study the problem of  testing of the couple of
functions $(T (\cdot),V (\cdot))$ in a  class of parametric
functions. Another words, let
$$\mathcal{M}=\left\{\left(m({\rho},\cdot), \sigma({\theta},\cdot)\right)
,~ ({\rho}^\top,{\theta}^\top)^\top \in \Theta_1 \times \Theta_2
\right\},
$$ $ \Theta_1 \times
\Theta_2\subset\mathbb{R}^\ell\times\mathbb{R}^p,$
 int$(\Theta_1)\neq \emptyset,$    int$(\Theta_2)\neq \emptyset,$
where for all set $A$, $int(A)$ denotes the interior of  the set
$A$ and the script $\top $ denotes the transpose . $\ell$ and $p$
are two positive integers, and each one of the two functions
$m({\rho},\cdot)$ and $\sigma({\theta},\cdot)$ has a known form
such that $\sigma({\theta},\cdot)>0.$ \noindent For a sample of
length $n$, we derive a test of $H_0\left[(T(\cdot),V(\cdot))\in
\mathcal{M}\right]$ against
$H_1\left[(T(\cdot),V(\cdot))\notin\mathcal{M}\right]$, one can
remark that the null hypothesis $H_0$ is equivalent
to:\begin{eqnarray*}
  H_0[(T(\cdot),V(\cdot)] &=&
  \Big(m({\rho}_{0},\cdot),\sigma({\theta}_0,\cdot)\Big),
\end{eqnarray*} for some $({\rho}_{0}^\top,{\theta}_{0}^\top)^\top \in \Theta_1
\times \Theta_2$ while the alternative hypothesis $H_1$ is
equivalent to\begin{eqnarray*} H_1[(T(\cdot),V(\cdot)] &\neq
&\Big(m({\rho}_0,\cdot),\sigma({\theta}_0,\cdot)\Big).
\end{eqnarray*}When we choose the alternative hypothesis like this  \\For all
integers $n\geq 1$ the alternative hypothesis  $H^{(n)}_1$ is
define by the following equality
\begin{eqnarray*}
H^{(n)}_1[(T(\cdot),V(\cdot)]&=& \Big(m({\rho}_0,\cdot)
+h\,n^{-\frac{1}{2}}G(\cdot),\sigma({\theta}_0,\cdot)+h'\,
n^{-\frac{1}{2}}S(\cdot)\Big),
\end{eqnarray*}$G$ and $S$ are two specified functions with values in
$\mathbb{R}$, $(h,h') \in K_1\times K_2$ where $K_1$ et $K_2$ are
two compacts of $\mathbb{R}$ and $hh'\neq 0.$\\
\noindent Under the null hypothesis  $(H_0),$ the time series
model (\ref{modelprincipal}) \begin{eqnarray}
 Y_i &=& m({\rho}_0,Z_i) + \sigma({\theta}_0,Z_i)\, \epsilon_i .
 \label{mnull}
 \end{eqnarray} And under the alternative hypothesis $H^{(n)}_1,$ the time series model (\ref{modelprincipal}) begin
\begin{eqnarray}
  Y_i &=&  m({\rho}_0,Z_i) +h\,n^{-\frac{1}{2}}G(Z_i) +\left(\sigma({\theta}_0,Z_i)
+h'\, n^{-\frac{1}{2}}S(Z_i)\right)\epsilon_i
.\label{malternative}
\end{eqnarray}Let $f_{0}$ and $f_{h,h'}$  denote the density function of the
random variable $Y_i$ corresponding to the time series model
(\ref{mnull}) and (\ref{malternative}) respectively, and let
$f_{n,0}$ and $f_{n,h,h'}$  denote the density function of the
random vector $(Y_1,...Y_n)$  corresponding to the time series
model (\ref{mnull}) and (\ref{malternative}) respectively.
\noindent Different specifications of $m({\rho}_0,\cdot)$ and
$\sigma({\theta}_0,\cdot)$ show that (\ref{mnull})  embodies a
large class of  time series models, for instance, we name  AR,
ARMA, SETAR, SETAR-ARCH and
$\beta$-ARCH.\\
 \noindent We consider  the  problem of testing the
null hypothesis $(H_0)$ against the alternative hypothesis
$(H^{(n)}_{1})$ such that
  $$(H_0) : m(\rho , Z_i) =m({\rho}_0 , Z_i)~~ and~~ \sigma(\theta ,
Z_i)=\sigma({\theta}_0,
  Z_i),$$ and,
 $$(H^{(n)}_{1}) :m(\rho , Z_i) =m({\rho}_0 ,
Z_i)+h\,{n}^{-\frac{1}{2}}G(Z_i) ~~ and~~ \sigma(\theta ,
Z_i)=\sigma({\theta}_0 , Z_i)+h'n^{-\frac{1}{2}}S(Z_i).$$
 We use  the  Neyman-Pearson test statistic based on the log-likelihood ratio
 $\Lambda_{n,h,h'}$ which is defined by the following equality\begin{equation}
   \Lambda_{n,h,h'}=\log\Big(\frac{f_{n,h,h'}}{f_{n,0}}\Big)=\sum_{i=1}^{n}\log(g_{n,i,h,h'}).\label{logarithm}
\end{equation}
 Our aim is to establish the normality of the test. Based on
 \cite[Theorem 1]{HB} and under some hypothesis and conditions
 and to a constant close, the log-likelihood
 ratio (\ref{logarithm}) is asymptotically equivalent to a sequence of random
 variables which is called the central sequence, therefore we
 obtain an optimal test in the
 case where the parameter $({\rho}_0,{\theta}_0)$ is specified.
 In a general case, the parameter $({\rho}_0,{\theta}_0)$ is
 unknown, so the propriety of the optimality of the test is not
 asserted. In order to estimate this parameter, we use
 locally discrete estimates, this kind of estimates was introduced
 by \cite{L}, and used by \cite{B} and \cite{K}.\\
\noindent The advantage of discrete estimates is the Lemma $(4.4)$
of (\cite{K}), This Lemma was among the fundamental tool used by
several authors to complete their research works, we can name the
articles of \cite{HP}, \cite{BH1,BH2} and \cite{DMD}.
\\
When we consider the difference between the two expressions of the
central sequence and an estimated central sequence, sometimes it
is possible to prove the optimality of  the test. In our case and
after the difference between the two central sequences, we get
asymptotically a non-degenerate term. In order to solve this very
problem and on the basis of  the discrete estimates, we introduce
a new estimator, the principle is to absorb the error of the
difference between the estimated central sequence and the central
sequence with the unknown parameter by modifying one component of
the discrete estimate, this method is presented in \cite[Section
$1$]{TL2012}. Consequently, under some assumptions, the optimality
of the constructed
test is proved.\\
 \noindent The paper is organized as follows\\
 In the forthcoming $(2),$ we establish some general
 assumptions and results  which are used in order to construct the
 test when the parameter is assumed known, the local asymptotic
 normality is established, an optimal  test is constructed and it's
 asymptotic power is derived. In section $(3),$ supplementary assumptions are given, the
discrete estimates were introduced and applied for the central
sequences. In section $(4)$, we prove the optimality of the test
when the parameter is unknown, the proof is based on the modified
estimate which is defined in the work of \cite{TL2012}. Section
$(5)$ concerned the generalization of our results in $\mathbb{Z}$.
In section $(6)$, we conduct a simulations in order to investigate
the performance of the proposed test. All mathematical
developments are relegated to the Section $(7)$.

\section{The construction of the test when the parameter is known}
\noindent Many results and assumptions are stated in the next
subsection in order to construct our test in the case when the
parameter of the study time series model is specified.

\subsection{Main results and assumption}
\noindent Throughout we assume that $i\in\mathbb{N}.$ An extension
on  $\mathbb{Z}$ will be made at
the end of this paper.\\
\noindent Consider the time series models\begin{eqnarray*}
 Y_i &=& m({\rho}_0,Z_i) + \sigma({\theta}_0,Z_i)\, \epsilon_i,
 \label{mii}
 \end{eqnarray*}and,
 \begin{eqnarray*}
  Y_i &=&  m({\rho}_0,Z_i) +h\,n^{-\frac{1}{2}}G(Z_i) +\left(\sigma({\theta}_0,Z_i)
+h'\, n^{-\frac{1}{2}}S(Z_i)\right)\epsilon_i . \label{miii}
\end{eqnarray*}  In order to establish the principle of local asymptotic
normality (LAN) for the log-likelihood ratio $\Lambda_{n,h,h'},$
we use  \cite[Theorem $(1)$]{HB}, so we check the three conditions
noted $(C.1) , (C.2) ~ and ~ (C.3)$
such that:\\
For a fixed step $(h,h')$ in $ K_{1}\times K_{2}$ where $hh'\neq
0,$ we have\begin{itemize}
    \item[(C.1)]$\max_{1\leq i\leq n}|g_{n,i,h,h'}
- 1| = o_{p}(1).$
    \item[(C.2)]There exist a positive constant
${\tau}^2 _{h,h'}$ such that \begin{eqnarray*} \sum_{i=1}^{n}
({g_{n,i,h,h'}} - 1)^2& =& {\tau}^2 _{h,h'} + o_{P}(1).
 \end{eqnarray*}
    \item[(C.3)]
   There exist a $\mathcal{F}_{n}$ measurable random  variable
$\mathcal{V}_{n,h,h'}$ such that\begin{eqnarray*} \sum_{i=1}^{n}
({g_{n,i,h,h'}} - 1)& =& \mathcal{V}_{n,h,h'}+ o_{P}(1).
\end{eqnarray*}
\end{itemize}
\noindent In order to establish our results, we need the following
assumptions and notations.\\
 For all $x \in \mathbb{R},$
let\begin{equation*} M_{f}(x)=\frac{\dot{f}(x)}{f(x)}.
\end{equation*}We assume that the function $x \longmapsto M_{f}(x)$ is
differentiable, we denote by $\dot{M}_{f}$ the derivative function
of $M_{f}.$ Consider the function $F$  defined by\begin{eqnarray*}
   F\left(x;a,b \right) &=& {\frac{1}{b}}\,f \left(\frac{x
   -a}{b}\right),\mbox{~~where~~}|a|<\infty \mbox{~~and~~}0<b<\infty.
\end{eqnarray*}We assume that the following assumptions are satisfied
:\begin{itemize}

    \item[($A_{1}$)] ($A_{1.1}$): There exist a measurable
    positive function $\varphi$, a real $p>1$ such that \\$\mathbf{E}({\varphi}^{p}({\epsilon}_{0}))<+\infty$ and a strictly
    positive real $\varsigma$, where $\varsigma>\max(|a|,|b-1|)$ such that
  $$\Big|\frac{{\partial}^2{F(x; a,b)}}{f(x)\,\partial
                            a^j\,
                            \partial b^k}\Big|\leq\varphi(x),$$
$j$ and $k$ are two positive integers such that
                $j + k = 2.$
\item[($A_{1.2}$)] There exist a  positive functions $V_1$
    and $V_2$ such that \begin{eqnarray*} \Big|\frac{{\partial}^2{F(x;a,b)}}{\partial a^j\,
                            \partial b^j} - \frac{{\partial}^2{F(x;a',b)}}{\partial
                            a^j\,
                            \partial b^k}\Big|\leq
                            V_1(x;a^\star,b)|a-a'|
                          \end{eqnarray*}
                          and
                          \begin{eqnarray*}
                            \Big|\frac{{\partial}^2{F(x; a,b)}}{\partial a^j\,
                            \partial b^k} - \frac{{\partial}^2{F(x;a,b')}}{\partial a^j
                            \partial b^k}\Big|\leq V_2(x;a,b^\star)|b-b'|,
                            \end{eqnarray*}
\noindent where $(a^\star,b^\star) \in[a,a']\times [b,b'],$ $j$
and $k$ are two positive integers such that
                $j + k = 2.$\\
\noindent There exist a measurable  positive function $\phi$ such
that $\mathbf{E}({\phi}({\epsilon}_{0}))<+\infty$ and a strictly
positive real ${\varsigma}',$ where
${\varsigma}'>\max(|\alpha|,|\beta-1|)$  such that
                            $$\Big|\frac{V_i(x;\alpha,\beta)}{f(x)}\Big|\leq\phi(x),~~i=1,2.$$
\end{itemize}
($A_2$)
    \begin{itemize}
        \item[($A_{2.1}$)]
                           ~~~~~~~~~~~~~~~~~~~~$\mathbf{E}\left\{ M_{f}(\epsilon_{0})\right\}=0.$
       \item[($A_{2.2}$)]~~~~~~~~~~~~~~~~~~~
        $\mathbf{E}\Big\{\epsilon_{0}M_{f}({\epsilon_{0}})
                         \}=-1.$

        \item[($A_{2.3}$)]~~~~~~~~~~~~~~~~~~~
        $\mathbf{E}\left \{ \dot{M}_{f}(\epsilon_{0}) + {M^2_{f}}(\epsilon_{0})\right\}=0.$

        \item[($A_{2.4}$)]~~~~~~~~~~~~~~~~~~~
        $\mathbf{E}\left \{ \epsilon_{0}(\dot{M}_{f}(\epsilon_{0}) + {{M^2_{f}}}(\epsilon_{0}))\right\}=0.$

        \item[($A_{2.5}$)]~~~~~~~~~~~~~~~~~~~
        $\mathbf{E}\left \{ \epsilon^2_{0}(\dot{M}_{f}(\epsilon_{0}) + {{M^2_{f}}}(\epsilon_{0}))\right\}=2.$
 \end{itemize}
    $(A_3)$ There exist $\lambda >0$  such that :
            \begin{itemize}
                \item[($A_{3.1}$)]~~~~~~~~~~~~~~~~~~~~ ${\mathbf{E}{\left|\frac{G(Z_{0})}{{\sigma}(\theta_0,Z_{0})}\right|}^{{\lambda} +
                2}<+\infty.}$
                \item[($A_{3.2}$)]~~~~~~~~~~~~~~~~~~~~
                $\mathbf{E}{\left|\frac{S(Z_{0})}{{\sigma}(\theta_0,Z_{0})}\right|}^{{\lambda} + 2}<+\infty.$
                 \item[($A_{3.3}$)]~~~~~~~~~~~~~~~~~~~~ $\mathbf{E}\Big|{M}_{f}(\epsilon_{0})\Big|^{{\lambda} + 2}<+\infty.$
                 \item[($A_{3.4}$)]~~~~~~~~~~~~~~~~~~~~$\mathbf{E}\Big|\epsilon_{0}{M}_{f}(\epsilon_{0})\Big|^{{\lambda} + 2}<+\infty.$
                 \item[($A_{3.5}$)]~~~~~~~~~~~~~~~~~~~~ $\mathbf{E}\Big|{\dot{M}}_{f}(\epsilon_{0})\Big|^{{\lambda} + 2}<+\infty.$
                 \item[($A_{3.6}$)]~~~~~~~~~~~~~~~~~~~~$\mathbf{E}\Big|\epsilon_{0}{\dot{M}}_{f}(\epsilon_{0})\Big|^{{\lambda} + 2}<+\infty.$
                 \item[($A_{3.7}$)]~~~~~~~~~~~~~~~~~~~~$\mathbf{E}|{\epsilon}_{0}|^{{2\,\lambda}+4}<+\infty. $
                 \item[($A_{3.8}$)]~~~~~~~~~~~~~~~~~~~~$\mathbf{E}\Big|\epsilon^2_{0}{M}_{f}(\epsilon_{0})\Big|^{{\lambda} + 2}<+\infty.$
\end{itemize}
\noindent A large class of the distribution functions satisfied
the condition $(A_2),$ we can, for instance, name the standard
normal distribution and the student distribution with  a degree of
freedom greater than $3$. The hypothesis ($A_{1.1}$) is similar to
the condition  $(A_3)$ fixed in \cite{HB}.\\
\noindent In order to get ($A_{1.2}$), we shall assume that the
partial derivatives with order $3$ exist and are locally bounded.
(The conditions$(A_{2.3}),$ ($A_{2.4}$) and ($A_{2.5}$) are
similar to the conditions ($A_{4.1}$)-($A_{4.5}$) fixed in
\cite{FL}).
\subsection{ Optimal test when the parameter is
known} \noindent In this subsection, we proceed to construct the
test in the case when the parameter $(\rho_0,\theta_0)$ is assumed
known, under the previous assumptions and conditions, we have the
following Theorem:\begin{Theorem}\label{firsttheorem} Under the
hypothesis  $(H_0)$, we have\begin{eqnarray}
 \Lambda_{n,h,h'}&=&\mathcal{V}_{n,h,h'} -\frac{{\tau}^2_{h,h'}}{2} +
 o_{P}(1),\label{lan}\mbox{~~ where~~}\mathcal{V}_{n,h,h'} \stackrel{\mathcal{D}}{\longrightarrow}
\mathcal{N}(0,{\tau}^2 _{h,h'}) \mbox{~~,~~}\\
 {\tau}^2_{h,h'}&=&h^2 I_0 \mathbf{E}\left(\frac{G(Z_{0})}
        {{\sigma}(\theta_0,Z{0})}\right)^2
+h'^2 (I_2 - 1) \mathbf{E}\left(\frac{S(Z_{0})}
{{\sigma}(\theta_0,
        Z_{0})}\right)^2 + 2hh'(I_1) \mathbf{E}\left(\frac{G(Z_{0})S(Z_{0})} {{\sigma^2}(\theta_0,
        Z_{0})}\right),~~~~~ \label{taux1}
 \end{eqnarray} and $$I_j
=\mathbf{E}\Big({\epsilon}^j_{0}{M^2_{f}}(\epsilon_{0})\Big),~~~~~~~j\in\{0,1,2\}.$$
\end{Theorem}
\subsection{Efficiency and power of the test}
\noindent In order to test the nul hypothesis  $(H_0)$ against the
alternative hypothesis  $(H^n_1)$ and for a fixed  step $(h,h')$
in $K_1\times K_2$, we use the Neuyman-Pearson statistics
$T_{n,h,h'}$ defined by\begin{eqnarray}
  T_{n,h,h'} &=& I{\Big\{{\frac{\mathcal{V}_{n,h,h'}}{{\tau}
_{h,h'}}\geq Z(u)}\Big\}},\label{test}
\end{eqnarray}Where $Z(u)$ is the quantile with order $ 1- u$ of the standard
normal distribution $(\Phi(Z(u))=1 -u )$.\\
We can deduce from the equality (\ref{lan}) that  $(H_0)$  and
$(H^n_1)$ are contiguous see for instance \cite[Corollary
(4.3)]{Droesbeke}. Under $(H^n_1)$ and from  Le Cam's third's
lemma \cite{HM}, we shall prove that the random variable $
\mathcal{V}_{n,h,h'}$ converges in distribution to
$\mathcal{N}({\tau}^2 _{h,h'} ,{\tau}^2 _{h,h'})$ as $n\rightarrow
+\infty$, therefore we obtain under the assumptions of the Theorem
(\ref{firsttheorem}) the following
statement:\begin{Theorem}\label{secondttheorem} The statistics
test is asymptotically optimal with a power function
 equal to $1 - \Phi(Z(1 -u)-{\tau}^2 _{h,h'})$ .
\end{Theorem}

\section{Estimation of the parameters and the link between the random local sequences.}
\noindent In practice the parameter $(\rho_{0},\theta_{0})$ is
unknown, so we can't assert the optimality of the test. For
estimating the unknown parameter, we use the discrete estimates.
Firstly, we begin by introducing the local random sequences
$\rho_n$ and $\theta_n$  of the parameters $\rho_{0}$ and
$\theta_{0}$ respectively, secondly we establish the difference
between the central sequences
$\mathcal{V}_{n,h,h'}(\rho_{0},\theta_{0})$ and
$\mathcal{V}_{n,h,h'}(\rho_n ,\theta_n),$ where
$\mathcal{V}_{n,h,h'}(\rho_n ,\theta_n)$ is the central sequence
obtained after replacing the parameter $(\rho_{0},\theta_{0})$ by
the parameter $(\rho_{n},\theta_{n})$ in the expression of
$\mathcal{{V}}_{n,h,h'}$, finally, and based on of \cite[Lemma
$(4.4)$]{K}, we introduce the discrete estimates. This kind of
estimator was introduced by \cite{L}, and
applied  by (\cite{HP}), (\cite{BH1,BH2} and  (\cite{DMD}). \\
We need in this work to remind some definitions and notations, and
we assume some supplementary assumptions.  The core of proof of
the optimality of the test is based on the instrumental
 Proposition (\ref{fondamental proposition}) which will be stated and proved later.
 \subsubsection*{Notations and definitions }
 \noindent Throughout,  $\|\cdot\|_{p}$
 and $\|\cdot\|_{\ell}$ are the euclidian norms in $ \mathbb{R}^\ell$ and
$\mathbb{R}^p$ respectively. \noindent We define the local
sequences  $\rho_n$ and $\theta_n$ of the parameters  $\rho_{0}$
and $\theta_{0}$ respectively by the following equalities
\begin{eqnarray*}
   \rho_n&=& \rho_{0}  +
   n^{-\frac{1}{2}}u^{(n)}\label{rholocal}
   \mbox{~,~} \theta_n=\theta_{0}  +
   n^{-\frac{1}{2}}v^{(n)}\label{thetalocal},\\&&~~n\mbox{~~is a strictly positive integer}
, (u^{(n)})^\top\times (v^{(n)})^\top \in
\mathbb{R}^\ell\times\mathbb{R}^p,
\end{eqnarray*}
such that
\begin{eqnarray*}
 (u^{(n)})^\top=(u^{(n)}_{1},\ldots,u^{(n)}_{\ell}),\quad
(v^{(n)})^\top =(v^{(n)}_{1},\ldots,v^{(n)}_{p}),\quad
{({\tau}^{(n)})}^\top =\Big({u^{(n)}}^\top
,v^{(n)})^\top\Big)^\top,\\ \mbox{~~and~~}
\sup_{n}[({{\tau}^{(n)})}^\top({\tau}^{(n)})]
<+\infty.\nonumber\\
  \label{c}
\end{eqnarray*}
For all $n\geq 1,$ we denote by\begin{eqnarray*}
  r_{n}=\|\rho_n - \rho_{0}\|_{\ell} \quad
\mbox{~and~}\quad r'_{n}=\|\theta_n - \theta_{0}\|_{p}.
\end{eqnarray*}
\noindent For all integers  $i$, we define the  residual
${\epsilon}_{i}$ by the following equation\begin{eqnarray}
 {\epsilon}_{i} &=& \frac{Y_{i} -
  m(\rho_{0}\,,\,Z_{i})}{\sigma(\theta_{0}\,,\,Z_{i})}.\label{residuel}
\end{eqnarray}
 \noindent By replacing in (\ref{residuel}) the parameters $\rho_{0}$ and $\theta_{0}$
  by the local sequences $\rho_n$ and $ \theta_n$ respectively,
 we obtained the expression  of the  natural estimate of the
 residuals $ {\epsilon}_{i}$
defined in the following equation
\begin{eqnarray}
  \tilde{\epsilon}_{i,n} &=& \frac{Y_{i} -
  m(\rho_{0}+n^{-\frac{1}{2}}u^{(n)}\,,\,Z_{i})}{\sigma(\theta_{0}+n^{-\frac{1}{2}}v^{(n)}\,,\,Z_{i})}.\label{residuelestimé}
\end{eqnarray}
 Let
\begin{eqnarray}
  r_{f,h,n}(\rho_0,\theta_0) = -n^{-\frac{1}{2}}\sum _{i=1}^{n} h M_{f}(\epsilon_i)\frac{G(Z_i)}
  {{\sigma}(\theta_0,
        Z_i)},\label{r1}\\
  \mbox{~~and~~}\nonumber\\
   q_{f,h',n}(\rho_0,\theta_0) = -n^{-\frac{1}{2}}\sum _{i=1}^{n} h'(1 +{ \epsilon_i
}M_{f}(\epsilon_i))\frac{S(Z_i)}
{{\sigma}(\theta_0,Z_i)}.\label{r2}
\end{eqnarray}
Clearly, we have:\begin{eqnarray}
{\mathcal{V}}_{n,h,h'}(\rho_0,\theta_0) &=&
r_{f,h,n}(\rho_0,\theta_0) + q_{f,h',n}(\rho_0,\theta_0).\label{r1
+ r2}
\end{eqnarray}
 \noindent By replacing  in  (\ref{r2}),
$\epsilon_i$ and $\theta_0$  by $\tilde{\epsilon}_{i,n}$ and $
\theta_n$ respectively, we get the following equalities
\begin{eqnarray}
 r_{f,h,n}(\rho_n,\theta_n) &=& -n^{-\frac{1}{2}}\sum _{i=1}^{n} h M_{f}(\tilde{\epsilon}_{i,n})\frac{G(Z_i)} {{\sigma}(\theta_{0}  +
 n^{-\frac{1}{2}}v^{(n)},
        Z_i)}, \label{r3} \\
 q_{f,h',n}(\rho_n,\theta_n) &=& -n^{-\frac{1}{2}}\sum _{i=1}^{n} h'(1 +\tilde{\epsilon}_{i,n}M_{f}(\tilde{\epsilon}_{i,n}))\frac{S(Z_i)}
   {{\sigma}(\theta_{0} + n^{-\frac{1}{2}}v^{(n)},
        Z_i)},\label{r4}\\
   \mathcal{V}_{n,h,h'}(\rho_n,\theta_n)&=&  r_{f,h,n}(\rho_n,\theta_n) +  q_{f,h',n}(\rho_n,\theta_n).\label{r5}
\end{eqnarray}
\subsubsection*{Assumptions} \noindent We suppose that the
conditions $(A)_{1}$--$(A)_{3}$ remains satisfied and we assume
that for all fixed  $x$,
  the functions  $\rho\rightarrow m(\rho,x)$ and $\theta\rightarrow \sigma(\theta,x)$
  are twice differentiable, we denote by
\begin{eqnarray*}
  \partial m(\rho,\cdot)^\top = (\frac{\partial
m(\rho,\cdot)}{\partial{\rho}_{1}},\cdot,\cdot,\cdot,\frac{\partial
m(\rho,\cdot)}{\partial
 {\rho}_{\ell}}),\quad \quad
  \partial{\sigma(\theta,\cdot)}^\top = (\frac{\partial
\sigma(\theta,\cdot)}{\partial{\theta}_{1}},\cdot,\cdot,\cdot,\frac{\partial
\sigma(\theta,\cdot)}{\partial
 {\theta}_{p}}),\\
 \partial^2m(\rho,\cdot)=\Big(\frac{\partial^2\,m(\rho,\cdot)}{\partial\rho_{i}\partial\rho_{j}}\Big)_{1\leq i ,j \leq \ell
 },\quad \quad \mbox{and}\quad \quad
\partial^2\sigma(\theta,\cdot)=\Big(\frac{\partial^2\,\sigma(\theta,\cdot)}{\partial\theta_{i}\partial_{j}}\Big)_{1\leq
i ,j \leq
  p}.
\end{eqnarray*}
\noindent $\partial^2m(\rho,\cdot)$ and $\partial^2
\sigma(\theta,\cdot)$ are  the hessian matrix of $m(\rho,\cdot)$
in $\rho$ and $\sigma(\theta,\cdot)$ in $\theta$ respectively.\\We
assume that the function $x \longmapsto M_{f}(x)$ is twice
differentiable with a \emph{ bounded second derivative},
 $\ddot{M}_{f}$ is the  seconde
derivative of $M_{f}$ (in this case we assume that the function
$f$ has a third derivative ). \noindent We define the function
$N_{f}$ by
$$N_{f} : x \longmapsto N_{f}(x)= 1 +
 x\, M_{f}(x).$$
 Note that the function $N_{f}$ is twice differentiable
with
\begin{eqnarray}
 \dot{N}_{f}(x) = M_{f}(x) + x\,\dot{
 M}_{f}(x), \label{derivéepremiereden} \mbox{~~and~~}
 \ddot{N}_{f}(x) = 2\,\dot{M}_{f}(x) + x\,
 \ddot{M}_{f}(x).\label{dsss}
\end{eqnarray}
 $\dot{N}_{f}$ and $\ddot{N}_{f}$ are respectively the derivative and the second derivative
 of ${N}_{f}$, \emph{we suppose that  $\ddot{N}_{f}$
 is bounded}.\\
\noindent  According to the notations of the previous subsection, we assume that the following conditions are satisfied:\\
($A_{4}$)
\begin{itemize}
    \item ($A_{4.1}$)\\ For all $n\geq 1,$  there exist two closed balls
     $ \overline{B_{1,n}}=\overline{B_{1,n}}\Big(\rho_{0},r_{1,n}\Big)\subset
     int(\Theta_{1})$ and \\$\overline{B_{2,n}}=\overline{B_{2,n}}(\theta_{0},r_{2,n})\subset
     int(\Theta_{2})$ where $r_{1,n}\geq r_{n}$ and $r_{2,n} \geq r'_{n}$  and a positive function
     $N_{1,n}$, such that $E\Big({\sup_{n\geq 1} N_{1,n}(Z_{0})}\Big)^{\mu
     +2}<\infty$, where  $\mu>0$ , such that, for all fixed $x$, we have
 $$ \max\Big(\sup_{(\rho,\,\theta) \in \overline{B_{1,n}} \times \overline{B_{2,n}}}\,\frac{\max_{1\leq i\leq\ell}
       |\frac{\partial\,m(\rho,x)}{\partial\rho_{i}}|}{\sigma(\theta,x)}\,,\,\sup_{(u,\theta) \in \overline{B_{2,n}} \times \overline{B_{2,n}}}\frac{\max_{1\leq j\leq
p}|\frac{\partial\,\sigma(u,x)}{\partial\theta_{j}}|}{\sigma(\theta,x)}\Big)\leq
     N_{1,n}(x).$$
 \item ($A_{4.2}$)\\ For all $n\geq 1,$  there exist two closed balls
     $ \overline{B'_{1,n}}=\overline{B'_{1,n}}\Big(\rho_{0},r'_{1,n}\Big)\subset
     int(\Theta_{1})$ and \\ $\overline{B'_{2,n}}=\overline{B'_{2,n}}(\theta_{0},r'_{2,,n})\subset
     int(\Theta_{2})$ where $r'_{1,n}\geq r_{n}$ et $r'_{2,n}\geq r'_{n}$, and a positive function $N_{2,n}$ ,
   such that $E\Big({\sup_{n\geq 1} N_{2,n}(Z_{0})}\Big)^{\mu' +3} < \infty$,
      where   $\mu'>0$, such that, for all fixed $x$, we have
$$ \max\Big(\sup_{(\rho,\,\theta) \in \overline{B'_{1,n}} \times \overline{B'_{2,n}}}\,\frac{\max_{1\leq i\leq\ell}
       |\frac{\partial\,m(\rho,x)}{\partial\rho_{i}}|}{\sigma(\theta,x)}\,,\, \sup_{(u,\theta) \in \overline{B'_{2,n}} \times \overline{B'_{2,n}}}\frac{\max_{1\leq j\leq
p}|\frac{\partial\,\sigma(u,x)}{\partial\theta_{j}}|}{\sigma(\theta,x)}\Big)\leq
      N_{2,n}(x).$$
\item ($A_{4.3}$)\\ For all $n\geq 1,$  there exist two closed
balls
     $ \overline{B^{(3)}_{1,n}}=\overline{B^{(3)}_{1,n}}\Big(\rho_{0},r^{(3)}_{1,n}\Big)\subset
     int(\Theta_{1})$ and \\ $\overline{B^{(3)}_{2,n}}=\overline{B^{(3)}_{2,n}}(\theta_{0},r^{(3)}_{2,n})\subset
     int(\Theta_{2})$ where  $r^{(3)}_{1,n}\geq r_{n}$ et $r^{(3)}_{2,n}\geq r'_{n}$ and a positive function $ N_{3,n}$  such
     that\\
     $E\Big({\sup_{n\geq 1} N_{3,n}(Z_{0})}\Big)^{{\mu_{3}} + 1}<\infty$, where
     $\mu_{3}>0$ , such that, for all fixed $x$, we have
$$ \max\Big(\sup_{(\rho,\,\theta) \in \overline{B^{(3)}_{1,n}} \times \overline{B^{(3)}_{3,n}}}\,\frac{\max_{1\leq i\leq\ell}
       |\frac{\partial\,m(\rho,x)}{\partial\rho_{i}}|}{\sigma(\theta,x)}\,,\, \sup_{(u,\theta) \in \overline{B^{(3)}_{2,n}} \times \overline{B^{(3)}_{2,n}}}\frac{\max_{1\leq j\leq
p}|\frac{\partial\,\sigma(u,x)}{\partial\theta_{j}}|}{\sigma(\theta,x)}\Big)\leq
     N_{3,n}(x).$$\\
 \item ($A_{4.4}$)\\ For all $n\geq 1,$  there exist two closed balls
     $ \overline{B^{(4)}_{1,n}}=\overline{B^{(4)}_{1,n}}\Big(\rho_{0},r^{(4)}_{1,n}\Big)\subset
     int(\Theta_{1})$ and \\ $\overline{B^{(4)}_{2,n}}=\overline{B^{(4)}_{2,n}}(\theta_{0},r^{(4)}_{2,n})\subset
     int(\Theta_{2})$ where $r^{(4)}_{1,n}\geq r_{n}$ et $r^{(4)}_{2,n}\geq r'_{n}$,  and a positive function $ N_{4,n}$  such
     that\\
     $E\Big({\sup_{n\geq 1} N_{4,n}(Z_{0})}^{{\mu_{4}} + 1}\Big)<\infty$, where
     ${\mu_{4}}>0$ , such that, for all fixed $x$, we have
 $$ \max\Big(\sup_{(\rho,\,\theta) \in \overline{B^{(4)}_{1,n}} \times \overline{B^{(4)}_{2,n}}}\,\frac{\max_{1\leq i,j\leq\ell}
       |\frac{\partial^2\,m(\rho,x)}{\partial\rho_{i}\partial\rho_{j}}|}{\sigma(\theta,x)}\,,\, \sup_{(u,\theta) \in \overline{B^{(4)}_{2,n}}
        \times \overline{B^{(4)}_{2,n}}}\,\frac{\max_{1\leq k,l\leq
p}|\frac{\partial^2\,\sigma(u,x)}{\partial\theta_{k}\partial_{l}}|}{\sigma(\theta,x)}\Big)\leq
     N_{4,n}(x).$$
\end{itemize}

\begin{Remark}\label{remarquederivéébornée}
\noindent Several families  of distribution assumed the condition
" \emph{$\ddot{M}_{f}$ is bounded }", we can for example cite the
case where  $f $ is a standard normal distribution, then
we have  $|\dot{M}_{f}(\epsilon_{0})|=1$ and $|\ddot{M}_{f}(\epsilon_{0})|=0$.\\
 When  $f$ is the student distribution with a degree of freedom greater
than $3$ , it is easy to prove with using simple calculation that
 the functions $x\longmapsto\dot{M}_{f}(x)$, $x \longmapsto \ddot{M}_{f}(x)$ and $x
\longmapsto x\ddot{M}_{f}(x)$ are bounded (see Appendix).
\end{Remark}

\subsubsection*{Locally asymptotic discrete estimates} \noindent
The great advantage of discrete estimates is \cite[Lemma
$(4.4)$]{K} who goes back to Le Cam and is also used by
(\cite{B}), (\cite{OL}), (\cite{HP}), (\cite{BH1,BH2}) and
(\cite{DMD}). The parameters $\rho_{0}$ and $\theta_{0}$ are
unknown, in order to estimate these parameters, we introduce the
discrete estimates $\hat{\rho}_{n}$ and $\hat{\theta}_{n}$
 of $\rho_{0}$ and $\theta_{0}$ respectively, such that these two
conditions $(D_{1})$ and $(D_{2})$ are satisfied :
\begin{itemize}
    \item[$(D_{1})$:]\quad  \emph{$\hat{\rho}_{n}$ is $\sqrt{n}$
    consistent, i.e for all $\epsilon>0$, there exist
    $\eta_{1}(\rho_{0},\epsilon)$ and $n_{1}(\rho_{0},\epsilon)$
    such that
   under $(H_{0}),$ we have  $\forall n\geq n_{1}(\rho_{0},\epsilon)$, $\mathbb{P}(\sqrt{n} \|\hat{\rho}_{n} - \rho_{0}\|_{\ell}
    >\eta_{1})\leq \epsilon .$}

 \noindent \emph{$\hat{\theta}_{n}$ is $\sqrt{n}$
    consistent, i.e for all $\epsilon>0$, there exist
    $\eta_{2}(\theta_{0},\epsilon)$ and $n_{2}(\theta_{0},\epsilon)$
    such that
    under $(H_{0}),$ we have\\$\forall n\geq n_{2}(\theta_{0},\epsilon),$ $\mathbb{P}(\sqrt{n} \|\hat{\theta}_{n}- \theta_{0}\|_{p}
    >\eta_{2})\leq \epsilon .$}
 \item[$(D_{2})$:]\quad \emph{$\hat{\rho}_{n}$, $\hat{\theta}_{n}$ are
 locally discrete, i.e for all fixed value  $c>0$ and under  $(H_{0})$ and as $n\rightarrow+\infty$, the number of possible values of
  $\hat{\rho}_{n}$ in $B_{1}=\{u\in {\mathbb{R}}^{\ell}, \sqrt{n} \|u - \rho_{0}\|_{\ell}
     \leq c \}$ and  $\hat{\theta}_{n}$ in $B_{2}= \{v\in {\mathbb{R}}^p,
\sqrt{n} \|v - \theta_{0}\|_{p}\leq c \}$ \\is bounded.}
\end{itemize}

\noindent Note that the condition $(D_{1})$ concerned the
appropriate rate of convergence in probability of the estimates,
this condition is satisfied by a several estimates such as the
maximum likelihood estimates, the Yule-Walker estimates, the
M-estimates and the least square estimates .\\
\noindent\emph{ We now may stat the fundamental proposition which
is the the core of the proof of the optimality. }
\begin{Proposition} \label{fondamental proposition}
For $(j,k) \in \{1,\dots,\ell\}\times \{1,\dots,p\}$, let
\begin{eqnarray}
  K^\top  &=&  (K_{1},\dots,K_{\ell})\mbox{,}\quad \quad
   K'^\top  =  (K'_{1},\dots,K'_{\ell})\nonumber\\
  K_{j}&=&  \mathbb{E}\Big[\frac{
\frac{\partial\,m(\rho_{0},\,Z_{0})}{\partial\,\rho_{j}
}}{\sigma(\theta_{0}\,,\,Z_{0})}\frac{\dot{M}_{f}({\epsilon}_{0})\,G(Z_{0})}
{\sigma(\theta_{0}\,,\,Z_{0})}\Big]\mbox{,}\quad \quad
 K'_{j}=  \mathbb{E}\Big[\frac{
\frac{\partial\,m(\rho_{0},\,Z_{0})}{\partial\,\rho_{j}
}}{\sigma(\theta_{0}\,,\,Z_{0})}\frac{\dot{N}_{f}({\epsilon}_{0})\,S(Z_{0})}
{\sigma(\theta_{0},\,Z_{0})}\Big],\label{first
constant}\label{second constant}\\
  J^\top  &=&  (J_{1},\dots,J_{p}),\quad \quad  J'^\top  =  (J'_{1},\dots,J'_{p}),\nonumber\\
\quad \quad J_{k}&=&  \mathbb{E}\Big[\frac{
\frac{\partial\,\sigma(\theta_{0},\,Z_{0})}{\partial\,\theta_{k}
}}{\sigma(\theta_{0}\,,\,Z_{0})}\frac{{\epsilon}_{0}\dot{M}_{f}({\epsilon}_{0})\,G(Z_{0})}
{\sigma(\theta_{0}\,,\,Z_{0})}\Big],\quad  \mbox{and}\quad
J'_{k}=\mathbb{E}\Big[\frac{\frac{\partial\,\sigma(\theta_{0},\,Z_{0})}{\partial\,\theta_{k}
}}{\sigma(\theta_{0}\,,\,Z_{0})}\frac{{\epsilon}_{0}\dot{N}_{f}({\epsilon}_{0})\,S(Z_{0})}
{\sigma(\theta_{0}\,,\,Z_{0})}\Big]\label{third
constant}\label{fourth constant}.
\end{eqnarray}
Then, we have the following equalities\begin{eqnarray}
  r_{f,h,n}(\rho_n,\theta_n) - r_{f,h,n}(\rho_0,\theta_0) &=& h\, (u^{(n)})^\top K^\top  + h\,(v^{(n)})^\top J^\top +o_{P}(1),\label{prop1}\\
 q_{f,h',n}(\rho_n,\theta_n) - q_{f,h',n}(\rho_0,\theta_0) &=&   h'\, (u^{(n)})^\top K'^\top  +h'\,(v^{(n)})^\top J'^\top +o_{P}(1),\label{prop2}\\
 \mathcal{V}_{n,h,h'}(\rho_n,\theta_n)-{\mathcal{V}}_{n,h,h'}(\rho_0,\theta_0)&=&\, (u^{(n)})^\top
 (h\,K^\top+h'\,K'^\top\,)+ (v^{(n)})^\top
 (h\,J^\top+h'\,J'^\top\,) + o_{P}(1).\label{prop3}
\end{eqnarray}
\end{Proposition}
\begin{Remark}
 The condition ``\emph{$\ddot{N}_{f}$ is bounded}'' is satisfied by
a large class of distribution functions.\\
 Based on the  remark (\ref{remarquederivéébornée}) and the equality (\ref{dsss}), we can
 deduce that, when  $f$ is the  density function of the standard normal
distribution, we have $|\ddot{N}_{f}(\epsilon_{0})|=2$,and when
$f$ is the  density of the student distribution with freedom
greater than $3$, $\ddot{N}_{f}$ is bounded (see appendix).
\end{Remark}
Using the estimator  $\hat{\rho}_{n}$ and $\hat{\theta}_{n}$ of
$\rho_{0}$ and $\theta_{0}$ respectively and such that the
conditions $(D_{1})$ and $(D_{2})$ are satisfied, with the
replacing  of the local sequences $\rho_n$ and $\theta_n$ by
$\hat{\rho}_{n}$ and $\hat{\theta}_{n}$ in  (\ref{prop1}}),
(\ref{prop2}) and (\ref{prop3}) respectively, and under the
assumptions of proposition (\ref{fondamental proposition}),  we
obtain the following statement:\begin{Proposition}\label{kreiss}
\begin{eqnarray}
  r_{f,h,n}(\rho_n,\theta_n) - r_{f,h,n}(\rho_0,\theta_0)  &=&  \sqrt{n}(\hat{\rho}_{n} - \rho_{0})^\top h\,K^\top +\sqrt{n}(\hat{\theta}_{n} - \theta_{0})^\top h\, J^\top + o_{P}(1),\label{prop1'}\\
 q_{f,h',n}(\rho_n,\theta_n) - q_{f,h',n}(\rho_0,\theta_0) &=&\sqrt{n}(\hat{\rho}_{n} - \rho_{0})^\top h'\,K'^\top+\sqrt{n}(\hat{\theta}_{n} - \theta_{0})^\top h'\, J'^\top+ o_{P}(1),\label{prop2'}\\
\mathcal{V}_{n,h,h'}(\rho_n,\theta_n)-{\mathcal{V}}_{n,h,h'}(\rho_0,\theta_0)&=&\sqrt{n}(\hat{\rho}_{n}
- \rho_{0})^\top ( h\,K^\top + h'\,K'^\top) +
  \sqrt{n}(\hat{\theta}_{n} - \theta_{0})^\top ( h\,J^\top + h'\,J'^\top)
  +o_{P}(1),\nonumber\\
  &=&D_{h,h'}(n) +o_{P}(1).~~~~~~\label{Link between the central sequences}
\end{eqnarray}
\end{Proposition}This last result, is a fundamental tool used later for the proof
of optimality of the test.\\
Consider  again the  equalities (\ref{Link between the central
sequences}), we remark that \begin{eqnarray*}
\Big(\hat{\rho}_{n},\hat{\theta}_{n}\Big)&=& \Big(\rho_{0}+
n^{-\frac{1}{2}}\,\sqrt{n}(\hat{\rho}_{n} -
\rho_{0})\,,\,\theta_{0}+
n^{-\frac{1}{2}}\,\sqrt{n}(\hat{\theta}_{n} - \theta _{0} )\Big),
\end{eqnarray*}
 with a probability  close to  $1$, the condition $(D_{1})$ gives the
 following condition\begin{eqnarray*}
\sup_{n}\Big\{\Big( \sqrt{n}(\hat{\rho}_{n} - \rho_{0})\,,\,
\sqrt{n}(\hat{\theta}_{n} - \theta _{0} )\Big)^\top
\Big(\sqrt{n}(\hat{\rho}_{n} - \rho_{0})\,,\,
\sqrt{n}(\hat{\theta}_{n} - \theta _{0} )\Big)\Big\} < +\infty.
\end{eqnarray*}
Since  $\sqrt{n}(\hat{\rho}_{n} - \rho_{0}) =O_{P}(1) $ and
$\sqrt{n}(\hat{\theta}_{n} - \theta_{0}) =O_{P}(1)$, we concluded
in a particular case corresponding to the equalities $K=K'=J=J=0,$
that the central sequences $\mathcal{V}_{n,h,h'}(\rho_n,\theta_n)$
and ${\mathcal{V}}_{n,h,h'}(\rho_0,\theta_0)$ are equivalent, in a
general case the right both side of the last previous equality is
not $o_{P}(1)$ as $n\rightarrow\infty,$ so it is not possible to
assert the optimality of the constructed test, in order to solve
this problem, we need to introduce another estimator which is
defined and described in the work of \cite[Section $1$]{TL2012}.
\section{Optimal test}
\noindent Throughout, we denote by
$\Omega^\prime_{n}=(\rho_n,\theta_n)$  the discrete estimate of
the unspecified parameter $\Omega^\prime=(\rho_0,\theta_0),$ with
the use of the results of \cite{TL2012}, we shall construct
another $\sqrt{n}$-consistency estimate $\bar{\Omega}_{n}$ of the
parameter $\Omega.$ According to the notations of \cite[Section
$(1)$]{TL2012}, we call this estimate the modified discrete
estimator which is denoted by M.D.E, under a supplementary
assumptions, we shall prove in the next subsection that with the
use of the M.D.E., it is possible to construct an
optimal test based on the Neyman-Pearson statistics.\\
\noindent\emph{We now may proceed to the proof of the optimality
of the test, we need that the conditions $(P.0)$ (or $(P'.0)$) and
$(P.1)$ (or $(P'.1)$ )  are fulfilled, such that: }
\begin{enumerate}
\item (P.0): $ \frac{\partial
{\mathcal{V}}_{n,h,h'}(\Omega_{n})}{\partial\rho_{j_{n}}}
  \neq0,$
\item(P'.0)  :$\frac{\partial
{\mathcal{V}}_{n,h,h'}(\Omega_{n})}{\partial\theta_{k_{n}}}
  \neq0,$
\item (P.1) :$ \frac{1}{\sqrt{n}} \frac{\partial
{\mathcal{V}}_{n,h,h'}(\Omega_{n})}{\partial\rho_{j_{n}}}
  \stackrel{P}{\longrightarrow}c_{1} \quad\mbox{as}\quad n\rightarrow\infty,$
   \item (P'.1):$
\frac{1}{\sqrt{n}}\frac{\partial\mathcal{V}_{n,h,h'}(\Omega_{n})}{\partial\theta_{k_{n}}}\stackrel{P}{\longrightarrow}c_{2}
\quad\mbox{as}\quad n\rightarrow\infty
,\label{Consistancegradient2}$ where $c_{1}$ and $c_{2}$ are two
constantes, such that $c_{1}\neq0$ and $c_{2}\neq0.$
\end{enumerate}
\begin{Remark}
\begin{itemize}
    \item The assumptions $ (P.0),$ $ (P'.0),$ $ (P.1)$ and $
    (P'.1)$ are fixed in \cite[ Section $1$]{TL2012}  in order to prove the
    existence and the $\sqrt{n}$-consistency of the modified  estimator.
\item Sufficient condition was stated for univariate time series
model, for more details  see \cite[ Lemma $3.1$ ]{TL2012}. A
generalization of this result concerned the $AR(m)$ model is
presented in the following subsection:
\end{itemize}
\end{Remark}

\subsubsection*{About a sufficient condition in $AR(m)$ model}
 Consider the following $AR(m)$ model:
\begin{eqnarray}
  Y_i = \sum_{j=1}^{m}(\rho_j  Y_{i-j})  + \epsilon_i, \quad \mbox{where} \quad \sum_{j=1}^{m} |\rho_j|  <1.\label{ARm model}
\end{eqnarray}

It will assumed that the model  (\ref{ARm model}) is  stationary
and ergodic with finite second and fourth moments,  in this case,
and according to the previous notations, we have
\begin{eqnarray}
m(\rho_0,Z_i)= \sum_{j=1}^{m}(\rho_j  Y_{i-j}), \quad
\mbox{,}\quad \sigma(\theta,Z_i)=1 \quad \mbox{and}\quad
\Omega^\top=\Big(\rho_1,\dots,\rho_m\Big)^\prime.
\end{eqnarray}

 We denote by
$\hat{\rho}_{n}=\Big(\hat{\rho}_{n,1},\dots,\hat{\rho}_{n,m})^\prime$
the estimator of the unknown parameter
$\rho=\Big(\rho_1,\dots,\rho_m\Big)^\prime$. Another estimator was
introduced in \cite[Section $1$]{TL2012}, its consistency is
satisfied under the following statement :
\begin{enumerate}
\item[(C.1)]
$$ \frac{1}{\sqrt{n}} \frac{\partial
{\mathcal{V}}_n(\hat{\rho}_{n})}{\partial\rho_{j_{n}}}
  \stackrel{P}{\longrightarrow}c_{1} \quad\mbox{as}\quad n\rightarrow\infty,
  \label{Consisatncegradient 1}$$ where $c_{1}$ is some constant
  no equal to $0$.
\end{enumerate}]. \\
Observe that, in practice, it is difficult to check this last
condition, therefore it is possible to give an equivalent
condition which is easier to establish. According to the previous
notations and assumptions, we have the following statement:
\begin{lemma}\label{about sufficuent condition}

$\epsilon_i$ are i.i.d. standard normal  distribution with
function density $f,$  Under $H_0,$  we have
\begin{eqnarray*}
\frac{1}{\sqrt{n}}\frac{\partial\mathcal{V}_n({\hat{\rho}}_n)}{\partial
\rho_j}=\frac{1}{\sqrt{n}}\frac{\partial\mathcal{V}_n(\rho_0)}{\partial
\rho_j} + o_{p}(1).
\end{eqnarray*}
\end{lemma}
\subsubsection*{Consequence} \noindent This lemma enables us to get
an equivalent condition for the consistency of the modified
estimator of the unknown parameter in $AR(m)$ model, the use of
the estimator of the unknown parameter in the stated condition
$(C.1)$ remains difficult, more precisely , it is possible to
calculate this limit with the unknown parameter. In this case, the
great advantage is that the result depends only on the
observations, under the condition of ergodicity and stationarity
of $AR(m)$ model, it is easy to prove that
$\frac{1}{\sqrt{n}}\frac{\partial\mathcal{V}_n(\rho_0)}{\partial
\rho_j} \stackrel{P}{\longrightarrow} \mathbb{E}(Y_{-j}G(Z_j)).$

In short, we shall replace in this case, the condition $(C'.1)$ by
the condition:
\begin{enumerate}
\item[(C'.1)]
$$ \frac{1}{\sqrt{n}} \frac{\partial
{\mathcal{V}}_n(\rho_{0})}{\partial \rho_j}
  \stackrel{P}{\longrightarrow}c_{1} \quad\mbox{as}\quad n\rightarrow\infty,
  \label{Consisatncegradient 1}$$ where $c_{1}$ is some constant
  no equal to $0$.
\end{enumerate}
Remark that, under $H_{0}$, with $o_{p}(1)$ close, the conditions
$(C.1)$ and $(C'.1)$, are equivalent.

\subsubsection*{Optimality} We assume that the conditions $(A.1)$
-$(A.4)$ are satisfied, now it is  obvious from the previous
results that we can state the following
theorem:\begin{Theorem}\label{optimality} Under LAN and the
conditions $ (P.0),$ (or $ (P'.0),$) and $ (P.1)$ or ($
    (P'.1)$),  the asymptotic power of $\bar{T}_n$ under $H^n_1$ is equal to to
$$1 - \Phi(Z(\alpha)-{\bar{\tau}}^2 ).$$ Furthermore,  $\bar{T}_n$ is
asymptotically optimal.
\end{Theorem}
\section{Generalization in Z}
\noindent Our results are established for $i \in \mathbb{N},$
doing an extension for $i \in \mathbb{Z},$ then, we process the  case where   $i\in{\mathbb{Z}}^{-}$.\\
Consider the following random variables $\mathcal{Y}$,
$\mathcal{Z}$ and $\varepsilon$ , such that, for
 all $i\in{\mathbb{Z}}^{-}$, we have \begin{eqnarray*}
  \mathcal{Y}_{-i} = Y_{i} \quad \mbox{,} \quad \mathcal{Z}_{-i} =
  {Z}_{i},\quad \mbox{and} \quad {\varepsilon_{-i}}= {\epsilon}_{i}.
\end{eqnarray*}

Clearly, $i'=-i \in{\mathbb{N}},$ therefore we obtain
\begin{eqnarray*}
   \mathcal{Y}_{i'} &=&T(\mathcal{Z}_{i'}) + V(\mathcal{Z}_{i'}){\varepsilon_{i'}}, \quad \mbox{where}\quad
   i'\in\mathbb{N}.
\end{eqnarray*}
\noindent The last time series model is similar to the model
$(1)$, by following the same previous reasoning in the case
corresponding to the model $(1)$, we shall construct a test $
T'_{n,h,h'}$ which is defined by the following equality
\begin{eqnarray*}
T'_{n,h,h'}&=&{I}_{\Big\{{\frac{\mathcal{V'}_{n,h,h'}}{{\tau}'
_{h,h'}}\geq Z(u)}\Big\}},\quad \mbox{where}\\
  {\tau'}^2 _{h,h'}&=&h^2 I'_0\mathbf{E}\left(\frac{G(\mathcal{Z}_{0})}
        {{\sigma}(\theta_0,\mathcal{Z}{0})}\right)^2
+h'^2 (I'_2 - 1)\mathbf{E}\left(\frac{S(\mathcal{Z}_{0})}
{{\sigma}(\theta_0,
        \mathcal{Z}_{0})}\right)^2 + 2hh'(I'_1) \mathbf{E}\left(\frac{G(\mathcal{Z}_{0})S(\mathcal{Z}_{0})} {{\sigma}(\theta_0,
        \mathcal{Z}_{0})}\right),\label{tauxRELATIF}\\
 U'_{n,i',h,h'}&=&-n^{-\frac{1}{2}}\left\{h
M_{f}(\varepsilon_{i'})\frac{G( \mathcal{Z}_{i'})}
{{\sigma}(\theta_0,
         \mathcal{Z}_i')}+h'( M_{f}(\varepsilon_{i'}){\varepsilon_{i'}}+1)\frac{S( \mathcal{Z}_{i'})} {{\sigma}(\theta_0,
         \mathcal{Z}_{i'})}\right\},\\
I'_j&=&\mathbf{E}\left({\varepsilon_{i'}}^j{{M}^2_{f}({\varepsilon_{i'}}}
)\right)\quad \mbox{,} \quad j\in\{0,1,2\}\quad \mbox{and} \quad
\mathcal{V'}_{n,h,h'}=\sum_{i'=1}^{n}{U'_{n,i',h,h'}}.
\end{eqnarray*}
\section{Simulations}
\noindent In order to investigate the  performance of the proposed
test, we conduct simulations,  the considering  time series models
are AR(1) and AR(2). We give simultaneously the power functions
with the true parameter, the estimated parameter and  the
estimated parameter by the M.D.E. respectively. The power relative
for each test estimated upon m = 1000 replicates, all those
representations use the discretized form of the modified estimate.
We devote a big importance about the choice of the functions $G$
and $S$ to aim to satisfied the stated conditions. In a sequel, we
assume that: $\epsilon_{i}$'s are centred iid and $\epsilon_{0}
\stackrel{\mathcal{D}}{\longrightarrow} \mathcal{N}(0,1),$ in this
case, we have
\begin{eqnarray*}
 \mathbb{E}({\epsilon_{i}})=0 \quad \mbox{,} \quad
 \mathbb{E}({\epsilon^2_{i}})=1 \quad \mbox{,} \quad
 \mathbb{E}({\epsilon^4_{i}})=3.
\end{eqnarray*}
\subsubsection*{Example1:\\ \emph{Nonlinear time series contiguous
to AR(1) processes} } \noindent Consider the sth order(nonlinear)
time series
\begin{eqnarray}
    Y_i =\rho_0 Y_{i -1} + \alpha \,G(Y(i -1)) + \epsilon_i  \quad \mbox{,} \quad |\rho_0| <1.\label{principal nonlinear}
\end{eqnarray}
\noindent It will be assumed that  the time series model
(\ref{principal nonlinear}) is stationary and ergodic with finite
second moments. Consider again the problem of testing the null
hypothesis $(H_0):\alpha=0$ (linearity of the AR(1) model) against
the alternative hypothesis $(H^{n}_1):\alpha =n^{-\frac{1}{2}}$
(nonlinearity of the AR(1) model).
 The purpose of this subsection is to treat this  problem of the testing  when
 $h=h'=1$, in this case, we have, for all integers $i$, the following equalities  :

 \begin{eqnarray}
D_{h,h'}(n)&=& D_n=-\Big(\sqrt{n}(\hat{\rho}_{n} - \rho_{0})^\top
( K^\top + K'^\top) +
  \sqrt{n}(\hat{\theta}_{n} - \theta_{0})^\top ( J^\top +J'^\top)\Big),\label{bounded ecart  ar1}\\
  m(\rho_0,Z_i)&=& \rho_0 Y_{i-1} \quad \mbox{,} \quad \sigma(\theta,Z_i)=1 \quad \mbox{,} \quad
  M_{f}({\epsilon_{i}})=-\epsilon_{i}  \quad \mbox{,} \quad  \dot{M}_{f}({\epsilon_{i}})=-1\quad \mbox{,} \quad
N_{f}({\epsilon_{i}})= 1-\epsilon^2_{i},\nonumber\\
\dot{N}_{f}({\epsilon_{i}})&=& -2\epsilon_{i} \quad \mbox{,} \quad
\Omega=\rho_0 \quad \mbox{,} \quad  \Omega_n=\hat{\rho}_n  \quad
\mbox{and} \quad Z_i = \Big(Y_{i-1}, Y_{i-2}, \cdot \cdot \cdot,
Y_{i-s},X_i,X_{i-1},\ldots,X_{i-q}\Big).\nonumber
\end{eqnarray}

\noindent We Choose $ G: \Big(x_1, x_{2}, \cdot \cdot \cdot,
x_{s},x_{s+1},x_{s+2},\cdot \cdot
\cdot,x_{s+q}\Big)\longrightarrow \frac{6a}{1+x^2_1}$,
$S(\cdot)=0$ and $a\neq0,$ clearly $G(Z_0)=\frac{6a}{1+Y^2_{-1}},$
note that this choice of the functions $G$ and $S$ enables us to
obey the conditions $(A_{3.1})$ and $(A_{3.2})$ .\\ The parameter
$\rho_0$ is estimated by the least square estimate $\hat{\rho}_n =
\frac{\sum_{i=1}^{n}Y_i Y_{i-1}}{\sum_{i=1}^{n} Y^2_{i-1}}$ and
the residual $\epsilon_i$ is estimated by $ \epsilon_{i,n}= Y_i -
\hat{\rho}_n Y_{i-1}.$ We have
$\mathcal{\dot{V}}_{n,h,h'}(\Omega)=\frac{-6a}{\sqrt{n}}\sum_{i=1}^{n}\frac{Y_{i-1}}{1+Y^2_{i-1}},$
Then, from the equalities (\ref{first constant}), (\ref{fourth
constant}), the ergodicity  and the stationarity of model
(\ref{principal nonlinear}), it follows that:
\begin{eqnarray*}
\frac{1}{\sqrt{n}\,}\mathcal{\dot{V}}_{n,h,h'}(\Omega)&=&\frac{-6a}{n}\sum_{i=1}^{n}\frac{Y_{i-1}}{1+Y^2_{i-1}}\quad
\mbox{,} \quad
\frac{1}{\sqrt{n}\,}\mathcal{\dot{V}}_{n,h,h'}(\Omega_n)=\frac{-6a}{n}\sum_{i=1}^{n}\frac{Y_{i-1}}{1+Y^2_{i-1}},\\
J&=&J'=K'=0 \quad \mbox{and} \quad
 K=-6a\mathbb{E}\Big[\frac{Y_{-1}}{1+Y^2_{-1}}\Big].
\end{eqnarray*}
We denote by $disrete(\hat{\rho}_{n})$  the discretization of the
least square estimator L.S.E. $\hat{\rho}_{n}.$ Note that from the
ergodicity and the stationarity of the model (\ref{principal
nonlinear}), it follows that the random variable
$\frac{1}{\sqrt{n}}\mathcal{\dot{V}}_{n,h,h'}(\Omega)
\stackrel{a.s.}{\longrightarrow}-6a\,\mathbb{E}\Big[\frac{Y_{-1}}{1+Y^2_{-1}}\Big]$
as $n\rightarrow\infty$. With the use of (\ref{Link between the
central sequences}) combined with equality (\ref{first constant}),
it follows that :

\begin{eqnarray}
\mathcal{\widehat{V}}_{n,h,h'}-\mathcal{V}_{n,h,h'}&=&-\sqrt{n}(disrete(\hat{\rho}_{n})
- \rho_{0})6a\,\mathbb{E}\Big[\frac{Y_{-1}}{1+Y^2_{-1}}\Big]
+o_{P}(1).\label{optimalité simulations}
\end{eqnarray}

Under the conditions $(P.0)$ and $(P.1)$, we have the
$\sqrt{n}$-consistency of the modified estimated M.D.E. which is
noted $\bar{\rho}_{n},$ with:

\begin{eqnarray*}
  \bar{\rho}_{n} &=&
  \frac{D_n}{\frac{\partial\mathcal{V}_{n,h,h'}(\Omega_{n})}{\partial\rho}}
  +(disrete(\hat{\rho}_{n}))\label{MDE},
\end{eqnarray*}
where the quantity $D_n$ is defined in the equality (\ref{bounded
ecart ar1}), it result  that:\begin{eqnarray}
 \bar{\rho}_n=\frac{\sqrt{n}(disrete(\hat{\rho}_{n})-
\rho_{0})6a\,\mathbb{E}\Big[\frac{Y_{-1}}{1+Y^2_{-1}}\Big]}{\mathcal{\dot{V}}_{n}(\hat{\rho}_{n})}
+(disrete(\hat{\rho}_{n})).\label{MDE}
\end{eqnarray}

\noindent For a fixed $\alpha = 0.05 $, the test proposed is
 $T_{n} = I{\Big\{{\frac{\mathcal{V}_{n}(\rho_0)}{\tau(\rho_0)}\geq
 Z(\alpha)}\Big\}},$ with the subsisting the parameter $\rho_0$  in the
expressions of the proposed test and the power function $1 -
\Phi(Z(\alpha)-{\tau^2( \rho_{0})} ),$ by it's  modified
 estimate $\bar{\rho}_n$ defined by the equality (\ref{MDE}), it result from the theorem
(\ref{optimality}) that the statistic test $\bar{T}_{n}$ is
asymptotically equivalent to $T_{n}$ and it's
power is equal to $1 -\Phi(Z(\alpha)-{\tau( \bar{\rho}_{n})} )$.\\
The true value of the parameter $\rho_0$ is fixed at $0.1$ and the
sample sizes are $n = 30, 40, 60$ and $80.$ We obtain the
following representations: \vspace{1cm}
\begin{figure}[h!]
\includegraphics[scale=0.2]{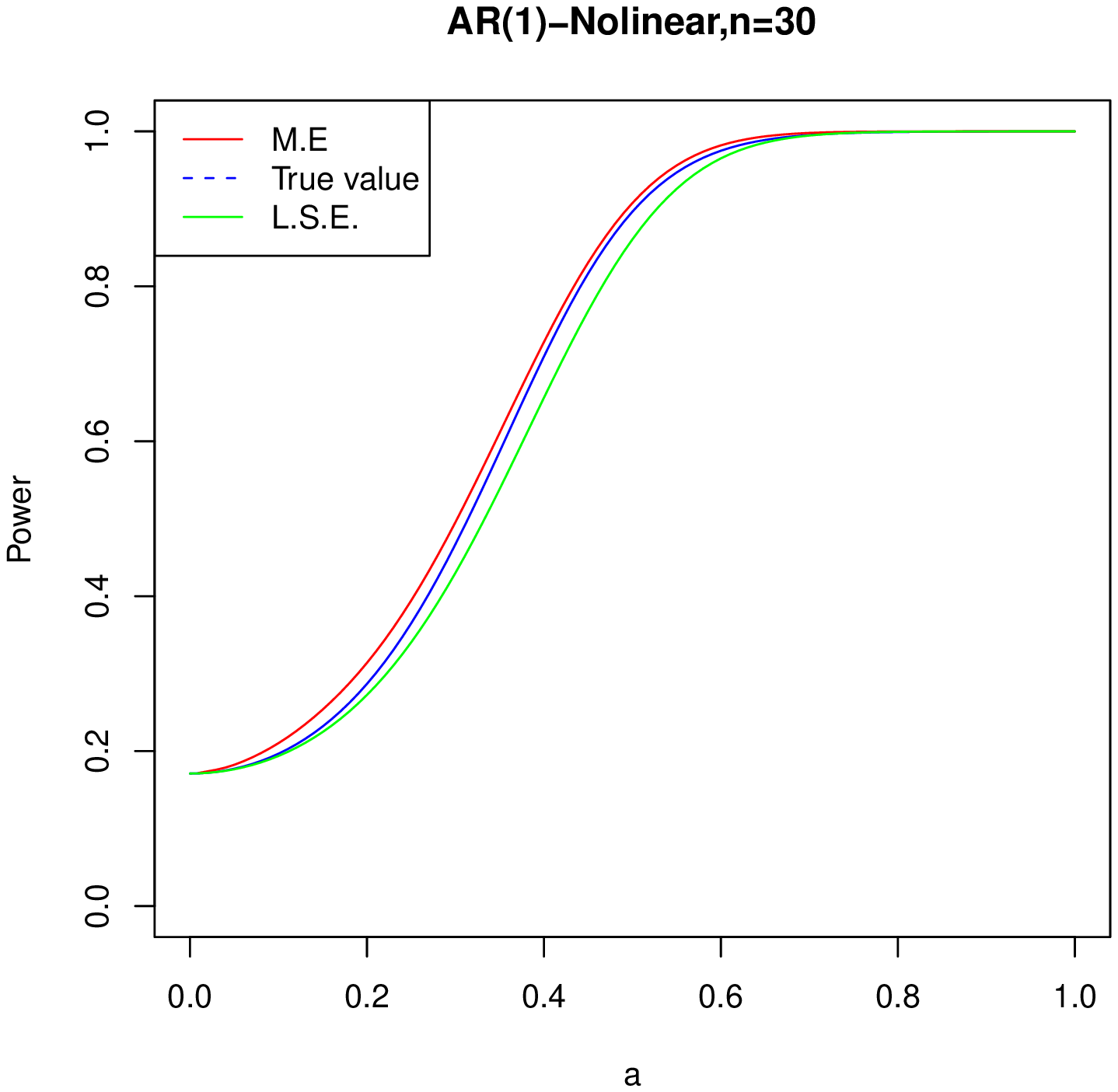}
\includegraphics[scale=0.2]{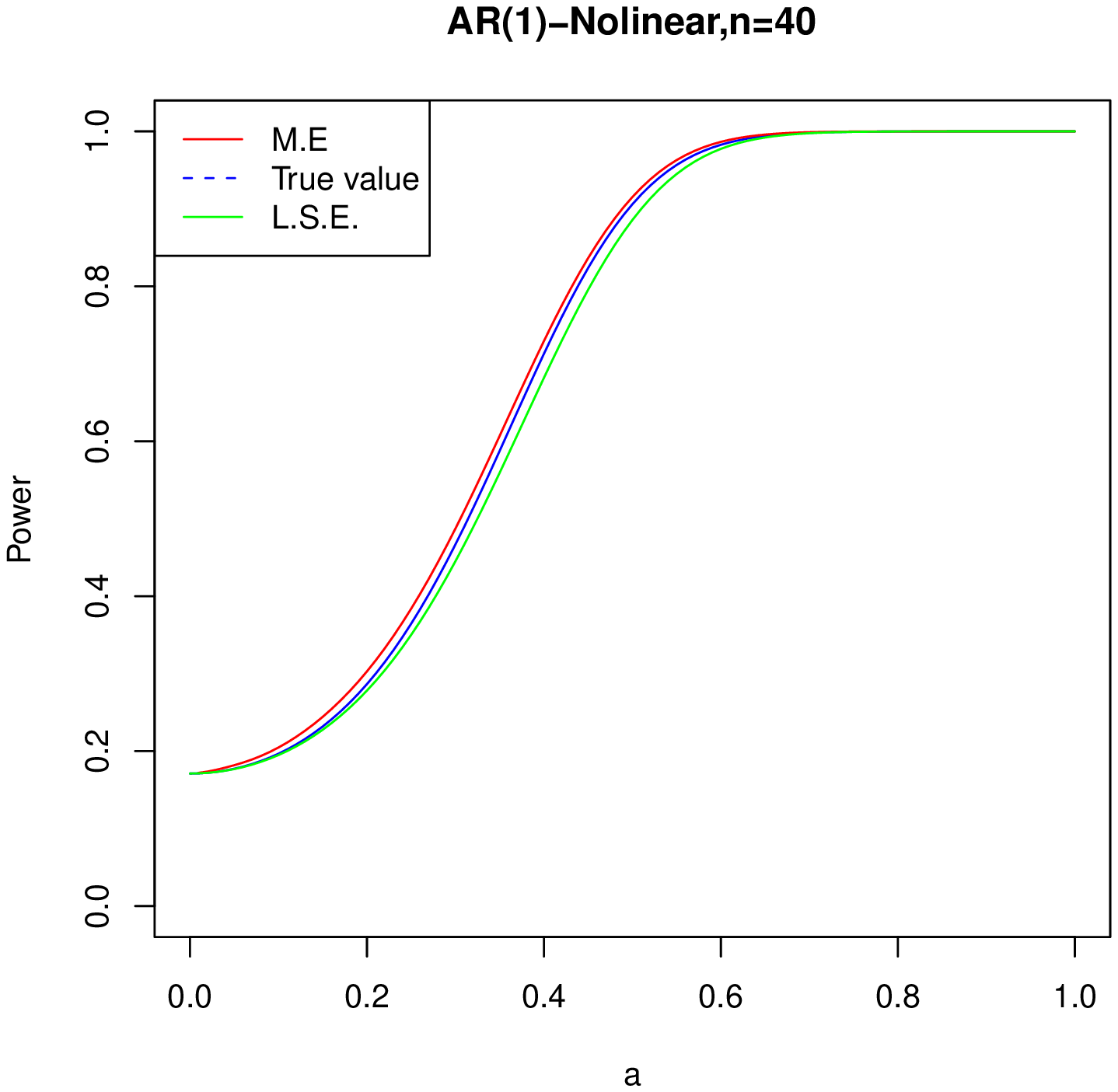}
\includegraphics[scale=0.2]{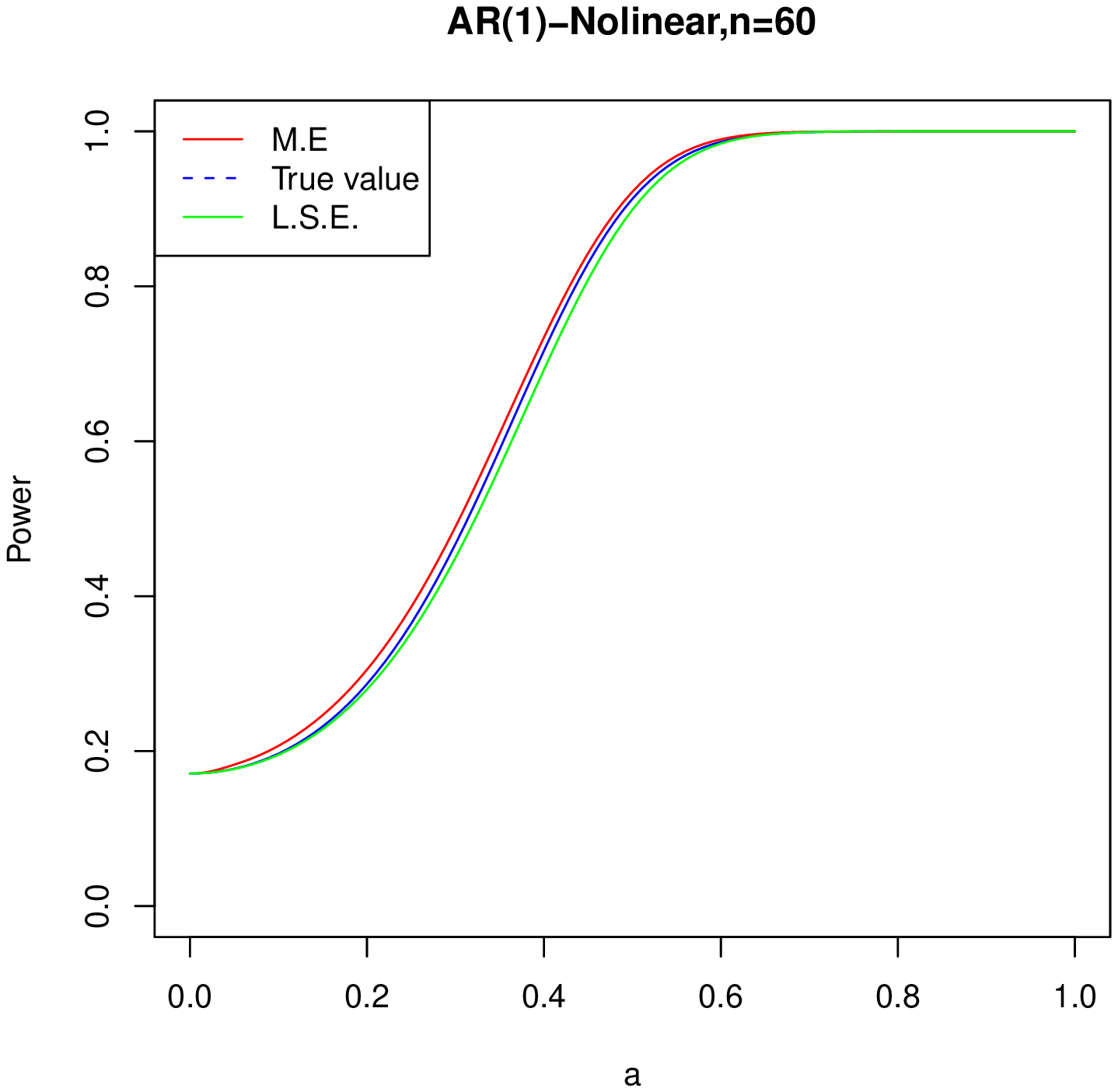}
\includegraphics[scale=0.2]{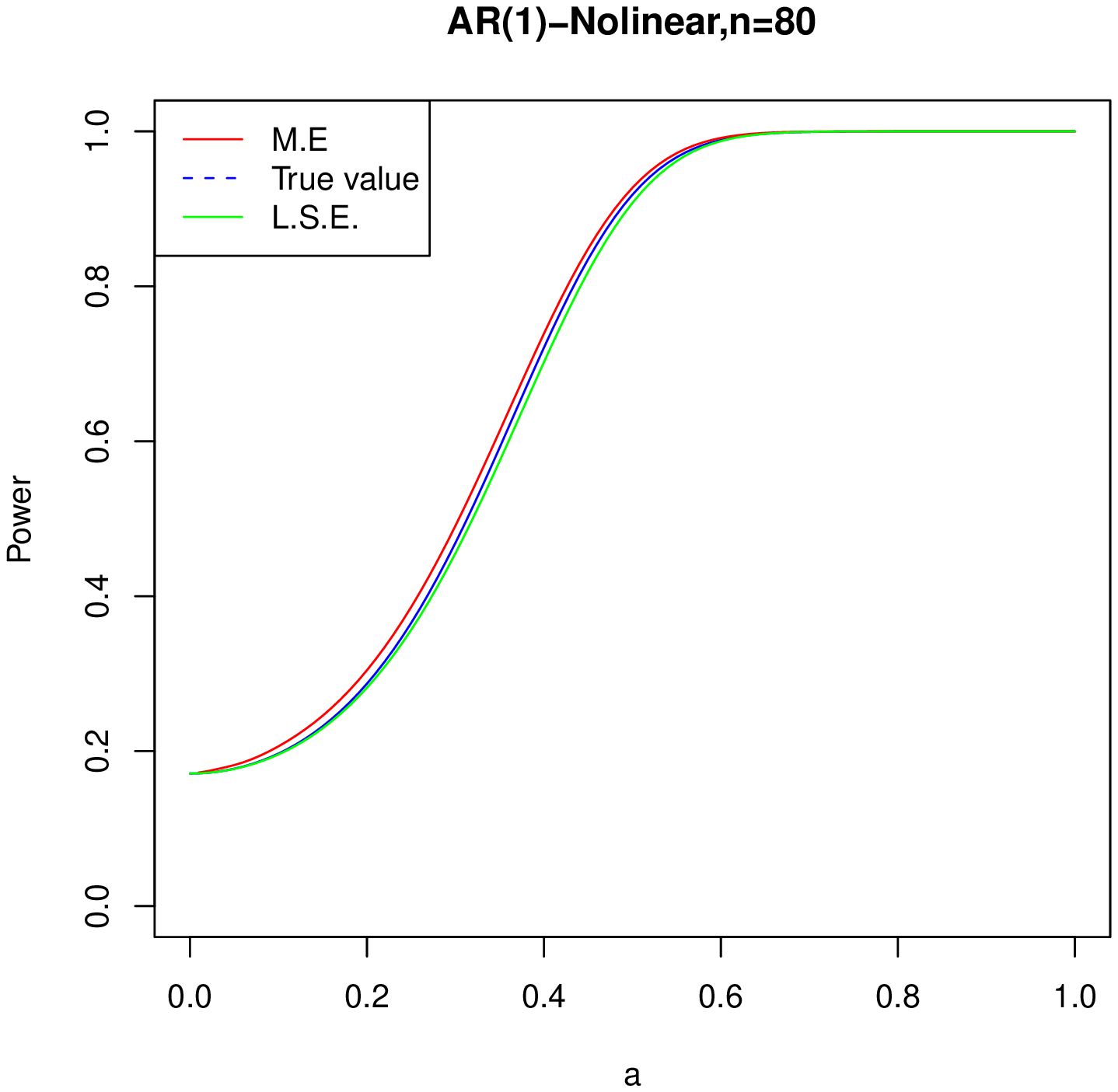}
\end{figure}

 We remark that, the power function  with true value and the
empirical power function with the M.D.E. are close as the value
$n$ is large.

\subsubsection*{ Example2:\\\emph{An extension
to ARCH processes}} Consider the following time series model with
conditional heteroscedasticity
\begin{eqnarray}
    Y_i =\rho_0 Y_{i -1} + \alpha \,G(Y(i -1)) + \sqrt{1  + \beta B(Y(i -1))} \,\epsilon_i, \quad
    i\in\mathbb{Z}.\label{model with conditional
heteroscedasticity}
\end{eqnarray}
 It is assumed that the model (\ref{model with
conditional heteroscedasticity}) is ergodic and stationary. We
conduct our simulation with the same method as the previous case,
we define the functions $G$ and $S$ by:
$$ G:  \Big(x_1,\dots,
x_{s},x_{s+1},x_{s+2},\dots,x_{s+q}\Big)\longrightarrow
  \frac{5a}{1+x^2_1} \quad \mbox{and} \quad  S=\frac{G}{4}.$$
Therefore,  we obtain the following equalities:
\begin{eqnarray*}
\frac{1}{\sqrt{n}\,}\frac{\partial\mathcal{V}_{n,h,h'}}{\partial\rho}(\Omega)&=&\frac{-5a}{n}\sum_{i=1}^{n}\frac{Y_{i-1}}{1+Y^2_{i-1}}
+\frac{-10a}{n}\sum_{i=1}^{n}\frac{Y_{i-1}}{1+Y^2_{i-1}}\,\epsilon_i,\\
\frac{1}{\sqrt{n}\,}\frac{\partial\mathcal{V}_{n,h,h'}}{\partial\rho}(\Omega_n)&=&\frac{-5a}{n}\sum_{i=1}^{n}\frac{Y_{i-1}}{1+Y^2_{i-1}}
+ \frac{-10a}{n}\sum_{i=1}^{n}\frac{Y_{i-1}}{1+Y^2_{i-1}}
(Y_i-\hat{\rho}_n Y_{i-1} ),\\
J&=&J'=K'=0,\\
K&=&-5a\,\mathbb{E}\Big[\frac{Y_{-1}}{1+Y^2_{-1}}\Big].
\end{eqnarray*}
Then we obtain :
\begin{eqnarray*}
\mathcal{\widehat{V}}_{n,h,h'}-\mathcal{V}_{n,h,h'}&=&-\sqrt{n}(\hat{\rho}_{n}
-\rho_{0})h\,5a\,\mathbb{E}\Big[\frac{Y_{-1}}{1+Y^2_{-1}}\Big]+o_{P}(1).
\end{eqnarray*}
For a fixed $\alpha = 0.05 $, the test proposed is
 $T_{n} = I{\Big\{{\frac{\mathcal{V}_{n}(\rho_0)}{\tau(\rho_0)}\geq
 Z(\alpha)}\Big\}},$ with the subsisting the parameter $\rho_0$ by it's estimator
$\bar{\rho}_n$ in the expressions of the proposed test and the
power function $1 - \Phi(Z(\alpha)-{\tau^2( \rho_{0})} ),$ we
obtain from theorem (\ref{optimality}) an optimal equivalent test
$\bar{T}_{n}$ with a power $1 -\Phi(Z(\alpha)-{\tau^2(
\bar{\rho}_{n})} ),$ the true value of the parameter  $\rho_0$ is
fixed at $0.1$ and the sample sizes are $n = 30, 40, 50$ and
$80$.\\ We obtain the following representations:
\begin{figure}[h!]
\includegraphics[scale=0.2]{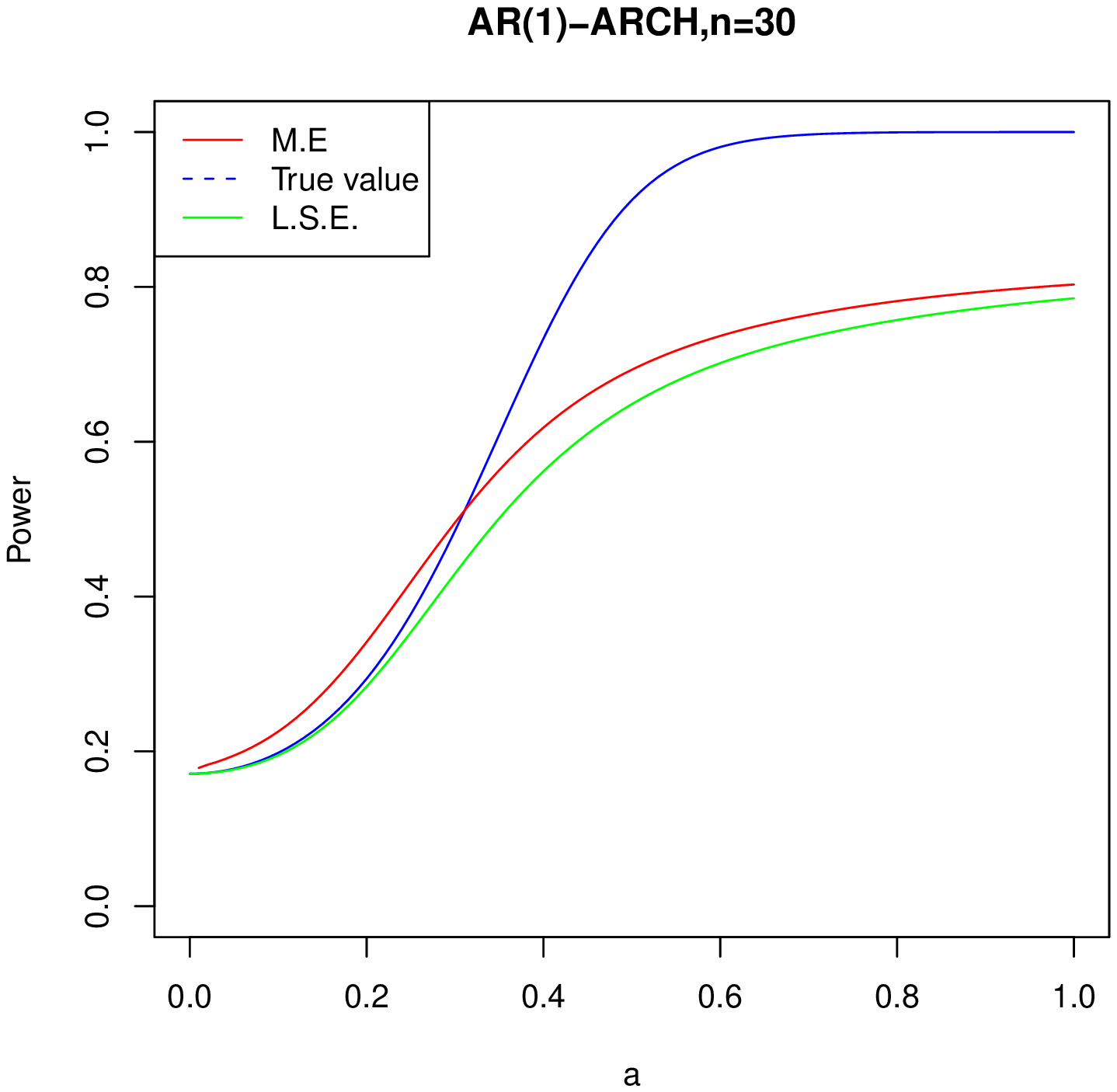}
\includegraphics[scale=0.2]{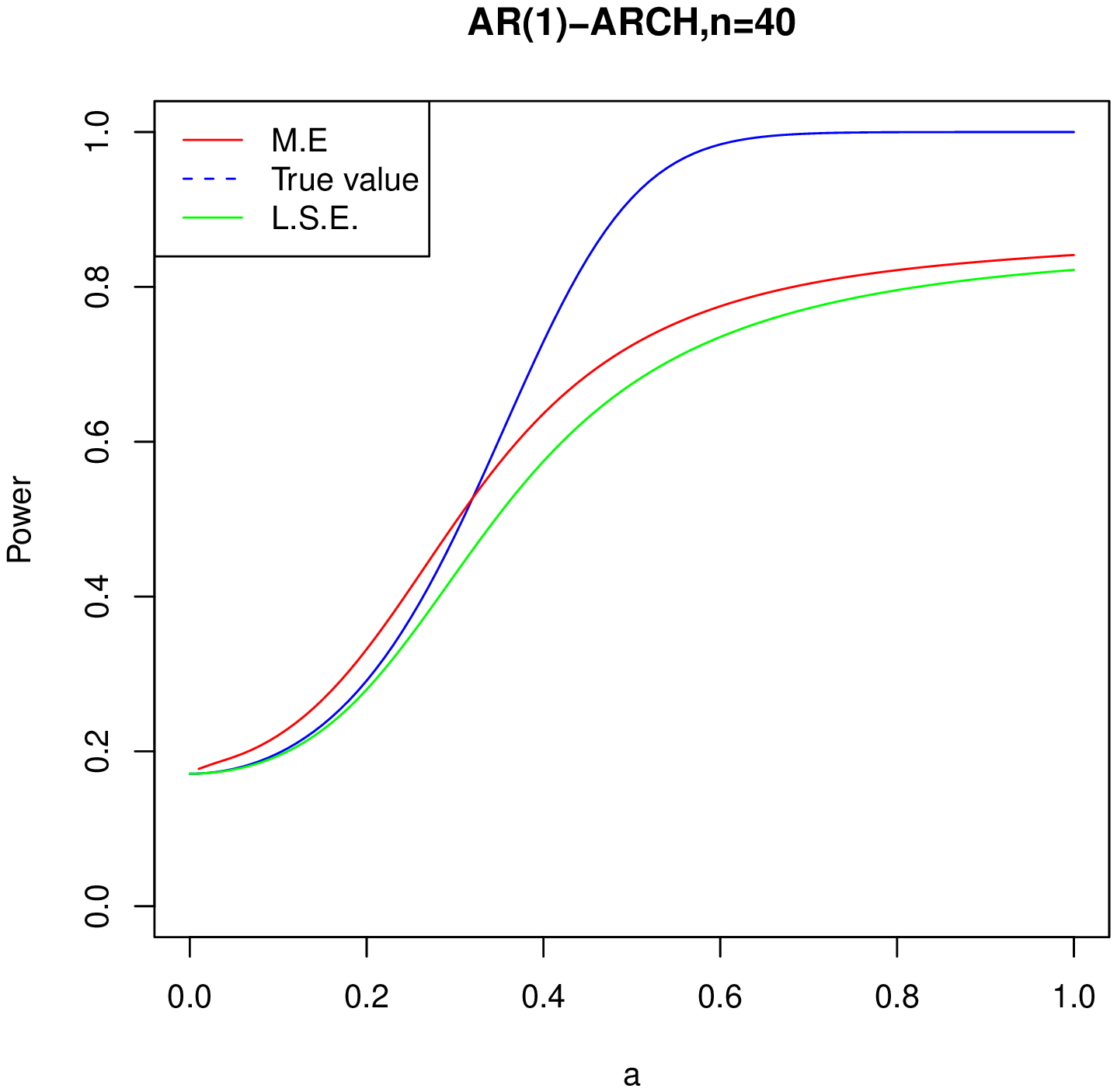}
\includegraphics[scale=0.2]{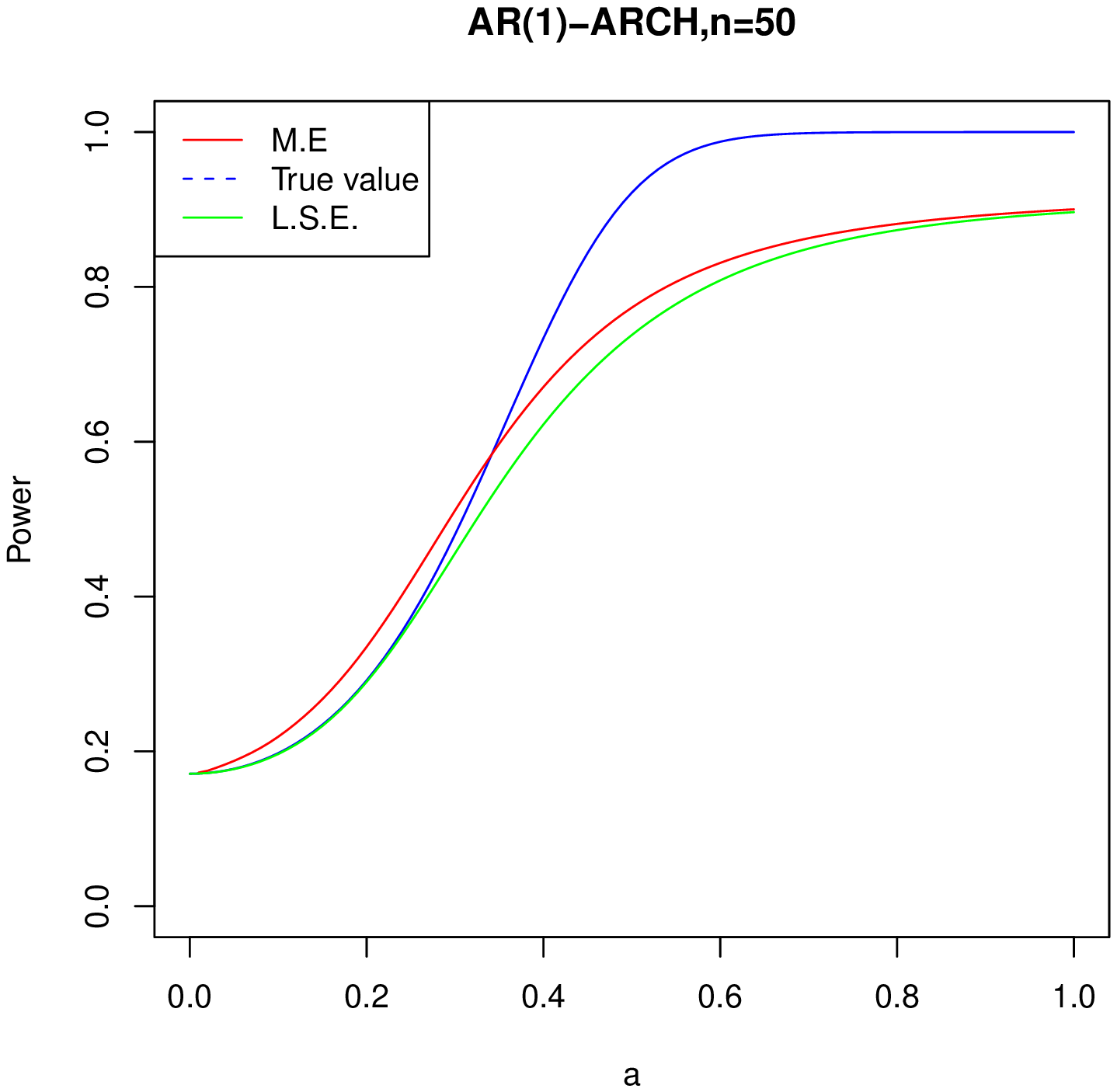}
\includegraphics[scale=0.2]{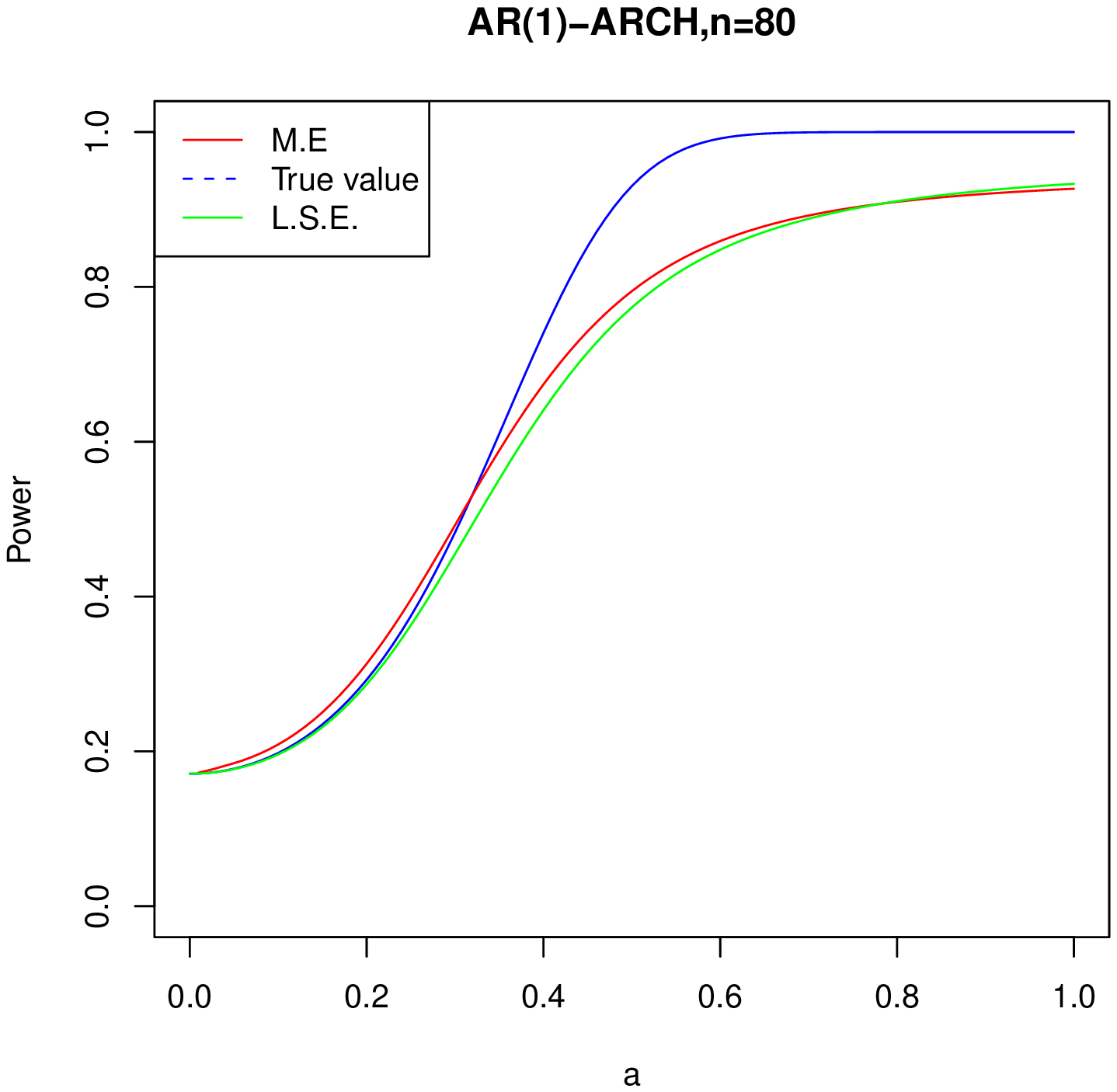}
\end{figure}

\subsubsection*{Example3:AR(2) model} \noindent Consider the
following AR($2$) model:
\begin{eqnarray}
  Y_i = \rho_1  Y_{i-1} + \rho_2  Y_{i-2} + \epsilon_i, \quad \mbox{where} \quad |\rho_1| +  |\rho_2| <1.\label{AR2 model}
\end{eqnarray}

It will assumed that the model  (\ref{AR2 model}) is  stationary
and ergodic,  in this case, we have
\begin{eqnarray}
m(\rho_0,Z_i)= \rho_1 Y_{i-1} + \rho_2 Y_{i-2}, \quad
\mbox{,}\quad \sigma(\theta,Z_i)=1 \quad \mbox{and}\quad
\Omega^\top=(\rho_1,\rho_2).
\end{eqnarray}
 We choose \\$ S,G: \Big(x_1, x_{2},
\cdot \cdot \cdot, x_{s},x_{s+1},x_{s+2},\cdot \cdot
\cdot,x_{s+q}\Big)\longrightarrow \frac{8a}{1+x^2_1+x^2_2}$, where
$a\neq0$, clearly, we obtain:
$$S(Z_0)=G(Z_0)=\frac{8a}{1+Y^2_{-1} + Y^2_{-2}}.$$
Note that the choice of the functions
 $G$ et $S$ enables us to obey the conditions
 $(A_{3.1})$ and $(A_{3.2})$ .We
denote by \\${\Omega^\top}_n=(\rho_{1,n} , \rho_{2,n})$ the least
square estimate of the parameter $\Omega^\top=(\rho_1 , \rho_2)$
such that:
\begin{eqnarray}
  \Omega_n &=& [X^\top X]^{-1}X^\top Y ,\label{estimateur du moindre  carre}
\end{eqnarray}
~~ $Y=\left(%
\begin{array}{c}
  Y_1 \\
  \cdot \\
  \cdot \\
  \cdot \\
  Y_n \\
\end{array}%
\right)$ ~~,~~ $X=\left(%
\begin{array}{cc}
   Y_0  &  Y_{-1}  \\
  \cdot & \cdot \\
  \cdot & \cdot \\
  \cdot & \cdot \\
  Y_{n-1}  &  Y_{n-2}  \\
\end{array}%
\right) $ ~~and~~ $X^\top=\left(%
\begin{array}{ccccc}
   Y_0 & \cdot & \cdot &\cdot &   Y_{n-1}\\
  Y_{-1}& \cdot & \cdot & \cdot & Y_{n-2} \\
\end{array}%
\right).$ \\
Recall that for each $i,$ the residual $\epsilon_i$ is estimated
by the following random variable\begin{eqnarray}
\hat{\epsilon}_{i,n}=Y_i-\rho_{1,n}\,Y_{i-1}
-\rho_{2,n}\,Y_{i-2}.\label{residual in AR2 model}
\end{eqnarray}
We have:
\begin{eqnarray*}
\frac{\partial\mathcal{V}_{n,h,h'}}{\partial\rho_1}(\Omega)=-\frac{8a}{\sqrt{n}}\sum_{i=1}^{n}\frac{Y_{i-1}}{1+Y^2_{i-1}
+Y^2_{i-2}} - \frac{16a}{\sqrt{n}}\sum_{i=1}^{n}\frac{Y_{i-1}}{1+Y^2_{i-1}+Y^2_{i-2}}(Y_i-\rho_1Y_{i-1} -\rho_2  Y_{i-2}),\nonumber\\
\frac{\partial\mathcal{V}_{n,h,h'}}{\partial\rho_2}(\Omega)=-\frac{8a}{\sqrt{n}}\sum_{i=1}^{n}\frac{Y_{i-2}}{1+Y^2_{i-1}
+Y^2_{i-2}} - \frac{16a}{\sqrt{n}}\sum_{i=1}^{n}\frac{Y_{i-2}}{1+Y^2_{i-1}+Y^2_{i-2}}(Y_i-\rho_1Y_{i-1}-\rho_2  Y_{i-2}).\nonumber\\
\end{eqnarray*}
We obtain :
\begin{eqnarray}
\frac{1}{\sqrt{n}}\frac{\partial\mathcal{V}_{n,h,h'}}{\partial\rho_1}(\Omega)=-\frac{8a}{n}\sum_{i=1}^{n}\frac{Y_{i-1}}{1+Y^2_{i-1}+Y^2_{i-2}} - \frac{16a}{n}\sum_{i=1}^{n}\frac{Y_{i-1}}{1+Y^2_{i-1}+Y^2_{i-2}}\epsilon_i,\label{df true parameter}\\
\frac{1}{\sqrt{n}}\frac{\partial\mathcal{V}_{n,h,h'}}{\partial\rho_2}(\Omega)=-\frac{8a}{n}\sum_{i=1}^{n}\frac{Y_{i-2}}{1+Y^2_{i-1}+Y^2_{i-2}}-\frac{16a}{n}\sum_{i=1}^{n}\frac{Y_{i-2}}{1+Y^2_{i-1}+Y^2_{i-2}}\epsilon_i,\label{ds
true parameter}
\end{eqnarray}
then:
\begin{eqnarray}
\frac{1}{\sqrt{n}}\frac{\partial\mathcal{V}_{n,h,h'}}{\partial\rho_1}(\Omega_n)=-\frac{8a}{n}\sum_{i=1}^{n}\frac{Y_{i-1}}{1+Y^2_{i-1}+Y^2_{i-2}} - \frac{16a}{n}\sum_{i=1}^{n}\frac{Y_{i-1}}{1+Y^2_{i-1}+Y^2_{i-2}}\hat{\epsilon}_{i,n},\label{dfestimated parameter}\\
\frac{1}{\sqrt{n}}\frac{\partial\mathcal{V}_{n,h,h'}}{\partial\rho_2}(\Omega_n)=-\frac{8a}{n}\sum_{i=1}^{n}\frac{Y_{i-2}}{1+Y^2_{i-1}+Y^2_{i-2}}-\frac{16a}{n}\sum_{i=1}^{n}\frac{Y_{i-2}}{1+Y^2_{i-1}+Y^2_{i-2}}\hat{\epsilon}_{i,n}.\label{dsestimated
parameter}
\end{eqnarray}
\subsubsection*{Correction with respect the first parameter
$\rho_1$:} The combinaison of the equalities  (\ref{AR2 model})
with (\ref{residual in AR2 model})  enables us to deduce that
\begin{eqnarray}
\hat{\epsilon}_{i,n} - \epsilon_i= - Y_{i-1}(\rho_{n,1}-\rho_1)
-Y_{i-2}(\rho_{n,2}-\rho_2).\label{ecart residuel in AR2 model }
\end{eqnarray}
From the difference between the equalities (\ref{dfestimated
parameter}) and (\ref{df true parameter}) combined with
(\ref{ecart residuel in AR2 model }), it follows that
\begin{eqnarray}
|\frac{1}{\sqrt{n}}\frac{\partial\mathcal{V}_{n,h,h'}}{\partial\rho_1}(\Omega_n)-\frac{1}{\sqrt{n}}\frac{\partial\mathcal{V}_{n,h,h'}}{\partial\rho_1}
(\Omega)|=
\frac{16a}{n}|\sum_{i=1}^{n}\frac{Y_{i-1}}{1+Y^2_{i-1}+Y^2_{i-2}}(\hat{\epsilon}_{i,n}-
\epsilon_i)|,\nonumber\\
\leq \Big|\rho_{n,1}-\rho_1\Big|\times
\frac{16a}{n}\Big|\sum_{i=1}^{n}\frac{Y^2_{i-1}}{1+Y^2_{i-1}+Y^2_{i-2}}\Big|+\Big|\rho_{n,2}-\rho_2\Big|\times\frac{16a}{n}\Big|\sum_{i=1}^{n}
\frac{Y_{i-1}Y_{i-2}}{1+Y^2_{i-1}+Y^2_{i-2}}\Big|.\nonumber\\
\label{normailzed difference between central sequances}
\end{eqnarray}
Remark that:
\begin{eqnarray} \frac{1}{1+Y^2_{i-1}+Y^2_{i-2}}\leq 1 \quad \mbox{then}\quad
\frac{Y^2_{i-1}}{1+Y^2_{i-1}+Y^2_{i-2}}\leq Y^2_{i-1}\quad
\mbox{this implies that}\quad
\Big|\sum_{i=1}^{n}\frac{Y^2_{i-1}}{1+Y^2_{i-1}+Y^2_{i-2}}\Big|\leq
\sum_{i=1}^{n}Y^2_{i-1}.\\\label{first inequality}
\end{eqnarray}
We can also remark that:
\begin{eqnarray} \Big|\sum_{i=1}^{n}
\frac{Y_{i-1}Y_{i-2}}{1+Y^2_{i-1}+Y^2_{i-2}}\Big|\leq
\Big|\sum_{i=1}^{n}Y_{i-1}Y_{i-2}\Big|\leq
\frac{1}{2}\sum_{i=1}^{n}(Y^2_{i-1} + Y^2_{i-2}).\label{Second
inequality}
\end{eqnarray}
From the ergodicity, the stationarity  and since the model is with
finite second moments, it follows the convergence almost surely of
the random variables $\frac{1}{n}\sum_{i=1}^{n}Y^2_{i-1}$ and
$\frac{1}{2}\sum_{i=1}^{n}(Y^2_{i-1} + Y^2_{i-2})$ to constants
$a_1$  and  $a_2$ respectively. The couples
$\Big(\rho_{n,1}-\rho_1,\frac{16a}{n}\sum_{i=1}^{n}Y^2_{i-1}\Big)$
and $\Big(\rho_{n,2}-\rho_2,\frac{8a}{n}\sum_{i=1}^{n}(Y^2_{i-1} +
Y^2_{i-2})\Big)$ converge in probability to $\Big(0,16a a_1\Big)$
and $\Big(0,8a a_2\Big)$ respectively, it follows from the
continuous mapping theorem (see for instance \cite{W}) applied on
the product and the sum of the functions that
$$\Big|\rho_{n,1}-\rho_1\Big|\times
\frac{16a}{n}\Big|\sum_{i=1}^{n}\frac{Y^2_{i-1}}{1+Y^2_{i-1}+Y^2_{i-2}}\Big|+\Big|\rho_{n,2}-\rho_2\Big|\times\frac{16a}{n}\Big|\sum_{i=1}^{n}
\frac{Y_{i-1}Y_{i-2}}{1+Y^2_{i-1}+Y^2_{i-2}}\Big|\stackrel{P}{\longrightarrow}0.$$
In connection with (\ref{normailzed difference between central
sequances}), it follows that, asymptotically, the quantities
$\frac{1}{\sqrt{n}}\frac{\partial\mathcal{V}_{n,h,h'}}{\partial\rho_1}(\Omega_n)$
and
$\frac{1}{\sqrt{n}}\frac{\partial\mathcal{V}_{n,h,h'}}{\partial\rho_1}(\Omega)$
have the same limit ( in probability sense). The random variables
$\frac{1}{\sqrt{n}}\frac{\partial\mathcal{V}_{n,h,h'}}{\partial\rho_1}(\Omega_n)$~~
and~~
$\frac{1}{\sqrt{n}}\frac{\partial\mathcal{V}_{n,h,h'}}{\partial\rho_2}(\Omega)$
converge to the constants
$-8a\,\mathbb{E}\Big[\frac{Y_{-1}}{1+Y^2_{-1}+Y^2_{-2}}\Big]$~~
and~~
$-8a\,\mathbb{E}\Big[\frac{Y_{-2}}{1+Y^2_{-1}+Y^2_{-2}}\Big]$
respectively. From the equalities (\ref{first constant}) and
(\ref{third constant}), it follows that:
\begin{eqnarray}
K^\top=-8a\Big(\mathbb{E}\Big[\frac{Y_{-1}}{1+Y^2_{-1}+Y^2_{-2}}\Big],\mathbb{E}\Big[\frac{Y_{-2}}{1+Y^2_{-1}+Y^2_{-2}}\Big]\Big),\\
J=J'=0,\quad   \mbox{and},  K'^\top=(0,0).
\end{eqnarray}
In sequel, we denote by $\bar{\Omega}_{1,n}$  the modified
estimate obtained after modifying the first component
$\rho_{1,n}$, under the assumptions $(P.0)$ and $(P.1)$,  we have
the following equalities:
\begin{eqnarray*}
  \bar{\rho}_{n,1} =
  \frac{D_{h,h'}(n)}{\frac{\partial\mathcal{V}_{n,h,h'}(\Omega_{n})}{\partial\rho_1}}
  +\hat{{\rho}}_{n,1} \quad
\mbox{~and~}\quad
 \bar{\rho}_{n,2}= \hat{\rho}_{n,2},\quad \mbox{with} \quad D_{h,h'}(n)=D_n= -\sqrt{n}(\hat{\Omega}_n
-\Omega).(\hat{K} +\hat{K}').
\end{eqnarray*}
For a fixed $\alpha = 0.05 $, the true value of the parameter
$(\rho_1,\rho_2)^\top$ is fixed at $(0.2,0.2)^\top$ and the sample
sizes are $n = 30, 40, 50,$ and $80.$\\
 We represent simultaneously the power test with a true
parameter $\rho_0,$ with the replace of the true parameter by its
least  square estimator L.S.E $\hat{\rho}_{n}$  and the empirical
power test which is obtained with the subsisting the true value
$\rho_0$ by it's estimate M.D.E. The correction of the estimation
is made with respect to the first parameter $\rho_1.$Throughout,
we denote by  $discrete(\Omega_n)$ the descritized form of the
estimator $\Omega_n,$ we obtain then
\begin{eqnarray*}
  \bar{\rho}_{n,1} =
  \frac{D_{h,h'}(n)}{\frac{\partial\mathcal{V}_{n,h,h'}(discrete(\Omega_{n}))}{\partial\rho_1}}
  +\hat{{\rho}}_{n,1} \quad
\mbox{~and~}\quad
 \bar{\rho}_{n,2}= discrete(\hat{\rho}_{n,2}),\quad \\\mbox{with} \quad D_{h,h'}(n)=D_n=
 -\sqrt{n}(\hat{discrete(\Omega}_n)
-\Omega).(\hat{K} +\hat{K}').
\end{eqnarray*}
By the replacing of the parameter $\Omega$  by it's estimator
$\Omega_n$ in the expression \ref{test}, we obtain the following
sequence of the test ${\hat{T}}_{1,n,h,h'}$, such that:
\begin{eqnarray}
  \hat{T}_{1,n,h,h'} &=& I{\Big\{{\frac{{\mathcal{\bar{V}}}_{n,h,h'}}{{\hat{\tau_{1}}}
_{h,h'}}\geq Z(u)}\Big\}},\label{testavecME}
\end{eqnarray}
where
\begin{eqnarray}
  {\hat{\tau_{1}}}^2 _{h,h'} &=&64a^2 \mathbf{E}\left(\frac{1}{1+Y^2_{-1}+Y^2_{-2}}\right)^2\Big[ h^2 \hat{I}_{n,0}
+h'^2 (\hat{I}_{n,2} - 1) + 2hh'(\hat{I}_{n,1}) \Big],~~~~~
\label{tauxdiscret}
 \end{eqnarray}
and
 \begin{eqnarray}
\hat{I}_{n,j}=\mathbf{E}\Big({\epsilon}^{j+2}_{n,0}\Big)=\mathbf{E}\Big({(Y_0  - \rho_{n,1}  Y_{-1} - \rho_{n,2}  Y_{-2})}^{j+2}\Big),\\
j=0,1,2.\nonumber
\end{eqnarray}
By the replacing of the estimate $\Omega_n$ by it's M.E.
$\bar{\Omega}_{n,1}$ in the expression  \ref{test}, we obtain the
following sequence of the test ${\bar{T}}_{1,n,h,h'}$, such that:
\begin{eqnarray}
  \bar{T}_{1,n,h,h'} &=& I{\Big\{{\frac{{\mathcal{\bar{V}}}_{n,h,h'}}{{\bar{\tau_{1}}}
_{h,h'}}\geq Z(u)}\Big\}},\label{testavecME}
\end{eqnarray}
where
\begin{eqnarray}
  {\bar{\tau_{1}}}^2 _{h,h'} &=&64a^2 \mathbf{E}\left(\frac{1}{1+Y^2_{-1}+Y^2_{-2}}\right)^2\Big[ h^2 \bar{I}_{n,0}
+h'^2 (\bar{I}_{n,2} - 1) + 2hh'(\bar{I}_{n,1}) \Big],~~~~~
\label{tauxdiscret}
 \end{eqnarray}
and
 \begin{eqnarray}
\bar{I}_{n,j}=\mathbf{E}\Big({\epsilon}^{j+2}_{n,0}\Big)=\mathbf{E}\Big({(Y_0  - \bar{\rho}_{n,1}  Y_{-1} - discrete(\hat{\rho}_{n,2})  Y_{-2})}^{j+2}\Big),\\
j=0,1,2.\nonumber
\end{eqnarray}
we give the representations of the power functions in terms to the
value of the constant $a$, the first representation (blue color)
corresponded to the power function with the true value of the
parameter, the second  corresponded (green color)  to the power
function with the least square estimator of the parameter and the
third (red color) corresponded to the power function with the
modified estimator M.D.E, then we obtain:
\begin{figure}[h!]
\includegraphics[scale=0.2]{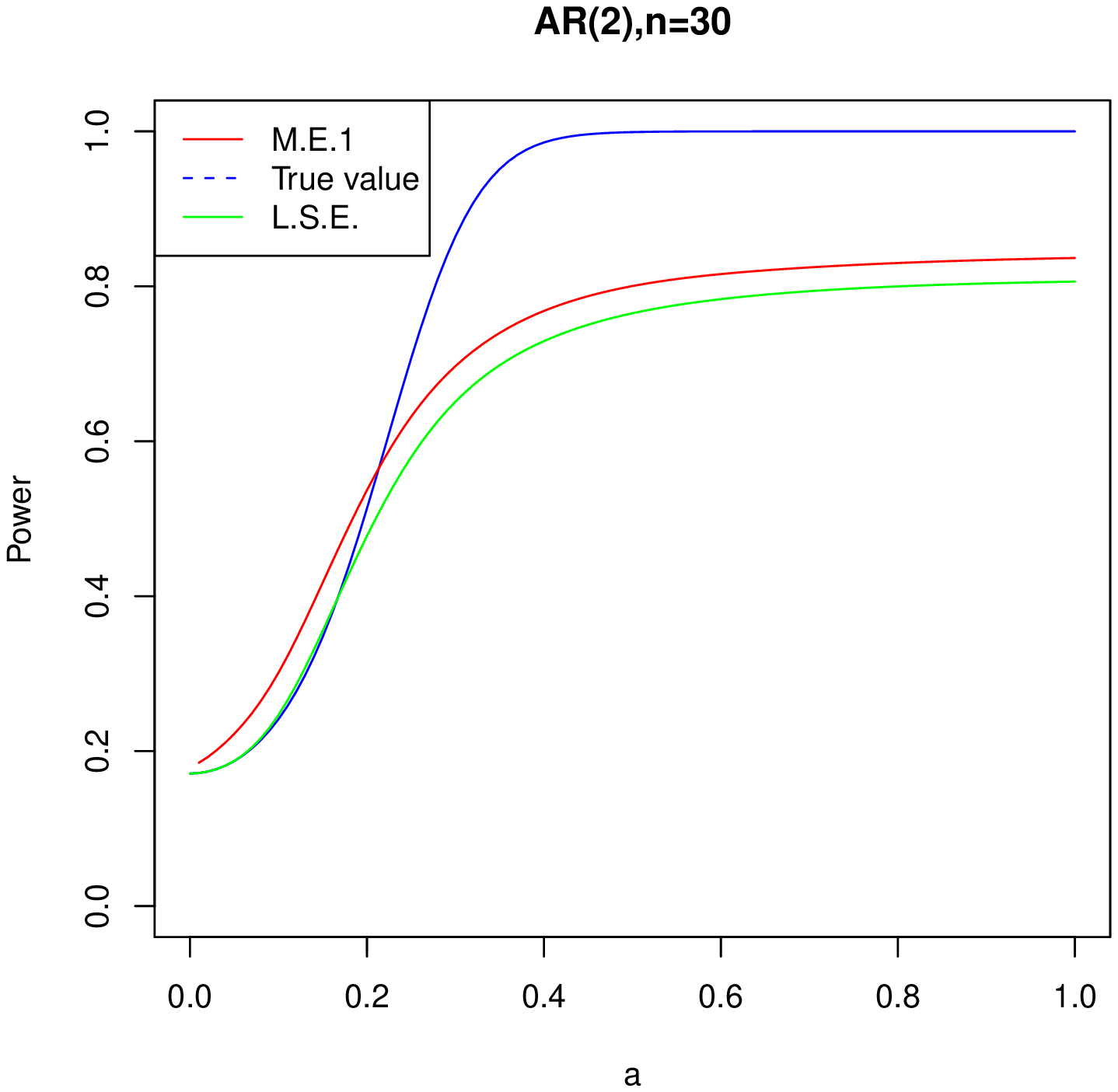}
\includegraphics[scale=0.2]{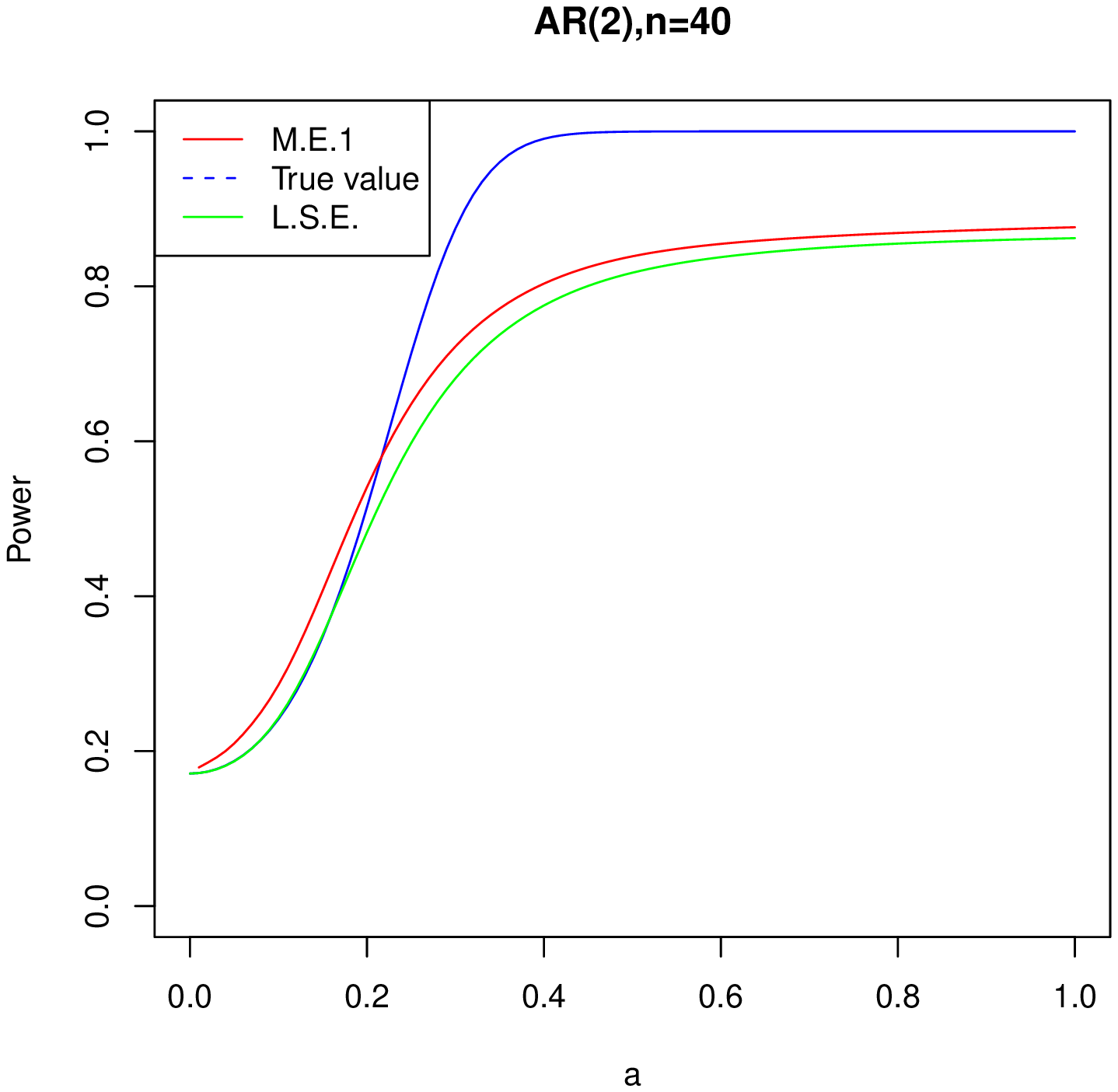}
\includegraphics[scale=0.2]{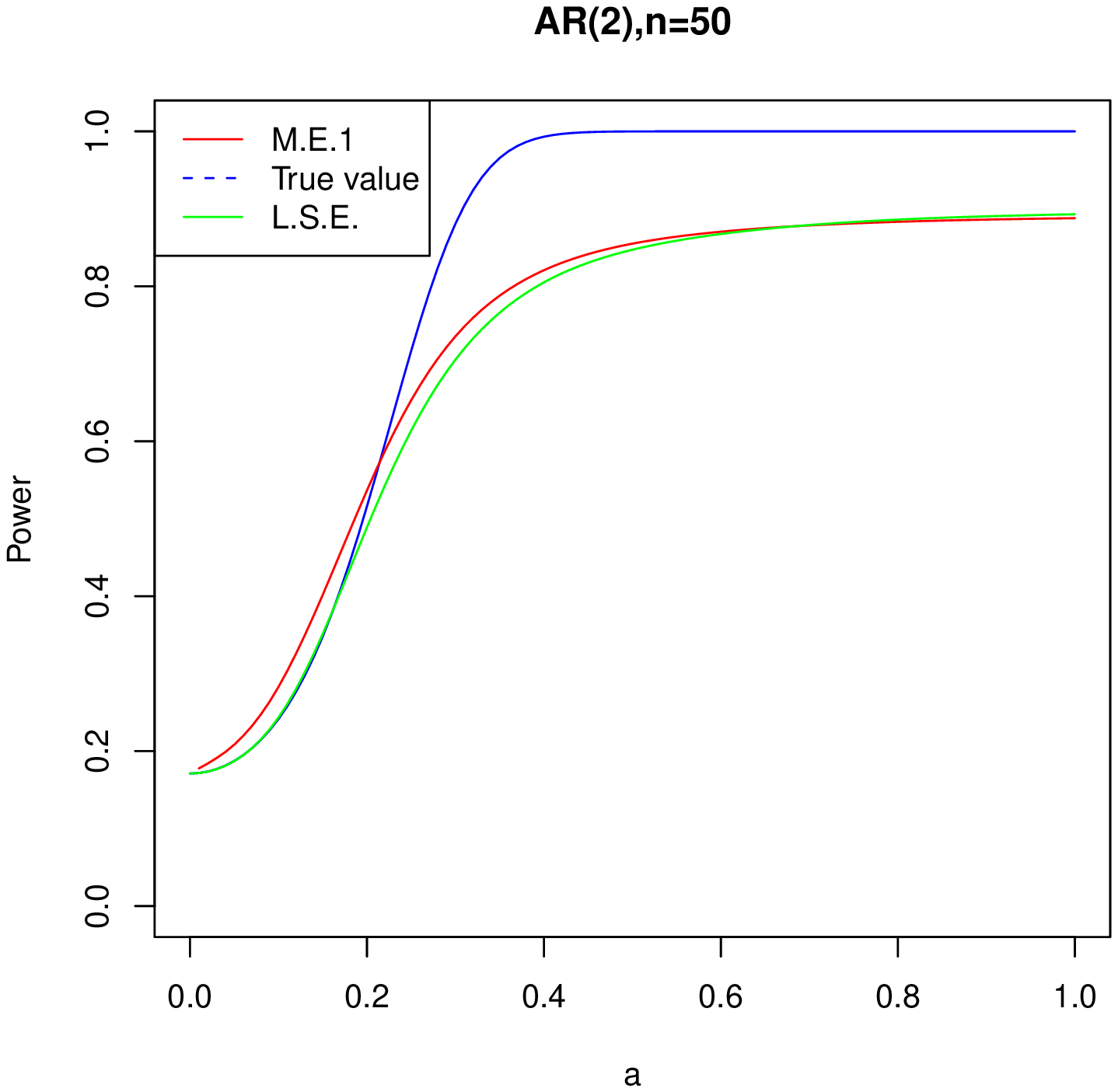}
\includegraphics[scale=0.2]{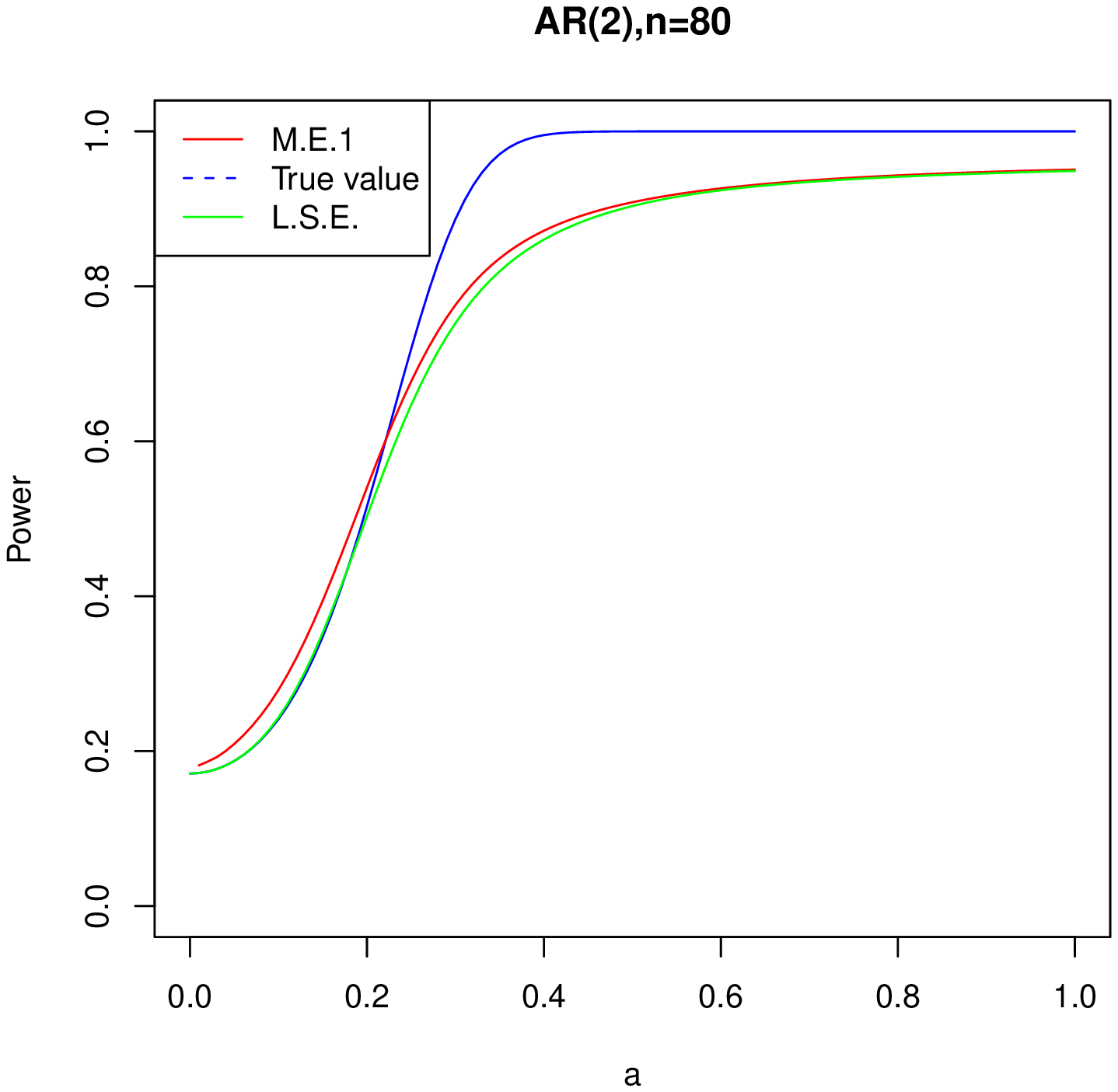}
\end{figure}

\textbf{\textbf{Correction with respect to $\rho_2$} }\\
With the same reasoning as the previous case and with the use of
the estimate $\bar{\Omega}_{n,2},$ we obtain following sequence of
the test ${\bar{T}}_{2,n,h,h'}$, such that:
\begin{eqnarray}
  \bar{T}_{n,h,h'} &=& I{\Big\{{\frac{{\mathcal{\bar{V}}}_{n,h,h'}}{{\bar{\tau_2}}
_{h,h'}}\geq Z(u)}\Big\}},\label{testavecME}
\end{eqnarray}
where
\begin{eqnarray*}
  \bar{\rho}_{n,2} = \frac{D_{h,h'}(n)}{\frac{\partial\mathcal{V}_{n,h,h'}(\Omega_{n})}{\partial\rho_2}}
  +\hat{{\rho}}_{n,2} \quad \mbox{~,~}\quad \bar{\rho}_{n,1}=
  \hat{\rho}_{n,1}, \\
 {\bar{\tau_{1}}}^2 _{h,h'} =a^2 \mathbf{E}\left(\frac{1}{1+Y^2_{-1}+Y^2_{-2}}\right)^2\Big[ h^2 \bar{I}_{n,0}
+h'^2 (\bar{I}_{n,2} - 1) + 2hh'(\bar{I}_{n,1}) \Big],\\
\bar{I}_{n,j}=\mathbf{E}\Big({\epsilon}^{j+2}_{n,0}\Big)=\mathbf{E}\Big({(Y_0  -  \hat{\rho}_{n,1} Y_{-1} - \bar{\rho}_{n,2}   Y_{-2})}^{j+2}\Big),\\
j=0,1,2.\nonumber
\end{eqnarray*}
therefore, we obtain the following representations:
\begin{figure}[h!]
\includegraphics[scale=0.2]{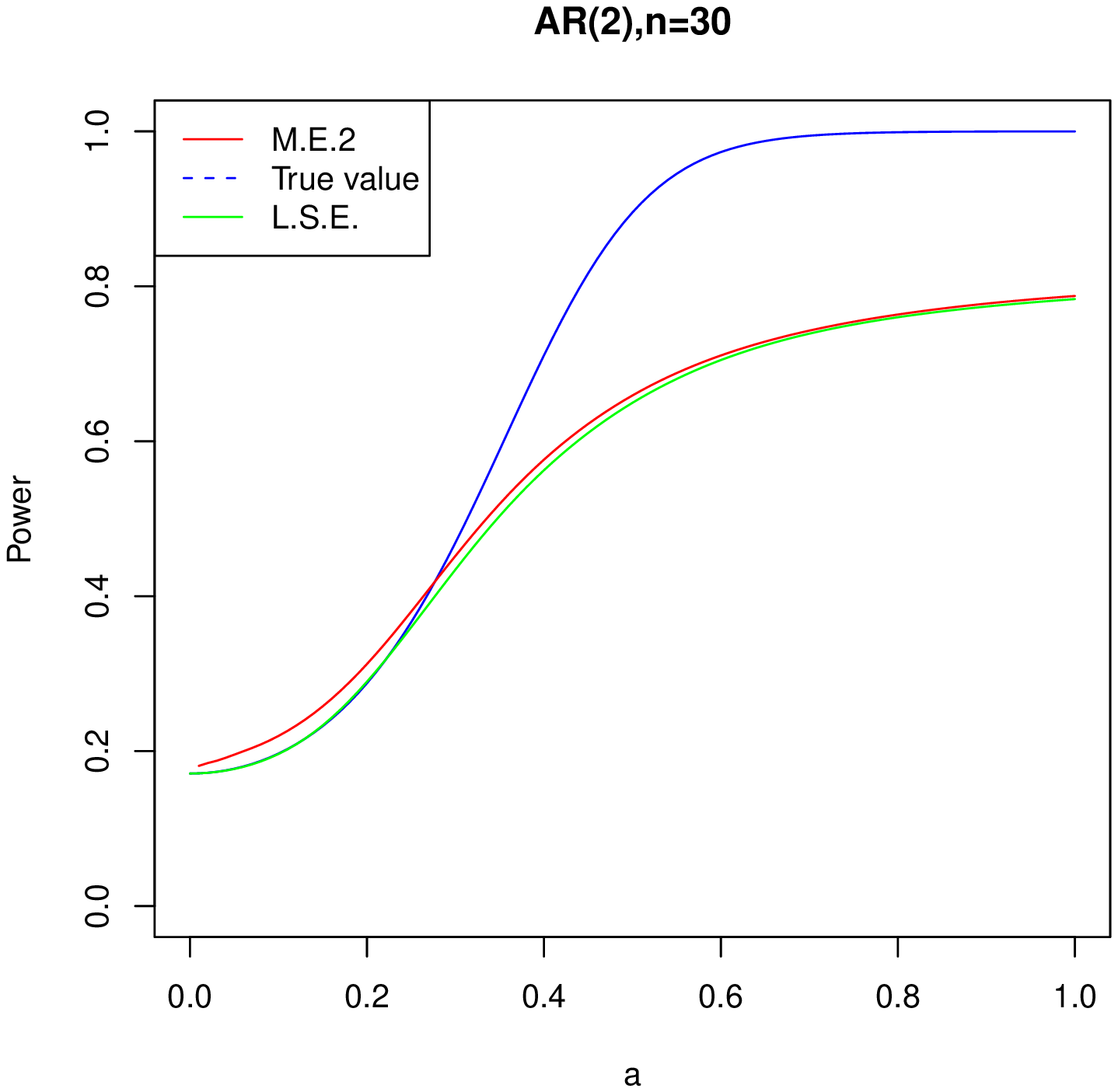}
\includegraphics[scale=0.2]{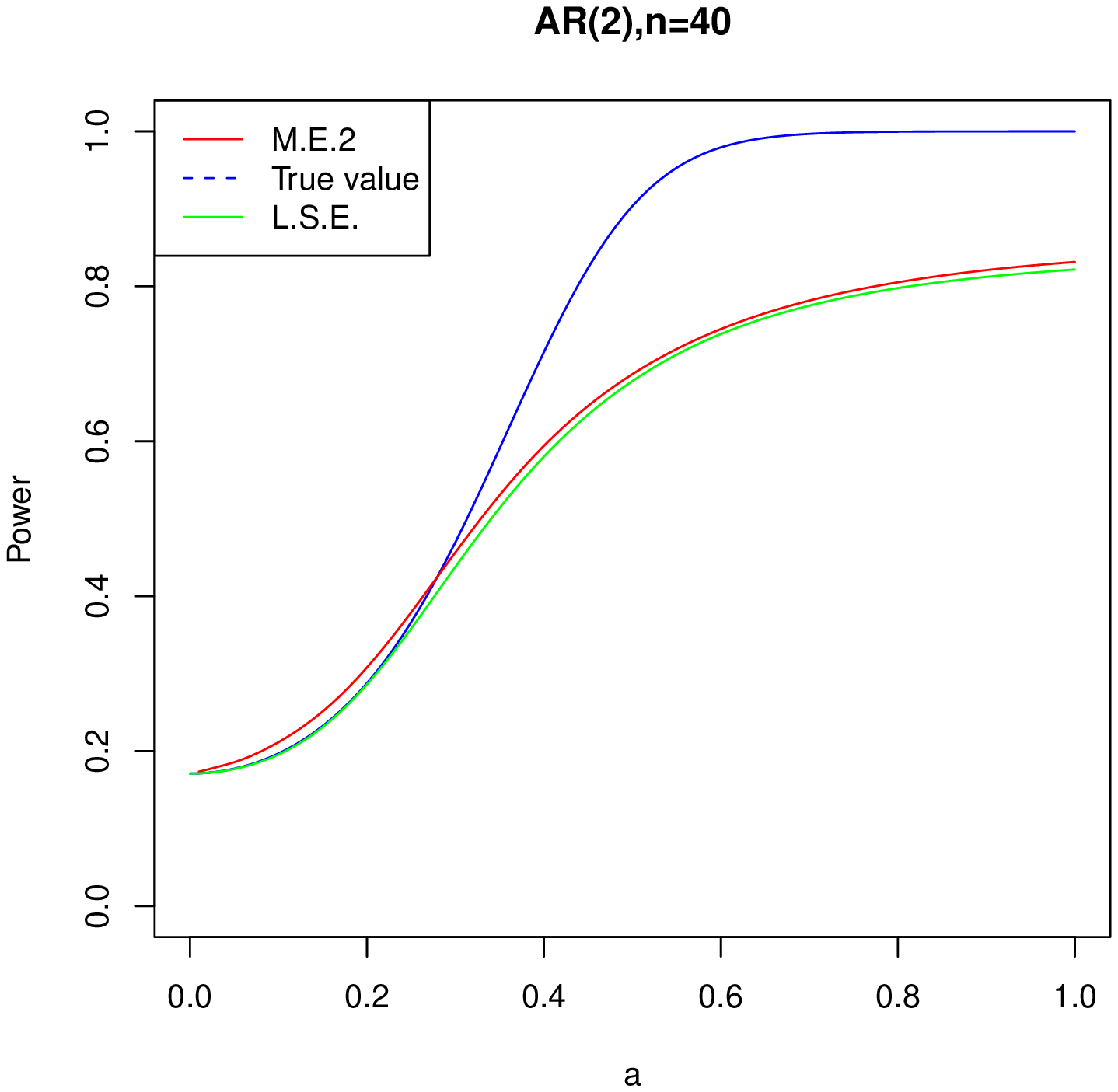}
\includegraphics[scale=0.2]{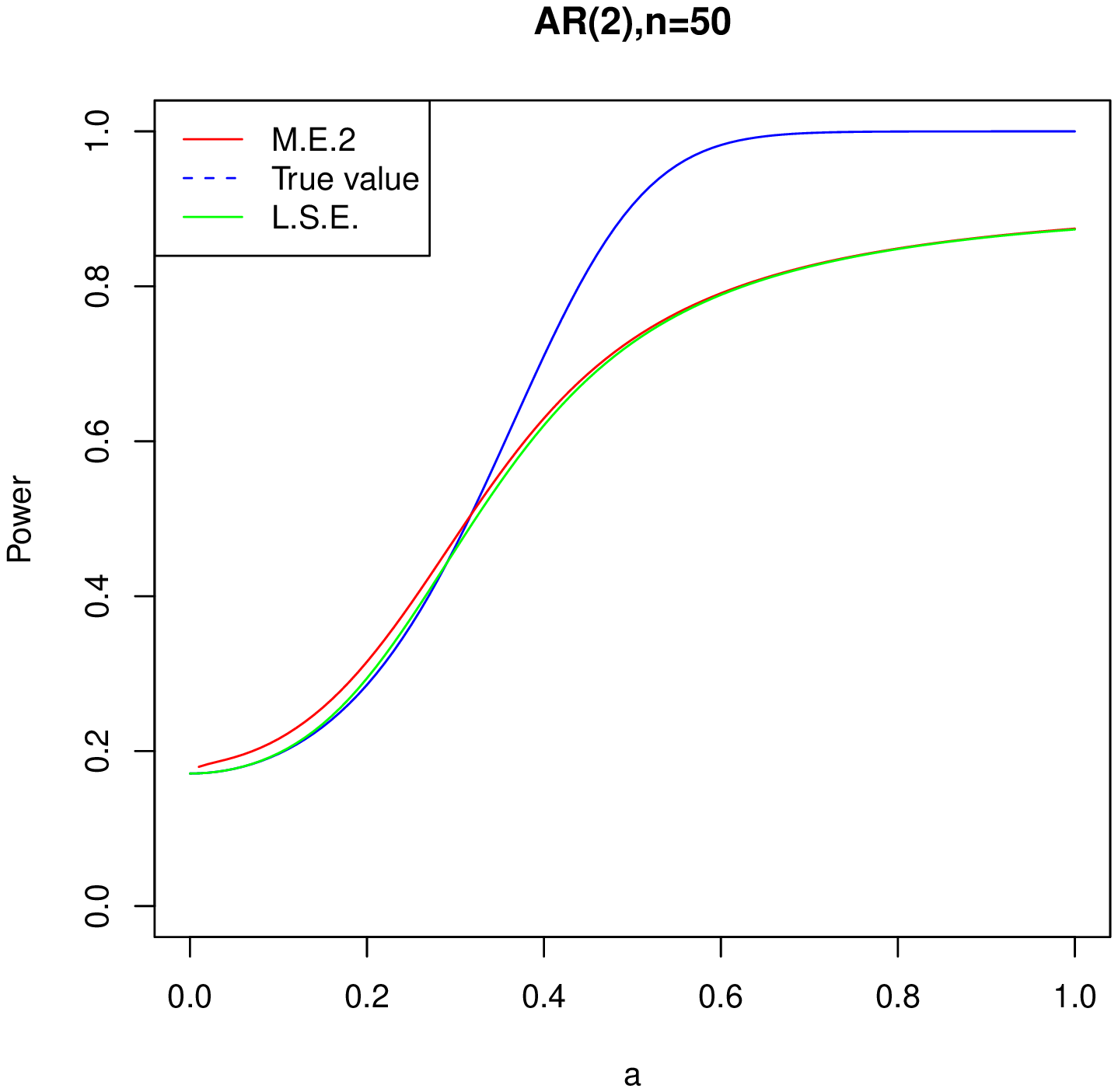}
\includegraphics[scale=0.2]{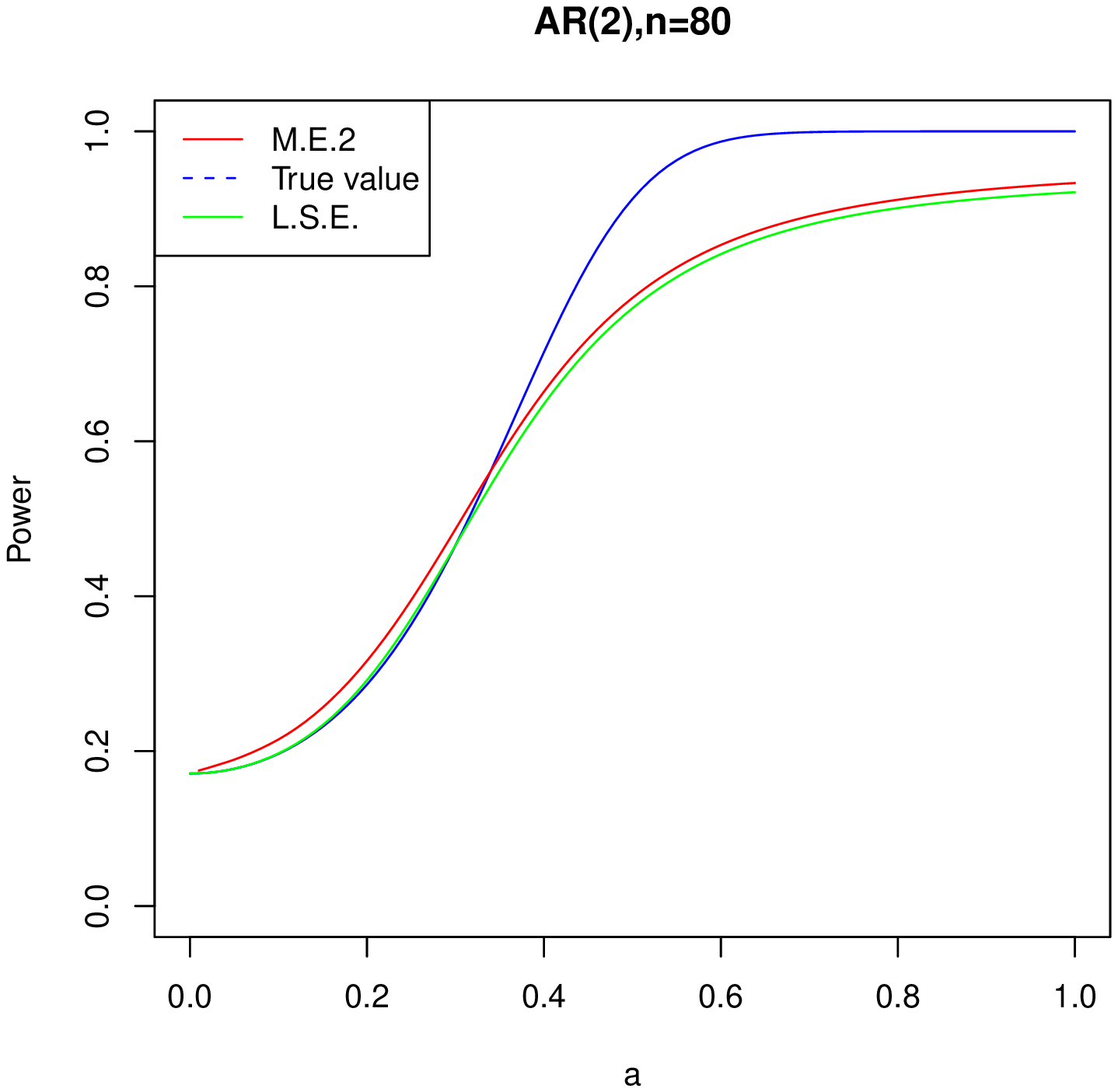}
\end{figure}
\section{Proofs of the results}
\emph{Throughout we fixe the step
 $(h,h')$ in the compact set $  K_{1} \times K_{2},$  where $hh'\neq
 0.$  $o_{P}(1) \in
\mathbb{R}\stackrel{\mathbf{P}}{\longrightarrow}0
\quad\mbox{as}\quad n\rightarrow\infty.$}\\
\noindent For some demonstrations, we need to prove the following
lemma :
\begin{lemma}\label{lemmainequality} Let $a$ and $ b$ are two positive reals
and   $\xi$ a real greater than $2$, then we have
\begin{eqnarray*}
(a+b)^{\xi} \leq 2^{\xi -1}(a^{\xi}+b^{\xi}).
\end{eqnarray*}
\end{lemma}
\subsection*{Proof of the Lemma \ref{lemmainequality}. }
 The function $\emph{d}: x \longmapsto x^{\xi} $ is twice
differentiable on $\mathbb{R}$, the second derivative function
$\ddot{\emph{d}}:{\xi}({\xi - 1})x^{\xi - 2} $ is positive on
${\mathbb{R}}^+$, therefore  $\emph{d}:x \longmapsto x^{\xi} $ is
a convex function on ${\mathbb{R}}^+$, then :\\$\forall (a,b) \in
{\mathbb{R}^+\times\mathbb{R}}^+$ and ~$\forall (\lambda_1 ,
\lambda_2) \in[0,1]\times[0,1]$ with $\lambda_1 + \lambda_2=1$, we
have $(\lambda_1 a + \lambda_2 b)^{\xi} \leq \lambda_1
a^{\xi}+\lambda_2 b^{\xi},$  By choosing  $\lambda_1
=\lambda_2=\frac{1}{2}$, we obtain the result.\\

\subsection*{Proof of the theorem \ref{firsttheorem}}
We check the three conditions $(C.1),$ $(C.2)$ and $(C.3)$  of
\cite[Theorem 1]{HB}.
\subsubsection*{Verification of the
condition $(C.1)$} \noindent Under $(H_{0}),$ and For
$i\in\{1,\dots,n\} $ and  we have : \begin{eqnarray*}
\Big|g_{n,i,h,h'} - 1\Big|= \Big|\frac{f_{h,h'}(Y_i)}{f_{0}(Y_i)}
-  1\Big|
 =\Big|\frac{f\Big(\frac{\epsilon_i - \alpha_{n,i,h}}{\beta_{n,i,h'}}\Big)}{f(\epsilon_i)} - 1\Big|
 =\Big|\frac{f\Big(\frac{\epsilon_i
 -\alpha_{n,i,h}}{\beta_{n,i,h'}}\Big)-f(\epsilon_i)}{f(\epsilon_i)}\Big|,
\end{eqnarray*}
where
\begin{eqnarray*}
\alpha_{n,i,h}= h\,n^{-\frac{1}{2}}
{\frac{G(Z_{i})}{{\sigma}(\theta_0 , Z_{i})}},\quad
\mbox{~and~}\quad
 \beta_{n,i,h'}=1+h'\,n^{-\frac{1}{2}}{\frac{S(Z_{i})}{{\sigma}(\theta_0 ,Z_{i})}}.
\end{eqnarray*}
\noindent Observe that
\begin{eqnarray*} \Big|g_{n,i,h,h'} - 1\Big|
&=&\Big|{F({\epsilon_i}};\alpha_{n,i,h},\beta_{n,i,h'})
-{F({\epsilon_i}};0,1)\Big|\Big|\frac{1}{f(\epsilon_i)}\Big|.
\end{eqnarray*}
\noindent By Taylor expansion of the function  $F({\epsilon_i}~;
\cdot,\cdot)$ around $(0,1)$, we obtain
\begin{eqnarray*}
  \Big|g_{n,i,h,h'} -1\Big|&=& \Big|\frac{\partial{F({\epsilon_i}};~0,1)}{f(\epsilon_i)\,\partial
a}\,{\alpha}_{n,i,h}
+\frac{\partial{F({\epsilon_i}};~0,1)}{f(\epsilon_i)\,\partial
b}\,({\beta}_{n,i,h'} - 1 )+\frac{
R_{n,i,h,h'}}{f(\epsilon_i)}\Big|,
\end{eqnarray*}
and,
\begin{eqnarray*}
 R_{n,i,h,h'}&=& \frac{1}{2}\left[{\alpha}_{n,i,h},\beta_{n,i,h'}
- 1\right]
\partial^2A_{n,i,h,h'}({\epsilon}_i;{\alpha}^\star_{n,i,h} ,
{\beta}^\star_{n,i,h'})\left[{\alpha}_{n,i,h},\beta_{n,i,h'} -
1\right]^\top,
\end{eqnarray*}
where, $({\alpha}^\star_{n,i,h} ,{\beta}^\star_{n,i,h'}) \in
[0,{\alpha}_{n,i,h}]\times[1, {\beta}_{n,i,h'}],$ and,
$\partial^2A_{n,i,h,h'}(\cdot~;{\alpha}^\star_{n,i,h},{\beta}^\star_{n,i,h'})$
is the hessian matrix of the function $F$ in
$(\cdot~;{\alpha}^\star_{n,i,h},{\beta}^\star_{n,i,h'})$.\\
Let
\begin{eqnarray*}
U_{n,i,h,h'}=\frac{\partial{F({\epsilon_i}};~0,1)}{f(\epsilon_i)\,\partial
a}\,{\alpha}_{n,i,h}
~~+\frac{\partial{F({\epsilon_i}};~0,1)}{f(\epsilon_i)\,\partial
b}\,({\beta}_{n,i,h'} - 1 )\mbox{~~and~~}
{R}^\ast_{n,i,h,h'}=\frac{ R_{n,i,h,h'}}{f(\epsilon_i)}.
\end{eqnarray*}
\noindent We have\begin{eqnarray*}
\frac{\partial{F({\epsilon_i}};~0,1)}{\partial
a}=-\dot{f}(\epsilon_i),\quad \mbox{~and,~}\quad
\frac{\partial{F({\epsilon_i}};~0,1)}{\partial b}=-\Big(
f(\epsilon_i) + \epsilon_i \,\dot{f}(\epsilon_i)\Big).
\end{eqnarray*}
\noindent Then\begin{eqnarray}
 U_{n,i,h,h'}&=&-n^{-\frac{1}{2}}\left\{h M_{f}(\epsilon_i)\frac{G(Z_i)}
 {{\sigma}(\theta_0,
        Z_i)}+h'( M_{f}(\epsilon_i){\epsilon_i}+1)\frac{S(Z_i)}
        {{\sigma}(\theta_0,
        Z_i)}\right\}.\label{centralepartiel}
 \end{eqnarray} We have:
\begin{center}
$g_{n,i,h,h'} - 1=U_{ n,i,h,h'}+{R}^\ast_{n,i,h,h'}.$
\end{center}
\noindent From $(A_{1.1}),$ there exist $p>1,$ a strictly positive
real $\varsigma,$ where
$\varsigma>\max(|{\alpha}^\star_{n,i,h}|,|{\beta}^\star_{n,i,h'}-1|)$
and a positive measurable function  $\varphi$ with
$\mathbf{E}({\varphi}^p(\epsilon_{0}))<+\infty$  such that
\begin{eqnarray}
|{R}^\ast_{n,i,h,h'}| &\leq &
\frac{1}{2}\left\{{{\alpha}^2_{n,i,h}}
+({\beta}_{n,i,h}-1)^2+2{\alpha}_{n,i,h}
({\beta}_{n,i,h}-1)\right\}\varphi(\epsilon_i)\nonumber\\
&\leq& \frac{1}{2}\left\{[{\alpha}_{n,i,h}+({\beta}_{n,i,h} -
1)]^2\varphi(\epsilon_i)\right\}\nonumber\\
&\leq&\frac{1}{2n}\left\{\frac{h\,G(Z_i) +
h'S(Z_i)}{\sigma(\theta_0,
        Z_i)}\right\}^2\varphi(\epsilon_i)\nonumber\\
&\leq&\frac{\delta}{n}\left\{
\frac{G(Z_i)+S(Z_i)}{\sigma(\theta_0,
        Z_i)}\right\}^2\varphi(\epsilon_i),\label{restemajoré}
\end{eqnarray}
\noindent where,  $\delta =\max({\delta_1}^2,{\delta_2}^2),$ and
${\delta_1}$ and ${\delta_2}$ are the diameters of the compact
sets  $K_{1}$ and $K_{2}$ respectively. \noindent Let $\nu
>1,$ by Markov 's inequality, we have for all $\gamma>0$
\begin{eqnarray*}
  \mathbf{P}\left(|{R}^\ast_{n,i,h,h'}|>\gamma\right) &=&\mathbf{P}\left(|{R}^\ast_{n,i,h,h'}|^{\nu}>{\gamma}^{\nu}\right)\leq
 \frac{1}{{\gamma}^{\nu}}\,\mathbf{E}|{R}^\ast_{n,i,h,h'}|^{\nu}.
\end{eqnarray*}

\noindent Then by the inequality (\ref{restemajoré}), we obtain
\begin{eqnarray*}
\mathbf{P}\left(|{R}^\ast_{n,i,h,h'}|>\gamma\right) &\leq &
\frac{1}{{\gamma}^{\nu}} \frac{{\delta}^{\nu}}{n^{\nu}}
\mathbf{E}\left\{\Big[\frac{G(Z_{i})~~+S(Z_{i})}{\sigma(\theta_0,
        Z_{i})}\Big]^{2{\nu}}{\varphi}^{\nu}(\epsilon_i)\right\}.
\end{eqnarray*}
\noindent It follows from the lemma (\ref{lemmainequality}), that
\begin{eqnarray*}
\left(\frac{G(Z_{i})+S(Z_i)}{\sigma(\theta_0,
        Z_i)}\right)^{2{\nu}}&\leq &
\left(\frac{|G(Z_i)|+|S(Z_i)|}{\sigma(\theta_0,
        Z_i)}\right)^{2{\nu}}\\
       &\leq & 2^{2{\nu} -
       1}\left\{\Big|\frac{G(Z_i)}{\sigma(\theta_0,
        Z_i)}\Big|^{2{\nu} }+\Big|\frac{S(Z_i)}{\sigma(\theta_0,
        Z_i)}\Big|^{2{\nu} }\right\}.
\end{eqnarray*}
\noindent Therefore by the stationarity, we have
\begin{eqnarray*}
 \mathbf{P}\Big(\Big|{R}^\ast_{n,i,h,h'}\Big|>\gamma\Big)&\leq & 2^{2{\nu}-1} \frac{1}{{\gamma}^{\nu}}
 \frac{\delta^\nu}{n^{\nu}}\Big\{\mathbf{E}\Big|\frac{G(Z_{0})}
 {\sigma(\theta_0,
        Z_{0})}\Big|^{2{\nu}
        }\mathbf{E}\Big[{\varphi}^{\nu}(\epsilon_{0})\Big]
        +\mathbf{E}\Big| \frac{S(Z_{0})}{\sigma(\theta_0,
        Z_{0})}\Big|^{2{\nu} }\mathbf{E}\Big[{\varphi}^{\nu}(\epsilon_{0})\Big]\Big\}\\
       &\leq & K 2^{2{\nu} - 1} \frac{1}{{\gamma}^{\nu}}\frac{{\delta}^{\nu}}{n^{\nu}}\left\{\mathbf{E}\Big|\frac{G(Z_{0})}{\sigma(\theta_0,
        Z_{0})}\Big|^{2{\nu} }+\mathbf{E}\Big|\frac{S(Z_{0})}{\sigma(\theta_0,
        Z_{0})}\Big|^{2{\nu}}\right\}.
\end{eqnarray*}

\noindent We have $2{\nu}>2$, then there exist $ \lambda>0$, such
that $2{\nu}={\lambda} +2$, we obtain
\begin{eqnarray*}
\mathbf{P}(\max_{i\in\{1,,\ldots,n\}}
\Big|{R}^\ast_{n,i,h,h'}\Big|>\gamma)&\leq&\sum_{i=1}^{n}\mathbf{P}\Big(
\Big|{R}^\ast_{n,i,h,h'}\Big|>\gamma\Big)\\
&\leq&
K2^{2{\nu}-1}\frac{{\delta}^\nu}{n^\nu{\gamma}^\nu}}\Big\{{\sum_{i=1}^{n}\mathbf{E}\Big|\frac{G(Z_{0})}
{\sigma(\theta_0,Z_{0})}\Big|^{2{\nu}}+\sum_{i=1}^{n}\mathbf{E}\Big|\frac{S(Z_{0})}{\sigma(\theta_0,Z_{0})}\Big|^{2{\nu} }\Big\}\\
       &\leq & K2^{{\lambda} +1
}\frac{{\delta}^{{\frac{\lambda}{2}}+1}}{n^{\frac{\lambda}{2}}{\gamma}^{\frac{\lambda}{2}+1}}\
\frac{1}{n}\Big\{\sum_{i=1}^{n}\mathbf{E}\Big|\frac{G(Z_{0})}{\sigma(\theta_0,
 Z_{0})}\Big|^{{\lambda} +2}+\sum_{i=1}^{n}\mathbf{E}
\Big|\frac{S(Z_{0})}{\sigma(\theta_0,
        Z_{0})}\Big|^{{\lambda} +2 }\Big\}\\
&\leq & K2^{{\lambda} +1
}\frac{{\delta}^{{\frac{\lambda}{2}}+1}}{n^{\frac{\lambda}{2}}{\gamma}^{\frac{\lambda}{2}+1}}\
\Big\{\mathbf{E}\Big|\frac{G(Z_{0})}{\sigma(\theta_0,
 Z_{0})}\Big|^{{\lambda} +2}+\mathbf{E}
\Big|\frac{S(Z_{0})}{\sigma(\theta_0,
        Z_{0})}\Big|^{{\lambda} +2 }\Big\}.
\end{eqnarray*}
\noindent It follows from $(A_{3.1})$ and $(A_{3.2})$ that
\begin{center}
$\mathbf{P}\left(\max_{i\in\{1,,\ldots,n\}}|{R}^\ast_{n,i,h,h'}|)>\gamma\right)\rightarrow
0 $ as $ n\rightarrow +\infty.$
\end{center} So we have
\begin{eqnarray}
 \max_{i\in\{1,,\ldots,n\}}\Big|{R}^\ast_{n,i,h,h'}\Big| &=&
 o_{P}(1) .\label{c1}
\end{eqnarray}
\noindent Now we have to show  that $$\max_{i\in\{1,,\ldots,n\}}
\Big|U_{n,i,h,h'}\Big|=o_{P}(1).$$ Remark that
\begin{eqnarray*}
  \mathbf{P}\left(\max_{i\in\{1,,\ldots,n\}}\Big|U_{n,i,h,h'}\Big|>\gamma\right)&\leq&\sum_{i=1}^{n}
  \mathbf{P}\Big(\Big|U_{n,i,h,h'}\Big|^{2{\nu}}>{\gamma}^{2{\nu}}\Big).
\end{eqnarray*}
\noindent It follows from Markov's inequality that, for all
 $\gamma>0$,  we have
\begin{eqnarray}
  \mathbf{P}\Big(\max_{i\in\{1,,\ldots,n\}}\Big|U_{n,i,h,h'}\Big|>\gamma\Big)&\leq&\frac{1}{{\gamma}^{2{\nu}}}
  \sum_{i=1}^{n}\mathbf{E}\Big|U_{n,i,h,h'}\Big|^{2{\nu}}.\label{markovpouru}
\end{eqnarray}
\noindent From the lemma (\ref{lemmainequality}), we can deduce
that
\begin{eqnarray*}
  \mathbf{E}\Big|U_{n,i,h,h'}\Big|^{2{\nu}} &\leq& n^{-{\nu}} {\delta }^{\nu} 2^{2{\nu}- 1}\left\{
  \mathbf{E}\left(\Big|M_{f}(\epsilon_i)\Big|^{2{\nu}}\Big|\frac{G(Z_i)}{{\sigma}(\theta_0,
        Z_i)}\Big|^{2{\nu}}\right)+\mathbf{E}\left(\Big|M_{f}(\epsilon_i){\epsilon_i}+1\Big|^{2{\nu}}\Big|\frac{S(Z_i)}{{\sigma}(\theta_0,
        Z_i)}\Big|^{2{\nu}}\right)\right\}.
\end{eqnarray*}
\noindent Combined this in connection with (\ref{markovpouru}), it
results that
\begin{eqnarray*}
 \mathbf{P}\left(\max_{i\in\{1,,\ldots,n\}}|U_{n,i,h,h'}|>\gamma\right)&\leq& \frac{{\delta }^{\nu} 2^{2{\nu}- 1}}
{n^{{\nu}-1}{\gamma}^{2{\nu}}}
\mathbf{E}\Big\{\frac{1}{n}\sum_{i=1}^{n}\Big|M_{f}(\epsilon_i)\Big|^{2{\nu}}\Big|\frac{G(Z_i)}{{\sigma}(\theta_0,
        Z_i)}\Big|^{2{\nu}}\Big\}\\
&+&\frac{{\delta }^{\nu} 2^{2{\nu}- 1}}
{n^{{\nu}-1}{\gamma}^{2{\nu}}}\mathbf{E}\Big\{\frac{1}{n}\sum_{i=1}^{n}\Big|M_{f}(\epsilon_i){\epsilon_i}+1\Big|^{2{\nu}}\Big|\frac{S(Z_i)}{{\sigma}(\theta_0,
        Z_i)}\Big|^{2{\nu}}\Big\}\\
        &\leq& \frac{{\delta }^{\frac{\lambda}{2}+1} 2^{{\lambda}+1}}
{n^{\frac{\lambda}{2}}{\gamma}^{{\lambda}+2}}
\Big\{\frac{1}{n}\sum_{i=1}^{n}\mathbf{E}\Big|M_{f}(\epsilon_i)\Big|^{{\lambda}+2}\mathbf{E}\Big|\frac{G(Z_i)}{{\sigma}(\theta_0,
        Z_i)}\Big|^{{\lambda}+2}\Big\}\\
&+&\frac{{\delta }^{\frac{\lambda}{2}+1} 2^{{\lambda}+1}}
{n^{\frac{\lambda}{2}}{\gamma}^{{\lambda}+2}}\Big\{\frac{1}{n}\sum_{i=1}^{n}\mathbf{E}\Big|M_{f}(\epsilon_i){\epsilon_i}+1\Big|^{{\lambda}+2}\mathbf{E}\Big|\frac{S(Z_i)}
{{\sigma}(\theta_0,
        Z_i)}\Big|^{{\lambda}+2}\Big\}.
\end{eqnarray*}
\noindent We can remark after using the lemma
(\ref{lemmainequality}) that\begin{eqnarray}
\Big|M_{f}(\epsilon_i){\epsilon_i}+1\Big|^{{\lambda}+2}&\leq&
2^{{\lambda}+1}
\Big|M_{f}(\epsilon_i){\epsilon_i}\Big|^{{\lambda}+2} +
2^{{\lambda}+1}. \label{majrés}
\end{eqnarray} It follows from $(A_{3.1})$, $(A_{3.2})$, $(A_{3.3}),$
$(A_{3.4})$ and the stationarity  of the model that
\begin{eqnarray}
  \max_{i\in\{1,\dots,n\}}
\Big|U_{n,i,h,h'}\Big|&=&o_{P}(1).\label{c2}
\end{eqnarray}
\noindent We deduce from the equalities  (\ref{c1}) and (\ref{c2})
that the condition $(CC1)$ is satisfied.

\subsubsection*{Verification of the condition $(C.2)$}
 We have\begin{eqnarray*}
  \sum_{i=1}^{n}({g_{n,i,h,h'}} - 1)^2 &=& \sum_{i=1}^{n}{U^2_{
n,i,h,h'}+\sum_{i=1}^{n}({R}^\ast_{n,i,h,h'}})^2 +
2\sum_{i=1}^{n}U_{ n,i,h,h'}{{R}^\ast_{n,i,h,h'}}.
\end{eqnarray*}
\noindent Using the inequality (\ref{restemajoré}) followed by a
simple majoration, we obtain
\begin{eqnarray*}
\sum_{i=1}^{n}({R}^\ast_{n,i,h,h'})^2&\leq&\max_{i\in\{1,,\ldots,n\}}\Big|{R}^\ast_{n,i,h,h'}\Big|\sum_{i=1}^{n}\mid{R}^\ast_{n,i,h,h'}\mid\\
&&\leq
\max_{i\in\{1,,\ldots,n\}}\Big|{R}^\ast_{n,i,h,h'}\Big|\left
\{\frac{\delta}{2n}\sum_{i=1}^{n}\Big[
\frac{G(Z_i)+S(Z_i)}{\sigma(\theta_0,
        Z_i)}\Big]^2\varphi(\epsilon_i)\right\}\\
&&\leq \max_{i\in\{1,,\ldots,n\}}\Big|{R}^\ast_{n,i,h,h'}\Big|
\left \{ \frac{\delta}{n} \sum_{i=1}^{n}\varphi(\epsilon_i)
\Big|\frac{G(Z_i)}{\sigma(\theta_0,Z_i)}\Big|^2+\frac{\delta}{n}\sum_{i=1}^{n}\varphi(\epsilon_i)
\Big|\frac{S(Z_i)}{\sigma(\theta_0, Z_i)}\Big|^2\right\}.
\end{eqnarray*}
 Let
  \begin{eqnarray*}
 A_{n,i,{\delta}}=\frac{\delta}{n}\, \sum_{i=1}^{n}
\varphi(\epsilon_i)\Big|\frac{G(Z_i)}{\sigma(\theta_0,Z_i)}\Big|^2
,\quad \mbox{~and~}\quad B_{n,i,{\delta}} = \frac{\delta}{n}\,
\sum_{i=1}^{n}
\varphi(\epsilon_i)\Big|\frac{S(Z_i)}{\sigma\theta_0,Z_i}\Big|^2.
\end{eqnarray*}
\noindent We consider the set of the events $\Omega_1$  such that
$\Omega_1=\{\omega,\varphi(\epsilon_i)\leq 1\}$, it is clear that
on the complementary  ${\Omega_1}^c$  of the set ${\Omega_1},$ we
have, for all real $p>1$,
~$\varphi(\epsilon_i)\leq{\varphi^p(\epsilon_i)}$  (In this case
we choose a value $p$  which is  corresponded to the condition
$(A_{1.1})$), therefore :
\begin{eqnarray*}
 | A_{n,i,{\delta}}|&\leq&\left\{\frac{\delta}{n} \sum_{i=1}^{n}
\varphi(\epsilon_i)I_{\Omega_1}\Big|\frac{G(Z_i)}{\sigma(\theta_0,
        Z_i)}\Big|^2 \right\}
 +\left\{\frac{\delta}{n} \sum_{i=1}^{n}
\varphi(\epsilon_i)I_{{\Omega_1}^c}\Big|\frac{G(Z_i)}{\sigma(\theta_0,
        Z_i)}\Big|^2 \right\}\\
        &\leq&\left\{\frac{\delta}{n} \sum_{i=1}^{n}
\Big|\frac{G(Z_i)}{\sigma(\theta_0,
        Z_i)}\Big|^2 \right\}
+\left\{\frac{\delta}{n} \sum_{i=1}^{n}
\varphi^p(\epsilon_i)\Big|\frac{G(Z_i)}{\sigma(\theta_0,
        Z_i)}\Big|^2 \right\},
\end{eqnarray*}
\noindent where  $I(\cdot)$ denotes the indicator function  .\\
 Let
\begin{eqnarray*}
A^\star_{n,i,{\delta}}&=&\left\{\frac{\delta}{n} \sum_{i=1}^{n}
\Big|\frac{G(Z_i)}{\sigma(\theta_0,
        Z_i)}\Big|^2 \right\} +\left\{\frac{\delta}{n} \sum_{i=1}^{n}
\varphi^p(\epsilon_i)\,\Big|\frac{G(Z_i)}{\sigma(\theta_0,
        Z_i)}\Big|^2 \right\}.
        \end{eqnarray*}
\noindent From the ergodic theorem and $(A_{1.1})$ and since the
second moments of the model are finite, it results that the random
variable $A^\star_{n,i,{\delta}}$ converges a.s. to some constant
$c_1$ as $n\rightarrow+\infty$. \\ Let
\begin{eqnarray*}
B^\star_{n,i,{\delta}}&=&\left\{\frac{\delta}{n} \sum_{i=1}^{n}
\Big|\frac{S(Z_i)}{\sigma(\theta_0,
        Z_i)}\Big|^2 \right\} +\left\{\frac{\delta}{n} \sum_{i=1}^{n}
\varphi^p(\epsilon_i)\,\Big|\frac{S(Z_i)}{\sigma(\theta_0,
        Z_i)}\Big|^2 \right\}.
 \end{eqnarray*}
\noindent With a same reasoning as $A^\star_{n,i,{\delta}},$ we
can show that the random variable $B^\star_{n,i,{\delta}}$
converges a.s. to some constant $c_2$  as $n\rightarrow+\infty,$
therefore the random variable $A^\star_{n,i,{\delta}}+
B^\star_{n,i,{\delta}}$ converges to $c=c_1 + c_2$  a.s. as
$n\rightarrow+\infty.$\\ The random vector
$\Big((A^\star_{n,i,{\delta}} + B^\star_{n,i,{\delta}})~,~
\max_{i\in\{1,,\ldots,n\}}|{R}^\ast_{n,i,h,h'}|\Big)$ converges in
probability to $\Big(c,0\Big)$. Since the function,
$(x,y)\longmapsto xy$ is continuous, it results from continuous
mapping theorem  (\cite{W}) that
$$\max_{i\in\{1,\dots,n\}}|{R}^\ast_{n,i,h,h'}|\,(A^\star_{n,i,{\delta}}+
B^\star_{n,i,{\delta}})\stackrel{\mathbf{P}}{\longrightarrow}0
\quad\mbox{a.s.}\quad n\rightarrow\infty,$$ which implies
\begin{eqnarray}
  \sum_{i=1}^{n}({R}^\ast_{n,i,h,h'})^2&=&o_{P}(1)\label{c3}.
\end{eqnarray}
We have
\begin{eqnarray*}
 \sum_{i=1}^{n}U_{ n,i,h,h'}{{R}^\ast_{n,i,h,h'}}&\leq&\sum_{i=1}^{n}\Big|U_{ n,i,h,h'}\Big|\Big|{{R}^\ast_{n,i,h,h'}}\Big|\\
 &&\leq\max_{i\in\{1,\dots,n\}}
    \Big|U_{n,i,h,h'}\Big|\sum_{i=1}^{n}\Big|{{R}^\ast_{n,i,h,h'}}\Big|\\
     &&\leq \max_{i\in\{1,\dots,n\}}
    \Big|U_{n,i,h,h'}\Big|(A^\ast_{n,i,{\delta}}+
    B^\ast_{n,i,{\delta}}).
\end{eqnarray*}
\noindent Using the same arguments as in the last case and
(\ref{c2}), we can show that,
\begin{eqnarray}
   \sum_{i=1}^{n}U_{n,i,h,h'}{{R}^\ast_{n,i,h,h'}}&=&
   o_{P}(1). \label{c4}
\end{eqnarray}
\noindent We have
\begin{eqnarray*}
  \sum_{i=1}^{n}{U^2_{n,i,h,h'}} &=&\sum_{i=1}^{n}{\left\{-n^{-\frac{1}{2}}\,\Big[h\, M_{f}(\epsilon_i)\frac{G(Z_i)} {{\sigma}(\theta_0,
        Z_i)} +h'(M_{f}(\epsilon_i){\epsilon_i}+1)\,\frac{S(Z_i)} {{\sigma}(\theta_0,
        Z_i)}\Big]\right\}}^2\\
&=&\frac{1}{n}\sum_{i=1}^{n}{{\Big(h
M_{f}(\epsilon_i)\frac{G(Z_i)} {{\sigma}(\theta_0,
        Z_i)}}\Big)^2 +{\Big(h'(M_{f}(\epsilon_i){\epsilon_i}+1)\frac{S(Z_i)} {{\sigma}(\theta_0,
        Z_i)}\Big)}^2}\\
&&+2hh'\frac{1}{n}\left\{\sum_{i=1}^{n}M_{f}(\epsilon_i)[M_{f}(\epsilon_i){\epsilon_i}+1]\frac{G(Z_i)S(Z_i)}
{{\sigma}^2(\theta_0,
        Z_i)}\right\}.
\end{eqnarray*}
\noindent  Note that
\begin{eqnarray*}
 \mathbf{E}\Big [\frac{G(Z_i)S(Z_i)} {{\sigma}^2(\theta_0,
        Z_i)}\Big] &\leq& \mathbf{E}\Big[\frac{|G(Z_i)S(Z_i)|} {{\sigma}^2(\theta_0,
        Z_i)}\Big]\\
&\leq& \frac{1}{2}\mathbf{E}\Big[\frac{G^2(Z_i)}
{{\sigma}^2(\theta_0,
        Z_i)}\Big]+\frac{1}{2}\mathbf{E}\Big[\frac{S^2(Z_i)} {{\sigma}^2(\theta_0,
        Z_i)}\Big]< +\infty .
\end{eqnarray*}
\noindent It follows from the ergodicity and stationarity of the
model, that the random variable $\sum_{i=1}^{n}{U^2_{n,i,h,h'}}$
converges  a.s. to a positive constant ${\tau}^2 _{h,h'}$ as
$n\rightarrow +\infty$, where
\begin{eqnarray}
  \nonumber{\tau}^2 _{h,h'} &=&h^2\mathbf{E}\left[{M^{2}_{f}}(\epsilon_i) \Big(\frac{G(Z_i)} {{\sigma}(\theta_0,
        Z_i)}\Big)^2\right] +h'^2\mathbf{E}\left[{[1+\epsilon_i M_{f}(\epsilon_i)]}^2\Big(\frac{S(Z_i)} {{\sigma}(\theta_0,
        Z_i)}\Big)^2\right]\\&&+  2hh' \mathbf{E}\left[[\epsilon_i{M^2_{f}(\epsilon_i)} + M_{f}(\epsilon_i)]\frac{G(Z_i)S(Z_i)} {{\sigma}(\theta_0,
        Z_i)}\right].\label{tauxequation}
\end{eqnarray}
Let  $I_j
=\mathbf{E}\Big({\epsilon}^j_{0}{M^2_{f}}(\epsilon_{0})\Big)$ and
$K_j =\mathbf{E}\Big({\epsilon}^j_{0}{M_{f}}(\epsilon_{0})\Big)$ ,
$j\in\{0,1,2\}.$ It results from ($A_{1.2}$) and ($A_{2.2}$), that
\begin{eqnarray}
  {\tau}^2 _{h,h'} &=& h^2 I_0\mathbf{E}\left(\frac{G(Z_{0})}
        {{\sigma}(\theta_0,Z{0})}\right)^2+h'^2 (I_2+2K_1+1)\mathbf{E}\left(\frac{S(Z_{0})}
{{\sigma}(\theta_0,
        Z_{0})}\right)^2+2hh'(I_1+K_0) \mathbf{E}\left(\frac{G(Z_{0})S(Z_{0})}
{{\sigma^2}(\theta_0,
        Z_{0})}\right)\nonumber\\
        &=&h^2 I_0 \mathbf{E}\left(\frac{G(Z_{0})}
        {{\sigma}(\theta_0,Z{0})}\right)^2
+h'^2 (I_2 - 1) \mathbf{E}\left(\frac{S(Z_{0})}
{{\sigma}(\theta_0,
        Z_{0})}\right)^2 + 2hh'(I_1) \mathbf{E}\left(\frac{G(Z_{0})S(Z_{0})} {{\sigma^2}(\theta_0,
        Z_{0})}\right).~~~~~ \label{taux}
 \end{eqnarray}
It follows from (\ref{c3}), (\ref{c4}) and (\ref{taux}), that the
condition $(C.2)$ is satisfied.\\
\subsubsection*{Verification of the  condition $(C.3)$}
 Let\begin{eqnarray}
 \mathcal{V}_{n,h,h'}(\rho_0,\theta_0)&=& \sum_{i=1}^{n}{U_{n,i,h,h'}}.\label{centrale}
  \end{eqnarray}
From ($A_{2.1}$) and ($A_{2.2}$), ${U_{n,i,h,h'}}$ is a $
\mathcal{F}_{n}$ centred martingale. In order to prove that the
random variable  $\mathcal{V}_{n,h,h'}(\rho_0,\theta_0)$ converges
in distribution to $ \mathcal{N}(0,{\tau}^2 _{h,h'})$ as
$n\rightarrow +\infty$, we use \cite[Theorem 3.2., Corollaries
3.1., and 3.2,]{PH} therefore we check the following conditions :
\begin{itemize}
\item (i) \emph{ Linderberg condition}: for all $\gamma>0$,
$$\sum_{i=1}^{n}\mathbf{E}\left(U^2_{n,i,h,h'}I_{\Big\{|U_{n,h,h'}|>\gamma\Big\}}/{\mathcal{F}_{i-1}}\right)
\stackrel{\mathbf{P}}{\longrightarrow}0 \quad\mbox{as}\quad
n\rightarrow\infty.$$
 \item (ii) \emph{Conditionally variance:}
$$\sum_{i=1}^{n}\mathbf{E}(U^2_{n,i,h,h'}/{\mathcal{F}_{i-1}})\stackrel{\mathbf{P}}{\longrightarrow}{\eta}^2
\quad\mbox{as}\quad n\rightarrow\infty.$$
  \item (iii) \emph{Measurability:}\\
 The random variable  $\eta$ is measurable on the field
 ${\mathcal{F}_{i-1}}$.
\end{itemize}
\subsubsection*{ Verification of the Linderberg condition }
\noindent By the conditionally  H\"{o}lder's inequality, there
exist $\nu>1$ and $p>1$, $\frac{1}{\nu}+\frac{1}{p}=1$ such
that:\begin{eqnarray*}
\mathbf{E}\left(U^2_{n,i,h,h'}I_{\Big\{|U_{n,i,h,h'}|>\gamma\Big\}}/{\mathcal{F}_{i-1}}\right)
&\leq&{\Big\{\mathbf{E}(|U_{n,i,h,h'}|^{2\nu}/{\mathcal{F}_{i-1}})}\Big\}^\frac{1}{\nu}\Big\{{\mathbf{P}(|U_{n,i,h,h'}|>\gamma/{\mathcal{F}_{i-1}})\Big\}^\frac{1}{p}}\\
&\leq&{\Big\{\mathbf{E}(|U_{n,i,h,h'}|^{2+\lambda}/{\mathcal{F}_{i-1}})}\Big\}^\frac{1}{\frac{\lambda}{2}+1}{\Big\{\mathbf{P}(|U_{n,i,h,h'}|>\gamma/{\mathcal{F}_{i-1}})\Big\}^\frac{1}{p}}\\
&\leq&{\Big\{\mathbf{E}(|U_{n,i,h,h'}|^{2+\lambda}/{\mathcal{F}_{i-1}})\Big\}}^\frac{1}{\frac{\lambda}{2}+1}{\Big\{\mathbf{P}(|U_{n,i,h,h'}|^{2+\lambda}>{\gamma}^{2+\lambda}/{\mathcal{F}_{i-1}})\Big\}^\frac{1}{p}},
\end{eqnarray*}
\noindent where $\nu=1+\frac{\lambda}{2}$ and $\lambda>0.$
\noindent Note that from the  lemma (\ref{lemmainequality}), it
follows :\begin{eqnarray*}
  |\epsilon_iM_{f}(\epsilon_i) +1)|^{2+\lambda} &\leq& 2^{1+\lambda}(|\epsilon_i M_{f}(\epsilon_i)|^{2+\lambda}
  +1).
\end{eqnarray*}
By $(A_{3.4}),$  we have\begin{eqnarray*}
 \mathbf{E}\Big|\epsilon_iM_{f}(\epsilon_i) +1)|^{2+\lambda} &\leq&
 2^{1+\lambda}\mathbf{E}(|\epsilon_i M_{f}(\epsilon_i)|^{2+\lambda})+2^{1+\lambda}
 <+\infty,
\end{eqnarray*} It follows from  Markov's  conditionally inequality that\begin{eqnarray*}
\mathbf{E}\left(U^2_{n,i,h,h'}I_{\Big\{|U_{n,i,h,h'}|>\gamma\Big\}}/{\mathcal{F}_{i-1}}\right)
&\leq&{\gamma}^{\frac{-{(2+\lambda)}}{p}}\left\{\mathbf{E}(|U_{n,i,h,h'}|^{2+\lambda}/{\mathcal{F}_{i-1}})\right\}^\frac{1}{\frac{\lambda}{2}+1}
\times\nonumber\\
&&\times{\Big\{\mathbf{E}(|U_{n,i,h,h'}|^{2+\lambda}/{\mathcal{F}_{i-1}})\Big\}^\frac{1}{p}},\\
&\leq&{\gamma}^{\frac{-{(2+\lambda)}}{p}}\Big\{\mathbf{E}(|U_{n,i,h,h'}|^{2+\lambda}/{\mathcal{F}_{i-1}})].
\end{eqnarray*} It results from The  lemma (\ref{lemmainequality}) followed by
the properties of the conditionally expectation that
\begin{eqnarray}
\sum_{i=1}^{n}\mathbf{E}(U^{2}_{n,i,h,h'}I_{\Big\{|U_{n,h,h'}|>\gamma\Big\}}/{\mathcal{F}_{i-1}})&\leq&2^{(1+\lambda)}{\gamma}^{\frac{-{(2+\lambda)}}{p}}
\{n^{-(1+\frac{\lambda}{2})}\sum_{i=1}^{n}\mathbf{E}[\Big|hM_{f}(\epsilon_i)\frac{G(Z_i)}
{{\sigma}(\theta_0,Z_i)}\Big|^{2+\lambda}/{\mathcal{F}_{i-1}}]\nonumber\\
&&+n^{-(1+\frac{\lambda}{2})}\sum_{i=1}^{n}\mathbf{E}[\Big|h'(\epsilon_iM_{f}(\epsilon_i)
+1)\frac{S(Z_i)}
{{\sigma}(\theta_0,Z_i)}\Big|^{2+\lambda}/{\mathcal{F}_{i-1}}]\}\nonumber\\
&\leq&2^{(1+\lambda)}\delta^{1+{\frac{\lambda}{2}}}{\gamma}^{\frac{-{(2+\lambda)}}{p}}
n^{-(1+\frac{\lambda}{2})}\Big\{\sum_{i=1}^{n}|\frac{G(Z_i)}
{{\sigma}(\theta_0,Z_i)}\Big|^{2+\lambda}\mathbf{E}\Big|M_{f}(\epsilon_{0})\Big|^{2+\lambda}\nonumber\\
&&+\sum_{i=1}^{n}|\frac{S(Z_i)}
{{\sigma}(\theta_0,Z_i)}\Big|^{2+\lambda}\mathbf{E}\Big|\epsilon_{0}
M_{f}(\epsilon_{0})
+1)\Big|^{2+\lambda}\Big\}\nonumber\\
&\leq&K2^{(1+\lambda)}\delta^{1+{\frac{\lambda}{2}}}{\gamma}^{\frac{-{(2+\lambda)}}{p}}n^{-\frac{\lambda}{2}}\times\nonumber\\
&&\times \Big\{\frac{1}{n}\sum_{i=1}^{n}|\frac{G(Z_i)}
{{\sigma}(\theta_0,Z_i)}\Big|^{2+\lambda}+\frac{1}{n}\sum_{i=1}^{n}|\frac{S(Z_i)}
{{\sigma}(\theta_0,Z_i)}\Big|^{2+\lambda}\Big\}.\label{inelinderberg}
\end{eqnarray}
\noindent Using the inequality (\ref {inelinderberg}) and from
 the ergodicity, the stationarity, ($A_{3.1}$)
and $(A_{3.2}),$  it results that
$$\sum_{i=1}^{n}\mathbf{E}(U^2_{n,i,h,h'}I_{\Big\{|U_{n,h,h'}|>\gamma\Big\}}/{\mathcal{F}_{i-1}})\stackrel{\mathbf{P}}{\longrightarrow}0
\quad\mbox{as}\quad n\rightarrow\infty.$$ Which implies that the
Linderberg condition is satisfied.

\subsubsection*{Conditionally variance }
\begin{eqnarray*}
\sum_{i=1}^{n}\mathbf{E}\Big(U^2_{n,i,h,h'}/{\mathcal{F}_{i-1}}\Big)
&=&\frac{1}{n}\left\{\sum_{i=1}^{n}\mathbf{E}\left(\Big[h
M_{f}(\epsilon_i)\frac{G(Z_i)}{\sigma(\theta_0,Z_i)}\Big]^2/{\mathcal{F}_{i-1}}\right)\right.\\
&&\left.
+\sum_{i=1}^{n}\mathbf{E}\left(\Big[h'(M_{f}(\epsilon_i)\epsilon_i+1)\frac{S(Z_i)}{\sigma(\theta_0,Z_i)}\Big]^2/\mathcal{F}_{i-1}\right)\right.\\
&&\left.+2hh'\sum_{i=1}^{n}\mathbf{E}\left(\Big[M_f(\epsilon_i)[M_{f}(\epsilon_i)\epsilon_i+1]\frac{G(Z_i)S(Z_i)}
{\sigma^2(\theta_0,Z_i)}\Big]/\mathcal{F}_{i-1}\right)\right\}.
\end{eqnarray*}
\noindent Using the properties of the conditionally expectation,
and since the random variables $\epsilon_i$ are independent of\\
$\mathcal{F}_{i} = \sigma(Z_j , j \leq i)$ and after the
application of the ergodic theorem, it follows the convergence of
$\sum_{i=1}^{n}\mathbf{E}(U^2_{n,i,h,h'}/{\mathcal{F}_{i-1}})$ to
   ${\eta}^2= {\tau}^2 _{h,h'}$ a.s. as   $n\rightarrow\infty$
   (so in  Probability).\\
\noindent $\emph{\textbf{Measurability}}$:\\
The random variable  $\eta$ is a  constant, so it is  measurable
on ${\mathcal{F}_{i-1}}$, therefore we obtain the
\emph{measurability}.\\
\emph{In summary, by collecting the conditions $(i,)$ $(ii)$ and
$(iii)$ , we deduce that the random variable
$\mathcal{V}_{n,h,h'}(\rho_0, \theta_0)$ converges in distribution
to} $ \mathcal{N}(0,{\tau}^2 _{h,h'})$ as $n \longrightarrow
+\infty$. \noindent It remains to  prove that
$\sum_{i=1}^{n}{R}^\ast_{n,i,h,h'}=o_{P}(1),$ where
\begin{eqnarray*}
 {R}^\ast_{n,i,h,h'} &=&\frac{1}{2f(\epsilon_i)}\Big[{\alpha}_{n,i,h},(\beta_{n,i,h'} -
1)\Big]\partial^2A_{n,i,h,h'}(\epsilon_i;{\alpha}^\star_{n,i,h},
{\beta}^\star_{n,i,h'})\Big[{\alpha}_{n,i,h},(\beta_{n,i,h'} -
1)\Big]^\top,
\end{eqnarray*}
and,
 $$
\partial^2A_{n,i,h,h'}(\epsilon_i;{\alpha}^\star_{n,i,h},
{\beta}^\star_{n,i,h'})=\left(%
\begin{array}{cc}
  D_{1,1}(\epsilon_i;{\alpha}^\star_{n,i,h},
{\beta}^\star_{n,i,h'}) ~~~~~~~~&
D_{1,2}(\epsilon_i;{\alpha}^\star_{n,i,h},
{\beta}^\star_{n,i,h'}) \\
  D_{2,1}(\epsilon_i;{\alpha}^\star_{n,i,h},
{\beta}^\star_{n,i,h'})~~~~~~~~
&D_{2,2}(\epsilon_i;{\alpha}^\star_{n,i,h},
{\beta}^\star_{n,i,h'}) \\
\end{array}%
\right).$$ We have
  \begin{eqnarray*}
 \sum
_{i=1}^{n}{R}^\ast_{n,i,h,h'}&=&\sum
_{i=1}^{n}\frac{D_{1,1}(\epsilon_i;{\alpha}^\star_{n,i,h},
{\beta}^\star_{n,i,h'})}{2f(\epsilon_i)}\,{\alpha}_{n,i,h}^2+\sum
_{i=1}^{n}\frac{D_{2,2}(\epsilon_i;{\alpha}^\star_{n,i,h},
{\beta}^\star_{n,i,h'})}{2f(\epsilon_i)}\,{({\beta}_{n,i,h'} -
1)}^2\\
&&+\sum _{i=1}^{n}{\alpha}_{n,i,h}{({\beta}_{n,i,h'} -
1)}\,\frac{D_{1,2}(\epsilon_i;{\alpha}^\star_{n,i,h},
{\beta}^\star_{n,i,h'})}{2f(\epsilon_i)}+\sum
_{i=1}^{n}{\alpha}_{n,i,h}{({\beta}_{n,i,h'} -
1)}\,\frac{D_{2,1}(\epsilon_i;{\alpha}^\star_{n,i,h},
{\beta}^\star_{n,i,h'})}{2f(\epsilon_i)}.
\end{eqnarray*}
\noindent We have
 \begin{eqnarray*}
\sum _{i=1}^{n}\frac{D_{1,1}(\epsilon_i;{\alpha}^\star_{n,i,h},
{\beta}^\star_{n,i,h'})}{2f(\epsilon_i)}\,{\alpha}_{n,i,h}^2&=&\frac{h^2}{n}\sum
_{i=1}^{n}\frac{D_{1,1}\,(\epsilon_i;{\alpha}^\star_{n,i,h},
{\beta}^\star_{n,i,h'})}{2f(\epsilon_i)}
\left[{\frac{G(Z_i)}{{\sigma}(\theta_0,Z_i)}}\right]^2.
\end{eqnarray*}
\noindent We have the following decomposition
\begin{eqnarray*}
 \frac{h^2}{n}\sum
_{i=1}^{n}\frac{D_{1,1}\,(\epsilon_i;{\alpha}^\star_{n,i,h},
{\beta}^\star_{n,i,h'})}{2f(\epsilon_i)}
\left[{\frac{G(Z_i)}{{\sigma}(\theta_0,Z_i)}}\right]^2
&=&\frac{h^2}{n}\Big\{\sum_{i=1}^{n}\frac{D_{1,1}(\epsilon_i;{\alpha}^\star_{n,i,h},
{\beta}^\star_{n,i,h'})-D_{1,1}(\epsilon_i;0,1)}{2f(\epsilon_i)}\,
\Big[{\frac{G(Z_i)}{{\sigma}(\theta_0,Z_i)}}\Big]^2 \\
 &&+\sum_{i=1}^{n}\frac{D_{1,1}(\epsilon_i;0,1)}{2f(\epsilon_i)}
\Big[{\frac{G(Z_i)}{{\sigma}(\theta_0,Z_i)}}\Big]^2\Big\}\\
 &\leq& \frac{h^2}{n}\sum_{i=1}^{n}\frac{\Big|D_{1,1}(\epsilon_i;{\alpha}^\star_{n,i,h},
{\beta}^\star_{n,i,h'})-D_{1,1}(\epsilon_i;0,1)\Big|}{2f(\epsilon_i)}
\Big[{\frac{G(Z_i)}{{\sigma}(\theta_0,Z_i)}}\Big]^2 \\
    &+&  \frac{h^2}{n}\sum_{i=1}^{n}\frac{D_{1,1}(\epsilon_i;0,1)}{2f(\epsilon_i)}
\Big[{\frac{G(Z_i)}{{\sigma}(\theta_0,Z_i)}}\Big]^2.
\end{eqnarray*}
\noindent From $(A_{1.2}),$ there exist a positive function $V_1$,
a strictly positive real  ${\varsigma}'$, whith\\
${\varsigma}'>\max(|{{\alpha}^{\star\star}_{n,i,h}}|,|{\beta}^\star_{n,i,h'}-1|)$
and a measurable positive function  $\phi$ such that
$\mathbf{E}({\phi}(\epsilon_{0}))<+\infty$ such that
\begin{eqnarray*}
\Big|D_{1,1}(\epsilon_i;{\alpha}^\star_{n,i,h},{\beta}^\star_{n,i,h'})-D_{1,1}(\epsilon_i;0,1)\Big|&\leq&|{\alpha}^\star_{n,i,h}|V_1({\epsilon}_i
;{\alpha}^{\star\star}_{n,i,h},{\beta}^\star_{n,i,h'}),
\end{eqnarray*}
\noindent  where\,${{\alpha}^{\star\star}_{n,i,h}}\in[0,{\alpha}^\star_{n,i,h}]$.\\
For all integers $n\geq1$, we have
\begin{eqnarray*}
  \mathbf{E}\Big\{\frac{h^2}{n}\sum
_{i=1}^{n}\frac{|D_{1,1}({\epsilon}_i;{\alpha}^\star_{n,i,h},
{\beta}^\star_{n,i,h'})-D_{1,1}({\epsilon}_i;0,1)|}{2f(\epsilon_i)}\Big
[{\frac{G(Z_i)}{{\sigma}(\theta_0,Z_i))}}\Big]^2\Big\}&&\leq\frac{h^2}{2n}\sum
_{i=1}^{n}\mathbf{E}\Big\{\phi(\epsilon_i)|{\alpha}^\star_{n,i,h}|
\Big[{\frac{G(Z_i)}{{\sigma}(\theta_0,Z_i))}}\Big]^2\Big\}.
\end{eqnarray*}
\noindent  Since $ {\alpha}^\star_{n,i,h}$ is in the interval
$[0,{\alpha}_{n,i,h}]$, therefore there
  exist a random sequence of parameter  $\left({\theta}_{n}\right)_{n\geq 1}$ with values in $[0,1]$ such
  that
\begin{eqnarray*}
{\alpha}^\star_{n,i,h}&=&{{\theta}_{n} {\alpha}_{n,i,h}}.
\end{eqnarray*}
Then we obtain
\begin{eqnarray*}
  \mathbf{E}\Big\{\frac{h^2}{n}\sum
_{i=1}^{n}\frac{\Big|D_{1,1}({\epsilon}_i;{\alpha}^\star_{n,i,h},
{\beta}^\star_{n,i,h'})-D_{1,1}({\epsilon}_i;0,1)\Big|}{2f(\epsilon_i)}
\Big[{\frac{G(Z_i)}{{\sigma}(\theta_0,Z_i)}}\Big]^2\Big\}&\leq&\frac{h^2}{n}\sum
_{i=1}^{n}{\theta}_{n}\mathbb{E}\Big\{\phi(\epsilon_{0}){
|{\alpha}_{n,i,h}}|\Big[{\frac{G(Z_i)}{{\sigma}(\theta_0,Z_i)}}\Big]^2 \Big\}\\
&\leq& K h^3
n^{-\frac{1}{2}}\mathbf{E}\Big\{\frac{1}{n}\sum_{i=1}^{n}\Big|{\frac{G(Z_i)}{{\sigma}(\theta_0,Z_i)}}\Big|^3\Big\}.
\end{eqnarray*}
\noindent By  Markov's inequality, for all $\gamma>0$, we have
\begin{eqnarray*}
   \mathbf{P}\Big(\frac{h^2}{n}\sum
_{i=1}^{n}\frac{\Big|D_{1,1}({\epsilon}_i;{\alpha}^\star_{n,i,h},
{\beta}^\star_{n,i,h'})-D_{1,1}({\epsilon}_i;0,1)\Big|}{2f(\epsilon_i)}
\Big[{\frac{G(Z_i)}{{\sigma}(\theta_0,Z_i)}}\Big]^2>\gamma\Big)&\leq&\frac{1}{{\gamma}}
K\,h^3
n^{-\frac{1}{2}}\mathbf{E}\Big\{\frac{1}{n}\sum_{i=1}^{n}\Big[{\frac{G(Z_i)}{{\sigma}(\theta_0,Z_i)}}\Big]^3\Big\}.
\end{eqnarray*}
From  the ergodicity and  the stationarity of the model and since
$n^{-\frac{1}{2}}\rightarrow 0$, it results  that
\begin{eqnarray*}
\frac{h^2}{n}\sum
_{i=1}^{n}\frac{|D_{1,1}({\epsilon}_i;{\alpha}^\star_{n,i,h},
{\beta}^\star_{n,i,h'})-D_{1,1}({\epsilon}_i;0,1)|}{2f(\epsilon_i)}
\Big[{\frac{G(Z_i)}{{\sigma}(\theta_0,Z_i)}}\Big]^2
\stackrel{\mathbf{P}}{\longrightarrow}0\quad \mbox{as}\quad
n\rightarrow+\infty.
\end{eqnarray*}
 \noindent Finally, we get
\begin{eqnarray}
R^{(1)}_{n,i,h,h'}&=&\sum
_{i=1}^{n}\frac{\Big|D_{1,1}({\epsilon}_i;{\alpha}^\star_{n,i,h},
{\beta}^\star_{n,i,h'}) -
D_{1,1}({\epsilon}_i;0,1)\Big|}{2f(\epsilon_i)}\,
{\alpha}_{n,i,h}^2=o_{P}(1)\label{firstremain}.
\end{eqnarray}
\noindent By following the same previous reasoning in the last
case, we shall prove that
\begin{eqnarray}
  R^{(2)}_{n,i,h,h'}&=&\sum
_{i=1}^{n}\frac{\Big|D_{2,2}({\epsilon}_i;{\alpha}^\star_{n,i,h},
{\beta}^\star_{n,i,h'}) -
D_{2,2}({\epsilon}_i;0,1)\Big|}{2f(\epsilon_i)}
\,{({\beta}_{n,i,h'} -
1)}^2 =o_{P}(1),\label{second remain}  \\
R^{(3)}_{n,i,h,h'}&=&\sum
_{i=1}^{n}\frac{\Big|D_{1,2}({\epsilon}_i;{\alpha}^\star_{n,i,h},
{\beta}^\star_{n,i,h'}) -
D_{1,2}({\epsilon}_i;0,1)\Big|}{2f(\epsilon_i)}\,
{\alpha}_{n,i,h}^2= o_{P}(1),\label{premierr}\\
R^{(4)}_{n,i,h,h'}&=&\sum
_{i=1}^{n}\frac{\Big|D_{1,2}({\epsilon}_i;{\alpha}^\star_{n,i,h},
{\beta}^\star_{n,i,h'}) -
D_{1,2}({\epsilon}_i;0,1)\Big|}{2f(\epsilon_i)}\,
{({\beta}_{n,i,h'} -
1)}^2= o_{P}(1),\label{secondr}\\
R^{(5)}_{n,i,h,h'}&=&\sum
_{i=1}^{n}\frac{\Big|D_{2,1}({\epsilon}_i;{\alpha}^\star_{n,i,h},
{\beta}^\star_{n,i,h'}) -
D_{2,1}({\epsilon}_i;0,1)\Big|}{2f(\epsilon_i)}\,
{\alpha}_{n,i,h}^2 = o_{P}(1),\nonumber \\
R^{(6)}_{n,i,h,h'}&=&\sum
_{i=1}^{n}\frac{\Big|D_{2,1}({\epsilon}_i;{\alpha}^\star_{n,i,h},
{\beta}^\star_{n,i,h'}) -
D_{2,1}({\epsilon}_i;0,1)\Big|}{2f(\epsilon_i)}\,
{({\beta}_{n,i,h'} - 1)}^2= o_{P}(1).\nonumber
\end{eqnarray} Let\begin{eqnarray}
 R^{(7)}_{n,i,h,h'}&=&\sum
_{i=1}^{n}\,\frac{\Big|D_{1,2}({\epsilon}_i;{\alpha}^\star_{n,i,h},
{\beta}^\star_{n,i,h'})
-D_{1,2}({\epsilon}_i;0,1)\Big|}{2f(\epsilon_i)}\,{\alpha}_{n,i,h}{({\beta}_{n,i,h'}
- 1)}, \quad \mbox{and}\\ R^{(8)}_{n,i,h,h'}&=&\sum
_{i=1}^{n}\,\frac{\Big|D_{2,1}({\epsilon}_i;{\alpha}^\star_{n,i,h},
{\beta}^\star_{n,i,h'})
-D_{2,1}({\epsilon}_i;0,1)\Big|}{2f(\epsilon_i)}\,{\alpha}_{n,i,h}{({\beta}_{n,i,h'}
- 1)}.
\end{eqnarray}From the following inequality\begin{eqnarray*}
 \Big |{\alpha}_{n,i,h}{({\beta}_{n,i,h'} - 1)}\Big| &\leq&{\frac{1}{2}\Big[{{\alpha}^2}_{n,i,h}+({\beta}_{n,i,h'}
 -1})^2\Big],
\end{eqnarray*} It results that\begin{eqnarray*}
  |R^{(7)}_{n,i,h,h'}|&\leq&\frac{1}{2}\sum
_{i=1}^{n}{\frac{\Big|D_{1,2}({\epsilon}_i;{\alpha}^\star_{n,i,h},
{\beta}^\star_{n,i,h'}) -D_{1,2}({\epsilon}_i;0,1)\Big|}{2f(\epsilon_i)}\,{\alpha}^2_{n,i,h}}\\
&&+\frac{1}{2}\sum_{i=1}^{n}{
\frac{\Big|D_{1,2}({\epsilon}_i;{\alpha}^\star_{n,i,h},
{\beta}^\star_{n,i,h'})
-D_{1,2}({\epsilon}_i;0,1)\Big|}{2f(\epsilon_i)}
\,{({\beta}_{n,i,h'} - 1)}^2}.
\end{eqnarray*}It follows from the  equalities (\ref{premierr}) and
(\ref{secondr}) that\begin{eqnarray} R^{(7)}_{n,i,h,h'} &=&
o_{P}(1).\label{remain 7}
\end{eqnarray}In a similar way, we can show that
\begin{eqnarray}
R^{(8)}_{n,i,h,h'} &=& o_{P}(1).\label{remain 8}
\end{eqnarray} We have
\begin{eqnarray*}
  \sum _{i=1}^{n}{R}^\ast_{n,i,h,h'} &=&\sum _{i=1}^{n}\frac{D_{1,1}({\epsilon}_i;{\alpha}^\star_{n,i,h},
{\beta}^\star_{n,i,h'})-D_{1,1}({\epsilon}_i;0,1)}{2f(\epsilon_i)} \,{\alpha}_{n,i,h}^2\\
&&+\sum
_{i=1}^{n}\frac{D_{2,2}({\epsilon}_i;{\alpha}^\star_{n,i,h},
{\beta}^\star_{n,i,h'})-
D_{2,2}({\epsilon}_i;0,1)}{2f(\epsilon_i)}\,{({\beta}_{n,i,h'} -
1)}^2\\
&&+\sum
_{i=1}^{n}\frac{D_{1,2}({\epsilon}_i;{\alpha}^\star_{n,i,h},
{\beta}^\star_{n,i,h'})-
D_{1,2}({\epsilon}_i;0,1)}{2f(\epsilon_i)}\,
{\alpha}_{n,i,h}{({\beta}_{n,i,h'} -
1)}\\
&&+\sum
_{i=1}^{n}\frac{D_{2,1}({\epsilon}_i;{\alpha}^\star_{n,i,h},
{\beta}^\star_{n,i,h'})-
D_{2,1}({\epsilon}_i;0,1)}{2f(\epsilon_i)}\,{\alpha}_{n,i,h}{({\beta}_{n,i,h'}
- 1)}\\
&&+ L_{n,i,h,h'},
\end{eqnarray*}
\noindent with
\begin{eqnarray*}
   L_{n,i,h,h'}&=&\sum_{i=1}^{n}\frac{D_{1,1}({\epsilon}_i;0,1)}{2f(\epsilon_i)}{\alpha}_{n,i,h}^2
  +\sum_{i=1}^{n}\frac{D_{2,2}({\epsilon}_i;0,1)}{2f(\epsilon_i)}{({\beta}_{n,i,h'}
- 1)}^2\\
&&+\sum _{i=1}^{n}{\alpha}_{n,i,h}{({\beta}_{n,i,h'} -
1)}\frac{D_{1,2}({\epsilon}_i;0,1))}{2f(\epsilon_i)}
+\sum_{i=1}^{n}{\alpha}_{n,i,h}{({\beta}_{n,i,h'} -
1)}\frac{D_{2,1}({\epsilon}_i;0,1)}{2f(\epsilon_i)}.
\end{eqnarray*}From the equalities (\ref{firstremain}), (\ref{second remain}),
(\ref{remain 7}) and (\ref{remain 8}), it results that
\begin{eqnarray*}
  \sum _{i=1}^{n}{R}^\ast_{n,i,h,h'} &\leq& R^{(1)}_{n,i,h,h'}
+R^{(2)}_{n,i,h,h'} +R^{(7)}_{n,i,h,h'} + R^{(8)}_{n,i,h,h'} +
L_{n,i,h,h'}.
\end{eqnarray*}
We have\begin{eqnarray*}
  D_{1,1}({\epsilon}_i;0,1) =\ddot{f}(\epsilon_i)\mbox{,}\quad
  D_{1,2}({\epsilon}_i;0,1)=D_{2,1}({\epsilon}_i;0,1)= \dot{f}(\epsilon_i)+ \epsilon_i \ddot{f}(\epsilon_i),
\end{eqnarray*}
  and
  \begin{eqnarray*}
 D_{2,2}({\epsilon}_i;0,1)= 2{\epsilon_i} \dot{f}(\epsilon_i) + {\epsilon_i}^2 \ddot{f}(\epsilon_i).
\end{eqnarray*}
\noindent By simple calculation, it is easy to prove
that:\begin{eqnarray*}
 \frac{\ddot{f}(x)}{f(x)} &=& \dot{M}_{f}(x)+ {M^2_{f}}(x).
\end{eqnarray*}
\begin{itemize}
    \item
By $(A_{2.3})$  combined with the ergodicity and the stationarity
of the model , it results that:\begin{eqnarray*}
  \sum_{i=1}^{n}\frac{D_{1,1}({\epsilon}_i;0,1)}{2f(\epsilon_i)}\,{\alpha}_{n,i,h}^2 = \frac{h^2}{n}\sum_{i=1}^{n}\frac{\ddot{f}
(\epsilon_i)}{2f(\epsilon_i)}{\Big[{\frac{G(Z_i)}{{\sigma}(\theta_0,Z_i)}}\Big]^2}\stackrel{a.s.}{\longrightarrow}
 h^2\mathbf{E}\Big\{\frac{\ddot{f}(\epsilon_i)}{2f(\epsilon_i)}{\Big[{\frac{G(Z_i)}{{\sigma}(\theta_0,Z_i)}}\Big]^2}
 \Big\},\quad \mbox{as}\quad  n\rightarrow\infty.
\end{eqnarray*}
 then
\begin{eqnarray*}
  \sum_{i=1}^{n}\frac{D_{1,1}({\epsilon}_i;0,1)}{2f(\epsilon_i)}\,{\alpha}_{n,i,h}^2
  \stackrel{a.s.}{\longrightarrow}\frac{h^2}{2}\mathbf{E}\left\{{\Dot{M}_{f}}(\epsilon_{0})+{M^2_{f}}(\epsilon_{0})\right\}
  \mathbf{E}{\Big[{\frac{G(Z_{0})}{{\sigma}(\theta_0,Z_{0})}}\Big]^2}=0  \quad \mbox{as}\quad  n\rightarrow\infty.
\end{eqnarray*}

\item    By  $(A_{2.2})$ and $(A_{2.5})$ combined with the
ergodicity and the stationarity of the model, it results
that:\begin{eqnarray*} \lefteqn{\frac{h'^2}{n}\sum
_{i=1}^{n}\frac{D_{2,2}({\epsilon}_i;0,1)}{2f(\epsilon_i)}{({\beta}_{n,i,h'}
- 1)}^2}\\ &=& \frac{h'^2}{n}\sum _{i=1}^{n}\left\{{\epsilon_{i}}
M_{f}({\epsilon_{i}}) +\frac{1}{2}
{\epsilon_{i}}^2\left({\Dot{M}_{f}}(\epsilon_{i})+
{M^2_{f}}(\epsilon_{i})\right)\right\}{\Big[{\frac{S(Z_i)}{{\sigma}(\theta_0,Z_i)}}\Big]^2}
\stackrel{a.s.}{\longrightarrow}0,\quad \mbox{as}\quad
n\rightarrow\infty.
\end{eqnarray*}
 \item It follows by $(A_{2.1})$ and $(A_{2.4})$
 and the  ergodicity and the stationarity  of the model that:$$\sum _{i=1}^{n}{\alpha}_{n,i,h}{({\beta}_{n,i,h'} -
1)}\frac{D_{1,2}({\epsilon}_i;0,1)}{2f(\epsilon_i)}+\sum
_{i=1}^{n}{\alpha}_{n,i,h}{({\beta}_{n,i,h'} -
1)}\frac{D_{2,1}({\epsilon}_i;0,1)}{2f(\epsilon_i)}$$
$$=\frac{hh'}{n}\sum _{i=1}^{n}
\frac{G(Z_i)S(Z_i)}{{\sigma}^2(\theta_0,Z_i)}\left[
M_{f}(\epsilon_i)+{\epsilon_i}({ \Dot{M}_{f}}(\epsilon_i) +{
M^2_{f}}(\epsilon_i))\right]$$
$$\stackrel{a.s.}{\longrightarrow}hh'\mathbf{E}(\frac{G(Z_i)S(Z_i)}{{\sigma}^2(\theta_0,Z_i)})\mathbf{E}\left[
M_{f}(\epsilon_{0})+{\epsilon_{0}}({ \Dot{M}_{f}}(\epsilon_{0}) +{
M^2_{f}}(\epsilon_{0}))\right]=0 \quad \mbox{as}\quad
n\rightarrow\infty.
$$
\end{itemize}
\noindent Consequently, the random variable ${L}_{n,i,h,h'}
\stackrel{a.s.}{\longrightarrow}0$ as $n\rightarrow+\infty.$\\
The random vector $\Big(R^{(1)}_{n,i,h,h'} +R^{(2)}_{n,i,h,h'} +
R^{(7)}_{n,i,h,h'}+ {R^{(8)}}_{n,i,h,h'}~,~{L}_{n,i,h,h'}\Big)
\stackrel{\mathbf{P}}{\longrightarrow}(0,0)$  as
$n\rightarrow+\infty.$ Since the function $(x,y) \longmapsto x +
y$ is continuous, it results that
$$\sum_{i=1}^{n}{R}^\ast_{n,i,h,h'}=o_{P}(1).$$
\subsubsection*{Conclusion}
\noindent  The conditions $(C.1)$,$(C.2)$ and $(C.3)$ are
established,from the \cite[Theorem 1 ]{HB}, it follows, under the
hypothesis $(H_0)$, that:\begin{eqnarray}
 \Lambda_{n,h,h'}&=&\mathcal{V}_{n,h,h'}(\rho_0,\theta_0) -\frac{{\tau}^2_{h,h'}}{2} +
 o_{P}(1).\label{LAN}
  \end{eqnarray}
\subsection*{Proof of the Theorem \ref{secondttheorem} }
\noindent The proof is similar as the proof of \cite[Theorem
3]{HB}.

\subsection*{Proof of the Proposition \ref{fondamental proposition}}
 Based on the equations (\ref{residuel}) and
(\ref{residuelestimé}), we have
\begin{eqnarray*}
  \sigma(\theta_{0}+n^{-\frac{1}{2}}v^{(n)}\,,\,Z_{i})\,\tilde{\epsilon}_{i,n}
-\sigma(\theta_{0}\,,\,Z_{i})\,{\epsilon}_{i} &=&
-\Big(m(\rho_{0}+n^{-\frac{1}{2}}u^{(n)}\,,\,Z_{i})-m(\rho_{0}\,,\,Z_{i})\Big),
\end{eqnarray*}
Then
\begin{eqnarray*}
\tilde{\epsilon}_{i,n}-{\epsilon}_{i}&=&-\frac{m(\rho_{0}+n^{-\frac{1}{2}}u^{(n)}\,,\,Z_{i})-m(\rho_{0}\,,\,Z_{i})}{\sigma(\theta_{0}+
n^{-\frac{1}{2}}v^{(n)}\,,\,Z_{i})}
-\frac{\sigma(\theta_{0}+n^{-\frac{1}{2}}v^{(n)}\,,\,Z_{i})-{\sigma(\theta_{0}\,,\,Z_{i})}}
{\sigma(\theta_{0}+n^{-\frac{1}{2}}v^{(n)}\,,\,Z_{i})}
\,{\epsilon}_{i},
\end{eqnarray*}
By Taylor's expansion with order $1$ of the functions
$\rho\rightarrow m(\rho,\cdot)$ and $\theta\rightarrow
\sigma(\theta,\cdot)$ around $\rho_{0}$ and $\theta_{0}$
respectively, we obtain the following equalities
\begin{eqnarray}
m(\rho_{0}+n^{-\frac{1}{2}}u^{(n)}\,,\,Z_{i})-m(\rho_{0}\,,\,Z_{i})&=&n^{-\frac{1}{2}}(u^{(n)})^\top\,\partial\,m(\tilde{\rho}_{n},\,Z_{i})^\top,\label{moyenne}\\
\sigma(\theta_{0}+n^{-\frac{1}{2}}v^{(n)}\,,\,Z_{i})-{\sigma(\theta_{0}\,,\,Z_{i})}&=&n^{-\frac{1}{2}}(v^{(n)})^\top\,
\partial\sigma(\tilde{\theta}_{n}\,,\,Z_{i})^\top,\label{variance}\\
\tilde{\epsilon}_{i,n}-{\epsilon}_{i}&=&-\frac{n^{-\frac{1}{2}}(u^{(n)})^\top\,\partial\,m(\tilde{\rho}_{n},\,Z_{i})^\top}{\sigma(\theta_{0}+
n^{-\frac{1}{2}}v^{(n)}\,,\,Z_{i})}
-\frac{n^{-\frac{1}{2}}(v^{(n)})^\top\,\partial\,
{\sigma(\tilde{\theta}_{n},Z_{i})^\top}}
{\sigma(\theta_{0}+n^{-\frac{1}{2}}v^{(n)}\,,\,Z_{i})}
\,{\epsilon}_{i},~~~~~~~ \label{epsilon}
\end{eqnarray}
\noindent The parameters  $\tilde{\rho}_{n}$ and
$\tilde{\theta}_{n}$ are
 between $\rho_{0}$ and  $\rho_n$ and $\theta_{0}$ and
 $\theta_n$ respectively.\\
\noindent By Taylor's expansion with order $2$ of the function
$u\longmapsto M_{f}(u)$ around $\epsilon_{i}$ combined with the
equality (\ref{variance}), we obtain
\begin{eqnarray}
  \frac{M_{f}(\tilde{\epsilon}_{i,n})}{\sigma(\theta_{0}+n^{-\frac{1}{2}}v^{(n)}\,,\,Z_{i})} - \frac{M_{f}({\epsilon}_{i})}{\sigma(\theta_{0},\,Z_{i})}&=&
  \frac{{\sigma(\theta_{0},\,Z_{i})}M_{f}(\tilde{\epsilon}_{i,n})-\sigma(\theta_{0}+n^{-\frac{1}{2}}v^{(n)},Z_{i})M_{f}({\epsilon}_{i})}
  {\sigma(\theta_{0},\,Z_{i})\sigma(\theta_{0}+n^{-\frac{1}{2}}v^{(n)},Z_{i})}\nonumber\\
&=&\frac{{\sigma(\theta_{0},\,Z_{i})}M_{f}(\tilde{\epsilon}_{i,n})-[{\sigma(\theta_{0}\,,\,Z_{i})}
+n^{-\frac{1}{2}}(v^{(n)})^\top\,\partial\sigma(\tilde{\theta}_{n}\,,\,Z_{i})^\top]M_{f}({\epsilon}_{i})}
  {\sigma(\theta_{0},\,Z_{i})\sigma(\theta_{0}+n^{-\frac{1}{2}}v^{(n)},Z_{i})}\nonumber\\
 &=&\frac{M_{f}(\tilde{\epsilon}_{i,n})  - M_{f}({\epsilon}_{i})}{\sigma(\theta_{0}+n^{-\frac{1}{2}}v^{(n)}\,,\,Z_{i})}-
  \frac{n^{-\frac{1}{2}} (v^{(n)})^\top \, \partial\, \sigma(\tilde{\theta}_{n},\,Z_{i})^\top}
  {\sigma(\theta_{0}+n^{-\frac{1}{2}}v^{(n)}\,,\,Z_{i})\sigma(\theta_{0}\,,\,Z_{i})}\,M_{f}({\epsilon}_{i}) \nonumber\\
&=&\frac{(\tilde{\epsilon}_{i,n} -
{\epsilon}_{i})\dot{M}_{f}({\epsilon}_{i})}{\sigma(\theta_{0}+n^{-\frac{1}{2}}v^{(n)}\,,\,Z_{i})}
+ \frac{(\tilde{\epsilon}_{i,n} - {\epsilon}_{i})^2
\ddot{M}_{f}(\tilde{\tilde{\epsilon}}_{i,n})}{2\sigma(\theta_{0}+n^{-\frac{1}{2}}v^{(n)}\,,\,Z_{i})}
\nonumber\\
&-& \frac{n^{-\frac{1}{2}} (v^{(n)})^\top \, \partial\,
\sigma(\tilde{\theta}_{n},\,Z_{i})^\top}
  {\sigma(\theta_{0}+n^{-\frac{1}{2}}v^{(n)}\,,\,Z_{i})\sigma(\theta_{0}\,,\,Z_{i})}\,M_{f}({\epsilon}_{i}),\nonumber\\
  \label{DIFFER}
 \end{eqnarray}
\noindent where  $\tilde{\tilde{\epsilon}}_{i,n}$  is between
 ${\epsilon}_{i}$ et $\tilde{\epsilon}_{i,n}.$\\
\noindent By a difference between the equalities  (\ref{r3}) and
(\ref{r1}), it follows that
\begin{eqnarray*}
  \tilde{ r}_{f,h,n}  - r_{f,h,n} &=& -n^{-\frac{1}{2}}\, h \,\sum _{i=1}^{n} [\frac{M_{f}(\tilde{\epsilon}_{i,n})}{\sigma(\theta_{0}+n^{-\frac{1}{2}}v^{(n)}\,,\,Z_{i})} -
  \frac{M_{f}({\epsilon}_{i})}{\sigma(\theta_{0},\,Z_{i})}]G(Z_{i}).
\end{eqnarray*}
Using the equality (\ref{DIFFER}), we obtain:
\begin{eqnarray*}
 \tilde{ r}_{f,h,n} - r_{f,h,n} &=& I_{n,h,1} + I_{n,h,2} + I_{n,h,3}.\label{f1}
\end{eqnarray*}
\noindent With
\begin{eqnarray}
  I_{n,h,1} &=& -n^{-\frac{1}{2}}\, h \,\sum_{i=1}^{n}\frac{(\tilde{\epsilon}_{i,n} -{\epsilon}_{i} )\dot{M}_{f}({\epsilon}_{i})}{\sigma(\theta_{0}+n^{-\frac{1}{2}}v^{(n)}\,,\,Z_{i})}\,G(Z_{i}),\\
I_{n,h,2} &=& -n^{-\frac{1}{2}}\, h \,\sum
_{i=1}^{n}\frac{(\tilde{\epsilon}_{i,n}
-{\epsilon}_{i})^2}{2\,\sigma(\theta_{0}+n^{-\frac{1}{2}}v^{(n)}\,,\,Z_{i})}
\ddot{M}_{f}(\tilde{\tilde{\epsilon}}_{i,n})\,G(Z_{i}),\\
\mbox{and} \quad I_{n,h,3} &=& {\frac{1}{n}}\, h \,\sum
_{i=1}^{n}\frac{(v^{(n)})^\top \, \partial\,
\sigma(\tilde{{\theta}_{n}},\,Z_{i})^\top}{\sigma(\theta_{0}+n^{-\frac{1}{2}}v^{(n)}\,,\,Z_{i})\sigma(\theta_{0}\,,\,Z_{i})}\,M_{f}({\epsilon}_{i})\,G(Z_{i}).
\end{eqnarray}
\noindent\emph{ Now we proceed  to evaluate the terms $I_{n,h,1}$,
$ I_{n,h,2}$ and $ I_{n,h,3}$, all the limits are calculated under the hypothesis $(H_{0}).$}\\
\subsubsection*{Evaluation of the term $I_{n,h,1}$ }
We have\begin{eqnarray*} I_{n,h,1}&=& -n^{-\frac{1}{2}}\, h
\,\sum_{i=1}^{n}\frac{(\tilde{\epsilon}_{i,n}
-{\epsilon}_{i})\dot{M}_{f}({\epsilon}_{i})}{\sigma(\theta_{0}+n^{-\frac{1}{2}}v^{(n)}\,,\,Z_{i})}\,G(Z_{i})\\
&+&n^{-\frac{1}{2}}\, h\sum_{i=1}^{n}\frac{(\tilde{\epsilon}_{i,n}
-{\epsilon}_{i})\dot{M}_{f}({\epsilon}_{i})}{\sigma(\theta_{0},,\,Z_{i})}\,G(Z_{i})\\
&-&n^{-\frac{1}{2}}\, h\sum_{i=1}^{n}\frac{(\tilde{\epsilon}_{i,n}
-{\epsilon}_{i})\dot{M}_{f}({\epsilon}_{i})}{\sigma(\theta_{0},,\,Z_{i})}\,G(Z_{i})\\
&=&n^{-\frac{1}{2}}\, h \,\sum
_{i=1}^{n}[\frac{1}{\sigma(\theta_{0}\,,\,Z_{i})}
-\frac{1}{\sigma(\theta_{0}+n^{-\frac{1}{2}}v^{(n)}\,,\,Z_{i})}]\,(\tilde{\epsilon}_{i,n}
- {\epsilon}_{i})\dot{M}_{f}({\epsilon}_{i})G(Z_{i})\\
&-&n^{-\frac{1}{2}}\, h\sum_{i=1}^{n}\frac{(\tilde{\epsilon}_{i,n}
-{\epsilon}_{i})\dot{M}_{f}({\epsilon}_{i})}{\sigma(\theta_{0},,\,Z_{i})}\,G(Z_{i})\\
 &=&I^{(1)}_{n,h,1} + I^{(2)}_{n,h,1}.\label{f2}
\end{eqnarray*}
\noindent From the equality (\ref{variance}), we have
\begin{eqnarray*}
I^{(1)}_{n,h,1}&=& n^{-\frac{1}{2}}\, h \,\sum
_{i=1}^{n}[\frac{1}{\sigma(\theta_{0}\,,\,Z_{i})}
-\frac{1}{\sigma(\theta_{0}+n^{-\frac{1}{2}}v^{(n)}\,,\,Z_{i})}]\,(\tilde{\epsilon}_{i,n}
- {\epsilon}_{i})\dot{M}_{f}({\epsilon}_{i})G(Z_{i})\\
&&=\frac{1}{n}\, h \,\sum _{i=1}^{n} \frac{(v^{(n)})^\top \,
\partial\,\sigma(\tilde{\theta}_{n},Z_{i})^\top}{\sigma(\theta_{0}+n^{-\frac{1}{2}}v^{(n)},\,Z_{i})}
 (\tilde{\epsilon}_{i,n} -
{\epsilon}_{i})\dot{M}_{f}({\epsilon}_{i})\frac{G(Z_{i})}{\sigma(\theta_{0}\,,\,Z_{i})}.
\end{eqnarray*}
\noindent Using the equality (\ref{epsilon}), we obtain
\begin{eqnarray*}
I^{(1)}_{n,h,1}&=&  n^{-\frac{1}{2}} h \, \frac{1}{n}\,\sum
_{i=1}^{n}\,A(\theta_{0},\tilde{\rho}_{n},\tilde{\theta}_{n},Z_{i})\,\dot{M}_{f}({\epsilon}_{i})\frac{G(Z_{i})}{\sigma(\theta_{0}\,,\,Z_{i})}
+  n^{-\frac{1}{2}} h \, \frac{1}{n}\,\sum
_{i=1}^{n}\,B(\theta_{0},\tilde{\theta}_{n},Z_{i})\,{\epsilon}_{i}\dot{M}_{f}({\epsilon}_{i})\frac{G(Z_{i})}{\sigma(\theta_{0}\,,\,Z_{i})},
\end{eqnarray*}
\noindent with
\begin{eqnarray}
A(\theta_{0},\tilde{\rho}_{n},\tilde{\theta}_{n},Z_{i})=-
\frac{(v^{(n)})^\top \,
\partial\,\sigma(\tilde{\theta}_{n},Z_{i})^\top}{\sigma(\theta_{0}+n^{-\frac{1}{2}}v^{(n)},\,Z_{i})}
\frac{(u^{(n)})^\top\,\partial\,m(\tilde{\rho}_{n},\,Z_{i})^\top}{\sigma(\theta_{0}+n^{-\frac{1}{2}}v^{(n)}\,,\,Z_{i})},\nonumber\\
\label{produit1}
\end{eqnarray}
\noindent and
\begin{eqnarray}
\quad B(\theta_{0},\tilde{\theta}_{n},Z_{i})= -\Big
[\frac{(v^{(n)})^\top \,
\partial\,\sigma(\tilde{\theta}_{n},Z_{i})^\top}{\sigma(\theta_{0}+n^{-\frac{1}{2}}v^{(n)},\,Z_{i})}\Big]^2.
\label{produit2}
\end{eqnarray}
\noindent The  parameters $\tilde{\rho}_{n}$ and
$\tilde{\theta}_{n}$ are into the  convex segments
$[\rho_{0},\rho_n]$ of ${\mathbb{R}}^\ell$ and
$[\theta_{0},\theta_n]$ of ${\mathbb{R}}^p$ respectively, then
there exist for all integers $n$, a sequence  ( $s_n,t_n$) with
values in $[0,1]\times [0,1]$, such that
\begin{eqnarray*}
\tilde{\rho}_{n} = s_n \rho_{0} + (1-s_n)\,\rho_n \quad
\mbox{and,}\quad \tilde{\theta}_{n} =t_n \theta_{0}+
(1-t_n)\,\theta_n.
\end{eqnarray*}
It  result  that
\begin{eqnarray}
  \|\tilde{\rho}_{n} -
\rho_{0}\|_{\ell} &\leq& (1-s_n)\|\rho_n - \rho_{0}\|_{\ell}
 \leq\|\rho_n - \rho_{0}\|_{\ell},\label{asyconvex1}
\end{eqnarray}
and,
\begin{eqnarray}
\| \tilde{\theta}_{n} - \theta_{0}\|_{p}&\leq& (1-t_n)\|\theta_n -
\theta_{0}\|_{p} \leq \|\theta_n -
\theta_{0}\|_{p}.\label{asyconvex2}
\end{eqnarray}
\noindent By applying Cauchy -Schwartz's inequality on each term
of the product (\ref{produit1}) and doing a  majoration, we obtain
\begin{eqnarray*}
|A(\theta_{0},\tilde{\rho}_{n},\tilde{\theta}_{n},Z_{i})|&\leq&
\Big\{\frac{\|(v^{(n)})\|_{p} \,
\|\partial\,\sigma(\tilde{\theta}_{n},Z_{i})\|_{p}}{\sigma(\theta_{0}+n^{-\frac{1}{2}}v^{(n)},\,Z_{i})}.\frac{(\|(u^{(n)})\|_{\ell}
\,\|\partial\,m(\tilde{\rho}_{n},\,Z_{i})\|_{\ell}}{\sigma(\theta_{0}+n^{-\frac{1}{2}}v^{(n)}\,,\,Z_{i})}\Big\}\\
& \leq& \frac{1}{2}\,\Big\{\Big[\frac{(\|(v^{(n)})\|_{p} \,
\|\partial\,\sigma(\tilde{\theta}_{n},Z_{i})\|_{p}}{\sigma(\theta_{0}+n^{-\frac{1}{2}}v^{(n)},\,Z_{i})}\Big]^2\,
+ \Big[\frac{(\|(u^{(n)})\|_{\ell}
\,\|\partial\,m(\tilde{\rho}_{n},\,Z_{i})\|_{\ell}}{\sigma(\theta_{0}+n^{-\frac{1}{2}}v^{(n)}\,,\,Z_{i})}\Big]^2\Big\}.
\end{eqnarray*}
It results that
\begin{eqnarray}
  \nonumber\lefteqn{|\frac{1}{n}\,\sum
_{i=1}^{n}\,A(\theta_{0},\tilde{\rho}_{n},\tilde{\theta}_{n},Z_{i})\,\dot{M}_{f}({\epsilon}_{i})\frac{G(Z_{i})}{\sigma(\theta_{0}\,,\,Z_{i})}|
 }\\&\leq&\frac{1}{2n}\sum
_{i=1}^{n}\Big[\frac{\|(v^{(n)})\|_{p}\,\|\partial\sigma(\tilde{\theta}_{n},Z_{i})\|_{p}}
{\sigma(\theta_{0}+n^{-\frac{1}{2}}v^{(n)},\,Z_{i})}\Big]^2|\dot{M}_{f}({\epsilon}_{i})\frac{G(Z_{i})}{\sigma(\theta_{0}\,,\,Z_{i})}|\nonumber\\
&+& \frac{1}{2n}\sum _{i=1}^{n}\Big[\frac{(\|(u^{(n)})\|_{\ell}
\|\partial\,m(\tilde{\rho}_{n},\,Z_{i})\|_{\ell}}{\sigma(\theta_{0}+n^{-\frac{1}{2}}v^{(n)}\,,\,Z_{i})}\Big]^2
|\dot{M}_{f}({\epsilon}_{i})\frac{G(Z_{i})}{\sigma(\theta_{0}\,,\,Z_{i})}|.\label{cs}
\end{eqnarray}
Since  that for all $x$, we have
\begin{eqnarray}
  \|\partial\,m(\rho,\,x)\|_{\ell}
 &\leq&\sqrt{\ell}\,\max_{1\leq i\leq\ell}|\frac{\partial\,m(\rho,x)}{\partial\rho_{i}}|,\label{normeetmoyenne}\\
\mbox{and,}\quad \|\partial\sigma(\theta,x)\|_{p}
&\leq&\sqrt{p}\,\max_{1\leq j\leq p}
|\frac{\partial\,\sigma(\theta,x)}{\partial\theta_{j}}|\label{normeetvariance}.
\end{eqnarray} Therefore, it follows from the inequalities
(\ref{asyconvex1}), (\ref{asyconvex2}), (\ref{cs}) and the conditions  ($A_{4.1}$), that \\
   There exist two closed balls
     $ \overline{B_{1,n}}=\overline{B_{1,n}}\Big(\rho_{0},r_{1,n}\Big)\subset
     int(\Theta_{1})$ and \\$\overline{B_{2,n}}=\overline{B_{2,n}}(\theta_{0},r_{2,n})\subset
     int(\Theta_{2})$ where $r_{1,n}\geq r_{n}$ and $r_{2,n} \geq r'_{n}$  and a positive function
     $N_{1,n}$, such that $E\Big({\sup_{n\geq 1} [N_{1,n}(Z_{0})]}\Big)^{\mu
     +2}<\infty$, where  $\mu>0$ , such that
\begin{eqnarray}
  |\frac{1}{n}\,\sum
_{i=1}^{n}\,A(\theta_{0},\tilde{\rho}_{n},\tilde{\theta}_{n},Z_{i})\,\dot{M}_{f}({\epsilon}_{i})\frac{G(Z_{i})}{\sigma(\theta_{0}\,,\,Z_{i})}|
&\leq&\max(\ell,p).\,\sup_{n}[({{\tau}^{(n)})}^\top({\tau}^{(n)})]\nonumber\\
&\times&\frac{1}{n}\sum _{i=1}^{n}{({\sup_{n\geq 1}
[N_{1,n}(Z_{i})]})}^2 \,\Big
|\dot{M}_{f}({\epsilon}_{i})\frac{G(Z_{i})}{\sigma(\theta_{0}\,,\,Z_{i})}\Big|.\nonumber\\
\label{evaluation1}
\end{eqnarray}
\noindent Note that the quantity
$E|\dot{M}_{f}({\epsilon}_{0})\frac{G(Z_{0})}{\sigma(\theta_{0}\,,\,Z_{0})}{({\sup_{n\geq
1}
[N_{1,n}(Z_{0})]})}^2|<+\infty$,\\
 In fact, by  H\"{o}lder's inequality, we have
\begin{eqnarray*}
\lefteqn{E|\dot{M}_{f}({\epsilon}_{0})\frac{G(Z_{0})}{\sigma(\theta_{0}\,,\,Z_{0})}{({\sup_{n\geq
1} [N_{1,n}(Z_{0})]})}^2|}\\&\leq&
\Big\{E|\dot{M}_{f}({\epsilon}_{0})\frac{G(Z_{0})}{\sigma(\theta_{0}\,,\,Z_{0})}|^{\lambda
+2}\Big\}^{\frac{1}{{\lambda +2}}}\Big\{E{({\sup_{n\geq 1}
[N_{1,n}(Z_{0})]})}^{2t}\Big\}^{\frac{1}{t}}\\
&\leq&\Big\{E|\dot{M}_{f}({\epsilon}_{0})|^{\lambda+2}\Big\}^{\frac{1}{{\lambda
+2}}}\Big\{E{|\frac{G(Z_{0})}{\sigma(\theta_{0}\,,\,Z_{0})}|}^{\lambda+2}\Big\}^{\frac{1}{{\lambda
+2}}}\Big\{E{({\sup_{n\geq 1}
[N_{1,n}(Z_{0})]})}^{2t}\Big\}^{\frac{1}{t}}.
\end{eqnarray*}
\noindent Since that $\frac{1}{{\lambda} +2}+ \frac{1}{t}=1,$ then
$t=1+\frac{1}{{\lambda} +1}$ then  $2\,t=2+\mu$, the conditions
$(A_{3.1})$, $(A_{3.5})$ and $(A_{4.1})$  enable us  to conclude
that,
$E|\dot{M}_{f}({\epsilon}_{0})\frac{G(Z_{0})}{\sigma(\theta_{0}\,,\,Z_{0})}{({\sup_{n\geq
1} [N_{1,n}(Z_{0})]})}^2|<+\infty$.\\
\noindent It follows from the stationarity and the ergodicity of
model that the random variable\\
$\frac{1}{n}\sum_{i=1}^{n}|\dot{M}_{f}({\epsilon}_{i})\frac{G(Z_{i})}{\sigma(\theta_{0}\,,\,Z_{i})}{({\sup_{n\geq
1} [N_{1,n}(Z_{i})]})}^2|$   converges a.s. to the  constant \\
$E|\dot{M}_{f}({\epsilon}_{0})\frac{G(Z_{0})}{\sigma(\theta_{0}\,,\,Z_{0})}{({\sup_{n\geq
1} [N_{1,n}(Z_{0})]})}^2|$ as $n\rightarrow +\infty .$ From
(\ref{c}) and the inequality (\ref{evaluation1}), it results that:
\begin{eqnarray}
 n^{-\frac{1}{2}} h \, \frac{1}{n}\,\sum
_{i=1}^{n}\,A(\theta_{0},\tilde{\rho}_{n},\tilde{\theta}_{n},Z_{i})\,\dot{M}_{f}({\epsilon}_{i})\frac{G(Z_{i})}{\sigma(\theta_{0}\,,\,Z_{i})}=
o_{P}(1).\label{ta1}
\end{eqnarray}
\noindent By following the same previous reasoning in the last
case and changing  $(A_{3.5})$ by $(A_{3.6})$, we shall prove that
\begin{eqnarray}
  |B(\theta_{0},\tilde{\theta}_{n},Z_{i})|\leq \Big [\frac{\|(v^{(n)}) \|_{p}\,
\|\partial\,\sigma(\tilde{\theta}_{n},Z_{i})\|_{p}}{\sigma(\theta_{0}+n^{-\frac{1}{2}}v^{(n)},\,Z_{i})}\Big]^2,
\quad \mbox{and}\quad
  n^{-\frac{1}{2}} h \, \frac{1}{n}\,\sum_{i=1}^{n}\,B(\theta_{0},\tilde{\theta}_{n},Z_{i})\,{\epsilon}_{i}\dot{M}_{f}({\epsilon}_{i})\frac{G(Z_{i})}{\sigma(\theta_{0}\,,\,Z_{i})}&=&
o_{P}(1).\nonumber\\
 \label{ta2}
\end{eqnarray}

\noindent From the equalities (\ref{ta1}) and (\ref{ta2}), we
deduce that
\begin{eqnarray}
I^{(1)}_{n,h,1} &=& o_{P}(1)\label{terme11}.
\end{eqnarray}
\noindent Using the equality (\ref{epsilon}), the expression
$I^{(2)}_{n,h,1}$ can also be written
\begin{eqnarray*}
I^{(2)}_{n,h,1}&=& -n^{-\frac{1}{2}}\, h \,\sum
_{i=1}^{n}\frac{(\tilde{\epsilon}_{i,n} -
{\epsilon}_{i})\dot{M}_{f}({\epsilon}_{i})
}{\sigma(\theta_{0}\,,\,Z_{i})}\,G(Z_{i})\\
&=&  h \,\frac{1}{n}\,\sum _{i=1}^{n}
\frac{(u^{(n)})^\top\,\partial\,m(\tilde{\rho}_{n},\,Z_{i})^\top}{\sigma(\theta_{0}+n^{-\frac{1}{2}}v^{(n)}\,,\,Z_{i})}\frac{\dot{M}_{f}({\epsilon}_{i})\,G(Z_{i})}
{\sigma(\theta_{0}\,,\,Z_{i})} \\&&+  h \,\frac{1}{n}\,\sum
_{i=1}^{n}
\frac{(v^{(n)})^\top\,\partial\,\sigma(\tilde{\theta}_{n},\,Z_{i})^\top}{\sigma(\theta_{0}+n^{-\frac{1}{2}}v^{(n)}\,,\,Z_{i})}\frac{{\epsilon}_{i}\dot{M}_{f}({\epsilon}_{i})\,G(Z_{i})}
{\sigma(\theta_{0}\,,\,Z_{i})}.
\end{eqnarray*}
We have the following decomposition :
\begin{eqnarray*}
 h \,\frac{1}{n}\,\sum _{i=1}^{n}
\frac{(u^{(n)})^\top\,\partial\,m(\tilde{\rho}_{n},\,Z_{i})^\top}{\sigma(\theta_{0}+n^{-\frac{1}{2}}v^{(n)}\,,\,Z_{i})}\frac{\dot{M}_{f}({\epsilon}_{i})\,G(Z_{i})}
{\sigma(\theta_{0}\,,\,Z_{i})}&=& h \,\frac{1}{n}\,\sum _{i=1}^{n}
\frac{(u^{(n)})^\top\,\partial\,m(\tilde{\rho}_{n},\,Z_{i})^\top}{\sigma(\theta_{0}+n^{-\frac{1}{2}}v^{(n)}\,,\,Z_{i})}\frac{\dot{M}_{f}({\epsilon}_{i})\,G(Z_{i})}
{\sigma(\theta_{0}\,,\,Z_{i})}\\
 &-&  h \,\frac{1}{n}\,\sum_{i=1}^{n}
\frac{(u^{(n)})^\top\,\partial\,m(\tilde{\rho}_{n},\,Z_{i})^\top}{\sigma(\theta_{0}\,,\,Z_{i})}\frac{\dot{M}_{f}({\epsilon}_{i})\,G(Z_{i})}
{\sigma(\theta_{0}\,,\,Z_{i})}\\
&+&h \,\frac{1}{n}\,\sum_{i=1}^{n}
\frac{(u^{(n)})^\top\,\partial\,m(\tilde{\rho}_{n},\,Z_{i})^\top}{\sigma(\theta_{0}\,,\,Z_{i})}\frac{\dot{M}_{f}({\epsilon}_{i})\,G(Z_{i})}
{\sigma(\theta_{0}\,,\,Z_{i})}\\
&=& I^{(2,1 )}_{n,h,1} + I^{(2,2 )}_{n,h,1}, \quad \mbox{where}\\
I^{(2,1 )}_{n,h,1}&=& h \,\frac{1}{n}\,\sum _{i=1}^{n}
\frac{(u^{(n)})^\top\,\partial\,m(\tilde{\rho}_{n},\,Z_{i})^\top}{\sigma(\theta_{0}+n^{-\frac{1}{2}}v^{(n)}\,,\,Z_{i})}\frac{\dot{M}_{f}({\epsilon}_{i})\,G(Z_{i})}
{\sigma(\theta_{0}\,,\,Z_{i})}\\
 &-&  h \,\frac{1}{n}\,\sum_{i=1}^{n}
\frac{(u^{(n)})^\top\,\partial\,m(\tilde{\rho}_{n},\,Z_{i})^\top}{\sigma(\theta_{0}\,,\,Z_{i})}\frac{\dot{M}_{f}({\epsilon}_{i})\,G(Z_{i})}
{\sigma(\theta_{0}\,,\,Z_{i})},\\
I^{(2,2 )}_{n,h,1}&=&h \,\frac{1}{n}\,\sum_{i=1}^{n}
\frac{(u^{(n)})^\top\,\partial\,m(\tilde{\rho}_{n},\,Z_{i})^\top}{\sigma(\theta_{0}\,,\,Z_{i})}\frac{\dot{M}_{f}({\epsilon}_{i})\,G(Z_{i})}
{\sigma(\theta_{0}\,,\,Z_{i})}.
\end{eqnarray*}
\noindent And
\begin{eqnarray*}
h \,\frac{1}{n}\,\sum _{i=1}^{n}
\frac{(v^{(n)})^\top\,\partial\,\sigma(\tilde{\theta}_{n},\,Z_{i})^\top}{\sigma(\theta_{0}+n^{-\frac{1}{2}}v^{(n)}\,,\,Z_{i})}\frac{{\epsilon}_{i}\dot{M}_{f}({\epsilon}_{i})\,G(Z_{i})}
{\sigma(\theta_{0}\,,\,Z_{i})}&=&h \,\frac{1}{n}\,\sum _{i=1}^{n}
\frac{(v^{(n)})^\top\,\partial\,\sigma(\tilde{\theta}_{n},\,Z_{i})^\top}{\sigma(\theta_{0}+n^{-\frac{1}{2}}v^{(n)}\,,\,Z_{i})}\frac{{\epsilon}_{i}\dot{M}_{f}({\epsilon}_{i})\,G(Z_{i})}
{\sigma(\theta_{0}\,,\,Z_{i})}\\
 &-& h \,\frac{1}{n}\,\sum _{i=1}^{n}
\frac{(v^{(n)})^\top\,\partial\,\sigma(\tilde{\theta}_{n},\,Z_{i})^\top}{\sigma(\theta_{0}\,,\,Z_{i})}\frac{{\epsilon}_{i}\dot{M}_{f}({\epsilon}_{i})\,G(Z_{i})}
{\sigma(\theta_{0}\,,\,Z_{i})}\\
&+& h \,\frac{1}{n}\,\sum _{i=1}^{n}
\frac{(v^{(n)})^\top\,\partial\,\sigma(\tilde{\theta}_{n},\,Z_{i})^\top}{\sigma(\theta_{0}\,,\,Z_{i})}\frac{{\epsilon}_{i}\dot{M}_{f}({\epsilon}_{i})\,G(Z_{i})}
{\sigma(\theta_{0}\,,\,Z_{i})}\\
&=& I^{(2,3)}_{n,h,1} + I^{(2,4)}_{n,h,1}, \quad \mbox{where}\\
I^{(2,3)}_{n,h,1}&=&h \,\frac{1}{n}\,\sum _{i=1}^{n}
\frac{(v^{(n)})^\top\,\partial\,\sigma(\tilde{\theta}_{n},\,Z_{i})^\top}{\sigma(\theta_{0}+n^{-\frac{1}{2}}v^{(n)}\,,\,Z_{i})}\frac{{\epsilon}_{i}\dot{M}_{f}({\epsilon}_{i})\,G(Z_{i})}
{\sigma(\theta_{0}\,,\,Z_{i})}\\
 &-& h \,\frac{1}{n}\,\sum _{i=1}^{n}
\frac{(v^{(n)})^\top\,\partial\,\sigma(\tilde{\theta}_{n},\,Z_{i})^\top}{\sigma(\theta_{0}\,,\,Z_{i})}\frac{{\epsilon}_{i}\dot{M}_{f}({\epsilon}_{i})\,G(Z_{i})}
{\sigma(\theta_{0}\,,\,Z_{i})},\\
I^{(2,4)}_{n,h,1}&=& h \,\frac{1}{n}\,\sum _{i=1}^{n}
\frac{(v^{(n)})^\top\,\partial\,\sigma(\tilde{\theta}_{n},\,Z_{i})^\top}{\sigma(\theta_{0}\,,\,Z_{i})}\frac{{\epsilon}_{i}\dot{M}_{f}({\epsilon}_{i})\,G(Z_{i})}
{\sigma(\theta_{0}\,,\,Z_{i})}.\\
\end{eqnarray*}
We have then
\begin{eqnarray*}
I^{(2)}_{n,h,1}&=&I^{(2,1)}_{n,h,1} +
I^{(2,2)}_{n,h,1}+I^{(2,3)}_{n,h,1} + I^{(2,4)}_{n,h,1}.
\end{eqnarray*}
We evaluate the terms $I^{(2,1)}_{n,h,1}$ , $I^{(2,2)}_{n,h,1}$,
$I^{(2,3)}_{n,h,1}$ and $I^{(2,4)}_{n,h,1}.$ \noindent From
(\ref{variance}) the expression $I^{(2,1)}_{n,h,1}$ can also be
written
\begin{eqnarray*}
I^{(2,1)}_{n,h,1}&=&  -n^{-\frac{1}{2}}h\,\frac{1}{n}\,\sum
_{i=1}^{n}\frac{(v^{(n)})^\top \,
\partial\,\sigma(\tilde{\theta}_{n},Z_{i})^\top}{\sigma(\theta_{0}+n^{-\frac{1}{2}}v^{(n)}\,,Z_{i})}
\frac{(u^{(n)})^\top\,\partial\,m(\tilde{\rho}_{n},\,Z_{i})^\top}{\sigma(\theta_{0}\,,\,Z_{i})}\,\frac{G(Z_{i})}{\sigma(\theta_{0}\,,\,Z_{i})}\dot{M}_{f}({\epsilon}_{i}).
\end{eqnarray*}
\noindent By  Cauchy -Schwartz's inequality, we obtain
\begin{eqnarray*}
|\frac{(v^{(n)})^\top\,\partial\,\sigma(\tilde{\theta}_{n},Z_{i})^\top}{\sigma(\theta_{0}
+n^{-\frac{1}{2}}v^{(n)},\,Z_{i})}\frac{(u^{(n)})^\top\,\partial\,m(\tilde{\rho}_{n},\,Z_{i})^\top}{\sigma(\theta_{0},\,Z_{i})}|&\leq&
\|v^{(n)}\|_{p}\frac{\|\partial\,\sigma(\tilde{\theta}_{n},Z_{i})\|_{p}}{\sigma(\theta_{0}+n^{-\frac{1}{2}}v^{(n)},\,Z_{i})}.
\|u^{(n)}\|_{\ell}\frac{\|\partial\,m(\tilde{\rho}_{n},\,Z_{i})\|_{\ell}}{\sigma(\theta_{0},Z_{i})}\nonumber\\
&\leq&\frac{1}{2}\Big\{\|v^{(n)}\|_{p}\frac{\|\partial\,\sigma(\tilde{\theta}_{n},Z_{i})\|_{p}}{\sigma(\theta_{0}+
n^{-\frac{1}{2}}v^{(n)},\,Z_{i})}\Big\}^2\nonumber \\
&&  + \frac{1}{2}\Big\{\|u^{(n)}\|_{\ell}\frac{\|\partial\,m(\tilde{\rho}_{n},\,Z_{i})\|_{\ell}}{\sigma(\theta_{0},Z_{i})}\Big\}^2\nonumber\\
&&\leq
\frac{1}{2}\sup_{n}[({{\tau}^{(n)})}^\top({\tau}^{(n)})]\Big[\Big\{\frac{\|\partial\,\sigma(\tilde{\theta}_{n},Z_{i})\|_{p}}{\sigma(\theta_{0}+
n^{-\frac{1}{2}}v^{(n)},\,Z_{i})}\Big\}^2 \nonumber\\
&& +
\Big\{\frac{\|\partial\,m(\tilde{\rho}_{n},\,Z_{i})\|_{\ell}}{\sigma(\theta_{0},Z_{i})}\Big\}^2
\Big].
\end{eqnarray*}
Then, we obtain
\begin{eqnarray}
|I^{(2,1)}_{n,h,1}|&\leq&n^{-\frac{1}{2}}|h|\,\sup_{n}[({{\tau}^{(n)})}^\top({\tau}^{(n)})]\frac{1}{n}\,\sum
_{i=1}^{n}|\frac{G(Z_{i})}{\sigma(\theta_{0}\,,\,Z_{i})}\dot{M}_{f}({\epsilon}_{i})|{({\sup_{n\geq
1} [N_{1,n}(Z_{i})]})}^2.\nonumber\\
\label{evaluation2}
\end{eqnarray}
\noindent Using the inequality (\ref{asyconvex1}),
(\ref{asyconvex2}), (\ref{normeetmoyenne}),
(\ref{normeetvariance}), (\ref{evaluation2}) and from $(A_{3.1})$,
$(A_{4.1})$, $(A_{3.5})$, (\ref{c}) and the ergodic theorem, it
results  that
\begin{eqnarray}
I^{(2,1)}_{n,h,1}&=&o_{P}(1)\label{terme21}.
\end{eqnarray}
\noindent With a same reasoning and changing $(A_{3.5})$ by
$(A_{3.6})$, we shall prove that :
\begin{eqnarray}
I^{(2,3)}_{n,h,1}&=&o_{P}(1)\label{terme23}.
\end{eqnarray}
It remains to evaluate the terms $I^{(2,2)}_{n,h,1}$ and
$I^{(2,4)}_{n,h,1}.$
\begin{eqnarray*}
I^{(2,2 )}_{n,h,1}&=&h \,\frac{1}{n}\,\sum_{i=1}^{n}
\frac{(u^{(n)})^\top\,\partial\,m(\tilde{\rho}_{n},\,Z_{i})^\top}{\sigma(\theta_{0}\,,\,Z_{i})}\frac{\dot{M}_{f}({\epsilon}_{i})\,G(Z_{i})}
{\sigma(\theta_{0}\,,\,Z_{i})}.
\end{eqnarray*}
\noindent By Taylor's expansion with order $2$ of the functions
$\rho\rightarrow m(\rho,\cdot)$ and $\theta\rightarrow
\sigma(\theta,\cdot)$ around $\rho_{0}$ and $\theta_{0}$
respectively, we obtain the following equalities
\begin{eqnarray}
n^{-\frac{1}{2}}(u^{(n)})^\top\,\partial\,m(\tilde{\rho}_{n},\,Z_{i})^\top&=&n^{-\frac{1}{2}}(u^{(n)})^\top\,\partial\,m({\rho}_{0},\,Z_{i})^\top
+\frac{1}{2}n^{-\frac{1}{2}}(u^{(n)})^\top\,\partial^2\,m(\tilde{\tilde{\rho}}_{n},\,Z_{i})n^{-\frac{1}{2}}(u^{(n)}),~~~~~~~~~~~~~~~~\label{moyenneseconde}\\
n^{-\frac{1}{2}}(v^{(n)})^\top\,\partial\,\sigma(\tilde{\theta}_{n},\,Z_{i})^\top&=&n^{-\frac{1}{2}}(v^{(n)})^\top\,\partial\,\sigma({\theta}_{0},\,Z_{i})^\top
+\frac{1}{2}n^{-\frac{1}{2}}(v^{(n)})^\top\partial^2\,\sigma(\tilde{\tilde{\theta}}_{n},\,Z_{i})n^{-\frac{1}{2}}(v^{(n)}),~~~~~~~~~~~~~~~~\label{varianceseconde}
\end{eqnarray}
\noindent where $\tilde{\tilde{\rho}}_n$ and
$\tilde{\tilde{\theta}}_{n}$ are between $\rho_0$ and $\rho_n$ and
${\theta}_{0}$ and ${\theta}_{n}$ respectively. \noindent From
(\ref{moyenneseconde}), the expression $I^{(2,2)}_{n,h,1}$ can
also be written
\begin{eqnarray*}
I^{(2,2)}_{n,h,1} &=& h \,\frac{1}{n}\,\sum _{i=1}^{n}
\frac{(u^{(n)})^\top\,\partial\,m(\rho_{0},\,Z_{i})^\top}{\sigma(\theta_{0},\,Z_{i})}\frac{\dot{M}_{f}({\epsilon}_{i})\,G(Z_{i})}
{\sigma(\theta_{0},Z_{i})}\\&& +\frac{h}{2}\,n^{-\frac{1}{2}}
\,\frac{1}{n}\,\sum _{i=1}^{n}
\frac{(u^{(n)})^\top\,\partial^2\,m(\tilde{\tilde{\rho}}_{n},\,Z_{i})(u^{(n)})}{\sigma(\theta_{0},\,Z_{i})}
\frac{\dot{M}_{f}({\epsilon}_{i})\,G(Z_{i})}{\sigma(\theta_{0},Z_{i})}\\
&=&I^{(2,2,1)}_{n,h,1} + I^{(2,2,2)}_{n,h,1}.
\end{eqnarray*}
\noindent We consider the following term
\begin{eqnarray}
I^{(2,2,2)}_{n,h,1}&=&\frac{h}{2}\,n^{-\frac{1}{2}}
\,\frac{1}{n}\,\sum _{i=1}^{n}
\frac{(u^{(n)})^\top\,\partial^2\,m(\tilde{\tilde{\rho}}_{n},\,Z_{i})(u^{(n)})}{\sigma(\theta_{0},\,Z_{i})}
\frac{\dot{M}_{f}({\epsilon}_{i})\,G(Z_{i})}{\sigma(\theta_{0},Z_{i})}.~~~~~~~~~~~~~~~~~~~~~~~~~~~~~~~~~~~~~~~~~~~~~~~~~~~~~
\end{eqnarray}
\noindent For fall integers $i$, we have
\begin{eqnarray}
(u^{(n)})^\top
\partial^2\,m(\tilde{\tilde{\rho}}_{n},\,Z_{i})(u^{(n)})
 &=&\sum_{k=1}^{\ell}
\frac{\partial^2\,m(\tilde{\tilde{\rho}}_{n},Z_{i})}{\partial\,
{\rho}_{k}^2} {({u_{k}}^{(n)})}^2 + \sum \sum_{1\leq k,j\leq \ell,
k\neq
j}\frac{\partial^2\,m(\tilde{\tilde{\rho}}_{n},Z_{i})}{\partial\rho_{k}\partial\rho_{j}}({u_{k}}^{(n)})({u_{j}}^{(n)}).~~~~~~~~~~~
\end{eqnarray}
We have the following inequalities
\begin{eqnarray}
({u_{k}}^{(n)})({u_{j}}^{(n)})&\leq&|{u_{k}}^{(n)})({u_{j}}^{(n)})|\nonumber \\
&\leq&{\frac{1}{2}}[{({u_{k}}^{(n)})}^2 + {({u_{j}}^{(n)})}^2]\nonumber\\
&\leq& {\frac{1}{2}} \|u^{(n)}\|^2_{\ell}.\label{majorationdeu}
\end{eqnarray}
Using the inequality (\ref{majorationdeu}), we obtain
\begin{eqnarray}
 (u^{(n)})^\top
\partial^2\,m(\tilde{\tilde{\rho}}_{n},\,Z_{i})(u^{(n)})&\leq& \max_{1\leq i,j\leq\ell}\Big|\frac{\partial^2\,m(\tilde{\tilde{\rho}}_{n},Z_{i})}{\partial\rho_{k}\partial\rho_{j}}\Big|\Big[\ell \| u^{(n)}\|_{\ell}^{2} +
 \frac{\ell(\ell - 1)}{2} \| u^{(n)}\|_{\ell}^{2} \Big]\nonumber\\
 &\leq& \max_{1\leq k,j\leq\ell}\Big|\frac{\partial^2\,m(\tilde{\tilde{\rho}}_{n},Z_{i})}{\partial\rho_{k}\partial\rho_{j}}\Big|\Big[\ell+\frac{\ell(\ell - 1)}{2}\Big] \|
 u^{(n)}\|_{\ell}^{2}\nonumber\\
&\leq& \max_{1\leq
k,j\leq\ell}\Big|\frac{\partial^2\,m(\tilde{\tilde{\rho}}_{n},Z_{i})}{\partial\rho_{k}\partial\rho_{j}}\Big|\Big[\frac{{\ell}^2
+
 \ell}{2}\Big]\sup_{n}[({{\tau}^{(n)})}^\top({\tau}^{(n)})].~~~~~~~~~~~~~~~~~~~~~~~~~~~~~~~~~~~~\label{majhessienne}
\end{eqnarray}
With a same reasoning as (\ref{asyconvex1}) and
(\ref{asyconvex2}), we shall prove that
\begin{eqnarray}
 \|\tilde{\tilde{\rho}}_{n} -\rho_{0}\|_{\ell}&\leq& \|\rho_{n}- \rho_{0} \|_{\ell}.\label{asyconvex3}\\
 \|\tilde{\tilde{\theta}}_{n} -\theta_{0}\|_{p}&\leq& \|\theta_{n}- \theta_{0} \|_{p}.\label{asyconvex4}
\end{eqnarray}
\noindent The inequality (\ref{majhessienne}) associated  with
(\ref{asyconvex3}), (\ref{asyconvex4}), $(A_{3.1})$,$(A_{3.5})$,
$(A_{4.4})$, $(\ref{c})$ and the ergodicity and the stationarity
of the model implies that, when $n\longrightarrow +\infty$, we
obtain
\begin{eqnarray}
I^{(2,2,2)}_{n,h,1}&=&o_{P}(1)\label{term222}.
\end{eqnarray}
It remains to treat the term $I^{(2,2,1)}_{n,h,1}$, such that
\begin{eqnarray*}
I^{(2,2,1)}_{n,h,1} &=& h \,\frac{1}{n}\,\sum _{i=1}^{n}
\frac{(u^{(n)})^\top\,\partial\,m(\rho_{0},\,Z_{i})^\top}{\sigma(\theta_{0},\,Z_{i})}\frac{\dot{M}_{f}({\epsilon}_{i})\,G(Z_{i})}
{\sigma(\theta_{0},Z_{i})}.
\end{eqnarray*}
We have for all integers  $i$
\begin{eqnarray*}
 (u^{(n)})^\top\,\partial\,m(\rho_{0},\,Z_{i})^\top &=&  \sum _{j=1}^{\ell}
 u_{j}^{(n)}\,\frac{\partial\,m(\rho_{0},\,Z_{i})}{\partial\,\rho_{j} }.
 \end{eqnarray*}
\noindent We  obtain
\begin{eqnarray*}
I^{(2,2,1)}_{n,h,1} &=&h \,\frac{1}{n}\,\sum _{i=1}^{n}
\frac{(u^{(n)})^\top\,\partial\,m(\rho_{0},\,Z_{i})^\top}{\sigma(\theta_{0}\,,\,Z_{i})}\frac{\dot{M}_{f}({\epsilon}_{i})\,G(Z_{i})}
{\sigma(\theta_{0}\,,\,Z_{i})}\\
 &=& h \,\frac{1}{n}\,\sum _{i=1}^{n} \frac{
u_{1}^{(n)}\,\frac{\partial\,m(\rho_{0},\,Z_{i})}{\partial\,\rho_{1}
}}{\sigma(\theta_{0}\,,\,Z_{i})}\frac{\dot{M}_{f}({\epsilon}_{i})\,G(Z_{i})}
{\sigma(\theta_{0}\,,\,Z_{i})} +\cdot\, \cdot\, \cdot\,+  h
\,\frac{1}{n}\,\sum _{i=1}^{n} \frac{
u_{\ell}^{(n)}\,\frac{\partial\,m(\rho_{0},\,Z_{i})}{\partial\,\rho_{\ell}
}}{\sigma(\theta_{0}\,,\,Z_{i})}\frac{\dot{M}_{f}({\epsilon}_{i})\,G(Z_{i})}
{\sigma(\theta_{0}\,,\,Z_{i})}\\
 &=&h\,u_{1}^{(n)}\,\frac{1}{n}\,\sum _{i=1}^{n} \frac{
\frac{\partial\,m(\rho_{0},\,Z_{i})}{\partial\,\rho_{1}
}}{\sigma(\theta_{0}\,,\,Z_{i})}\frac{\dot{M}_{f}({\epsilon}_{i})\,G(Z_{i})}
{\sigma(\theta_{0}\,,\,Z_{i})} +\cdot\, \cdot\, \cdot\,+ h
\,u_{\ell}^{(n)} \,\frac{1}{n}\,\sum _{i=1}^{n} \frac{
\,\frac{\partial\,m(\rho_{0},\,Z_{i})}{\partial\,\rho_{\ell}
}}{\sigma(\theta_{0}\,,\,Z_{i})}\frac{\dot{M}_{f}({\epsilon}_{i})\,G(Z_{i})}
{\sigma(\theta_{0}\,,\,Z_{i})}.
\end{eqnarray*}
\noindent It follows from $(A_{3.1})$, $(A_{3.5})$ and $(A_{4.3})$ that \\
\noindent For all $j \in \{1,\ldots,\ell\}$ and as
$n\rightarrow+\infty,$ we have
\begin{eqnarray*}
  \frac{1}{n}\,\sum _{i=1}^{n} \frac{
\frac{\partial\,m(\rho_{0},\,Z_{i})}{\partial\,\rho_{j}
}}{\sigma(\theta_{0}\,,\,Z_{i})}\frac{\dot{M}_{f}({\epsilon}_{i})\,G(Z_{i})}
{\sigma(\theta_{0}\,,\,Z_{i})} \stackrel{a.s.}{\longrightarrow}
\mathbb{E}\Big[\frac{
\frac{\partial\,m(\rho_{0},\,Z_{0})}{\partial\,\rho_{j}
}}{\sigma(\theta_{0}\,,\,Z_{0})}\frac{\dot{M}_{f}({\epsilon}_{0})\,G(Z_{0})}
{\sigma(\theta_{0}\,,\,Z_{0})}\Big] = K_{j} .
\end{eqnarray*}
\noindent Therefore, there exist for all $j \in \{1,\ldots,\ell\}$
a random variable $E_{j,n},$  where $E_{j,n}$ converges a.s to $0$
as $n \rightarrow +\infty$,  such that
\begin{eqnarray*}
\frac{1}{n}\,\sum _{i=1}^{n} \frac{
\frac{\partial\,m(\rho_{0},\,Z_{i})}{\partial\,\rho_{j}
}}{\sigma(\theta_{0}\,,\,Z_{i})}\frac{\dot{M}_{f}({\epsilon}_{i})\,G(Z_{i})}
{\sigma(\theta_{0}\,,\,Z_{i})} &=& K_{j} + E_{j,n}._{}
\end{eqnarray*}
\noindent So
\begin{eqnarray*}
u_{j}^{(n)}\frac{1}{n}\,\sum _{i=1}^{n} \frac{
\frac{\partial\,m(\rho_{0},\,Z_{i})}{\partial\,\rho_{j}
}}{\sigma(\theta_{0}\,,\,Z_{i})}\frac{\dot{M}_{f}({\epsilon}_{i})\,G(Z_{i})}
{\sigma(\theta_{0}\,,\,Z_{i})} &=& u_{j}^{(n)}\,K_{j}
+u_{j}^{(n)}\, E_{j,n}.
\end{eqnarray*}
\noindent We have
$$ u_{j}^{(n)}\leq \| u^{(n)} \|_{\ell}\leq
[\sup_{n}[({{\tau}^{(n)})}^\top({\tau}^{(n)})] <  +\infty .$$
\noindent Therefore
\begin{eqnarray*}
u_{j}^{(n)}\, E_{j,n}&=& o_{P}(1).
\end{eqnarray*}
\noindent It results that
\begin{eqnarray}
I^{(2,2,1)}_{n,h,1} &=& h\, (u^{(n)})^\top K^\top   +
o_{P}(1),\label{term221}
\end{eqnarray}
\noindent with
\begin{eqnarray*}
  K^\top  &=&  (K_{1},\cdot,\cdot,\cdot,K_{\ell}),\\
  K_{j}&=&  \mathbb{E}\Big[\frac{
\frac{\partial\,m(\rho_{0},\,Z_{0})}{\partial\,\rho_{j}
}}{\sigma(\theta_{0}\,,\,Z_{0})}\frac{\dot{M}_{f}({\epsilon}_{0})\,G(Z_{0})}
{\sigma(\theta_{0}\,,\,Z_{0})}\Big], \\
j &\in&\{1,\ldots,\ell\}.
\end{eqnarray*}
\noindent It follows from the equalities (\ref{term222}) and
(\ref{term221})
 that
\begin{eqnarray}
I^{(2,2)}_{n,h,1} &=& h\, (u^{(n)})^\top K^\top   +
o_{P}(1).\label{terme22}
\end{eqnarray}
\noindent It remain to process the term $I^{(2,4)}_{n,h,1}.$\\
With a similar method, we shall give a similar inequality as
(\ref{majhessienne}), therefore we obtain
\begin{eqnarray}
 (v^{(n)})^\top\,\partial^2\,\sigma(\tilde{\tilde{\theta}}_{n},\,Z_{i})(v^{(n)})&\leq& \max_{1\leq k,j\leq p}\Big|\frac{\partial^2\,
 \sigma(\tilde{\tilde{\theta}}_{n},\,Z_{i})}{\partial\theta_{k}\partial\theta_{j}}\Big|\Big[\frac{p^2+p}{2}\Big]
 \sup_{n}[({{\tau}^{(n)})}^\top({\tau}^{(n)})]\nonumber\\\label{majhessienne2}.
\end{eqnarray}
\noindent By changing  (\ref{moyenneseconde}),
(\ref{majhessienne}) and $(A_{3.5})$ by
 (\ref{varianceseconde}), (\ref{majhessienne2}) and  $(A_{3.6})$ respectively  and using the same reasoning as the term $I^{(2,2)}_{n,h,1}$
 , we obtain the following equation :
\begin{eqnarray}
I^{(2,4)}_{n,h,1} &=& h\, (v^{(n)})^\top J^\top  +
o_{P}(1),\label{terme24}\\
\mbox{where}\quad
 J^\top  &=&  (J_{1},\cdot,\cdot,\cdot,J_{p}),\\
\mbox{and}\quad J_{k}&=&  \mathbb{E}\Big[\frac{
\frac{\partial\,\sigma(\theta_{0},\,Z_{0})}{\partial\,\theta_{k}
}}{\sigma(\theta_{0}\,,\,Z_{0})}\frac{{\epsilon}_{0}\dot{M}_{f}({\epsilon}_{0})\,G(Z_{0})}
{\sigma(\theta_{0}\,,\,Z_{0})}\Big].
\end{eqnarray}
\textbf{In summary, we have }
\begin{eqnarray*}
I_{n,h,1} &=&I^{(1)}_{n,h,1} + I^{(2)}_{n,h,1}.\\
I^{(2)}_{n,h,1}&=&I^{(2,1)}_{n,h,1}+I^{(2,2)}_{n,h,1}+I^{(2,3)}_{n,h,1}+
I^{(2,4)}_{n,h,1}.
\end{eqnarray*}
It follows from the equalities  (\ref{terme11}), (\ref{terme21}),
(\ref{terme23}), (\ref{terme22}) and (\ref{terme24}), that:
\begin{eqnarray}
I_{n,h,1} &=&  h\, (u^{(n)})^\top K^\top  + h\, (v^{(n)})^\top
J^\top  + o_{P}(1).\label{terme1}
\end{eqnarray}
\subsubsection*{Evaluation of the term $I_{n,h,3}$ }
From the equality (\ref{varianceseconde}), we obtain
\begin{eqnarray*}
I_{n,h,3} &=& {\frac{1}{n}}\, h \,\sum
_{i=1}^{n}\frac{(v^{(n)})^\top \, \partial\,
\sigma(\tilde{\theta_{n},}\,Z_{i})^\top}{\sigma(\theta_{0}+n^{-\frac{1}{2}}v^{(n)}\,,\,Z_{i})}\,M_{f}({\epsilon}_{i})\,\frac{G(Z_{i})}{\sigma(\theta_{0}\,,\,Z_{i})}\\
&&=I^{(1)}_{n,h,3} + I^{(2)}_{n,h,3},\\
\mbox{ where} \quad I^{(1)}_{n,h,3}&=& {\frac{1}{n}}\, h \,\sum
_{i=1}^{n}\frac{(v^{(n)})^\top \,
\partial\,
\sigma({\theta}_{0},\,Z_{i})^\top}{\sigma(\theta_{0}+n^{-\frac{1}{2}}v^{(n)}\,
,\,Z_{i})}\,M_{f}({\epsilon}_{i})\,\frac{G(Z_{i})}{\sigma(\theta_{0}\,,\,Z_{i})},\\
\mbox{ and} \quad I^{(2)}_{n,h,3}&=& \frac{n^{-\frac{1}{2}}}{2}\,
h\,\frac{1}{n} \,\sum _{i=1}^{n}\frac{(v^{(n)})^\top \,
\partial^2\,
\sigma(\tilde{\tilde{\theta}}_{n},\,Z_{i})(v^{(n)})}{\sigma(\theta_{0}+n^{-\frac{1}{2}}v^{(n)}\,,\,Z_{i})}\,M_{f}({\epsilon}_{i})\,\frac{G(Z_{i})}{\sigma(\theta_{0}\,,\,Z_{i})}.\\
\end{eqnarray*}
 \noindent and $\tilde{\tilde{\theta}}_{n}$ is between  ${\theta}_{0}$
 and ${\theta}_{n}.$\\We have
\begin{eqnarray*}
|I^{(2)}_{n,h,3}|&\leq& \frac{n^{-\frac{1}{2}}}{2}\,
|h|\,\frac{1}{n} \,\sum _{i=1}^{n}\frac{|(v^{(n)})^\top \,
\partial^2\,
\sigma(\tilde{\tilde{\theta}}_{n},\,Z_{i})(v^{(n)})|}{\sigma(\theta_{0}+n^{-\frac{1}{2}}v^{(n)}\,,\,Z_{i})}\,|M_{f}({\epsilon}_{i})\,
\frac{G(Z_{i})}{\sigma(\theta_{0}\,,\,Z_{i})}|.
\end{eqnarray*}
\noindent It follows from (\ref{c}), (\ref{asyconvex3}),
(\ref{asyconvex4}) , (\ref{majhessienne2}), $(A_{3.1})$,
$(A_{3.3})$, $(A_{4.4})$  and the ergodicity and the stationarity
that, when $n \rightarrow +\infty$, we obtain
\begin{eqnarray}
I^{(2)}_{n,h,3}&=&o_{P}(1)\label{32}.
\end{eqnarray}
 With the use of  Taylor's expansion with order $1$ of the
function $\sigma(\theta,\cdot)$ around $\theta_{0}$, we have
\begin{eqnarray*}
I^{(1)}_{n,h,3}&=& {\frac{1}{n}}\, h \,\sum
_{i=1}^{n}\frac{(v^{(n)})^\top \, \partial\,
\sigma({\theta}_{0},\,Z_{i})^\top}{\sigma(\theta_{0}+n^{-\frac{1}{2}}v^{(n)}\,
,\,Z_{i})}\,M_{f}({\epsilon}_{i})\,\frac{G(Z_{i})}{\sigma(\theta_{0}\,,\,Z_{i})}\\
&=&{\frac{1}{n}}\, h \,\sum _{i=1}^{n}\frac{(v^{(n)})^\top \,
\partial\,
\sigma({\theta}_{0},\,Z_{i})^\top}{\sigma(\theta_{0}+n^{-\frac{1}{2}}v^{(n)}\,
,\,Z_{i})}\,M_{f}({\epsilon}_{i})\,\frac{G(Z_{i})}{\sigma(\theta_{0}\,,\,Z_{i})}\\
&&-{\frac{1}{n}}\, h \,\sum _{i=1}^{n}\frac{(v^{(n)})^\top \,
\partial\,
\sigma({\theta}_{0},\,Z_{i})^\top}{\sigma(\theta_{0},\,Z_{i})}\,M_{f}({\epsilon}_{i})\,\frac{G(Z_{i})}{\sigma(\theta_{0}\,,\,Z_{i})}\\
&&+{\frac{1}{n}}\, h \,\sum _{i=1}^{n}\frac{(v^{(n)})^\top \,
\partial\,
\sigma({\theta}_{0},\,Z_{i})^\top}{\sigma(\theta_{0},\,Z_{i})}\,M_{f}({\epsilon}_{i})\,\frac{G(Z_{i})}{\sigma(\theta_{0}\,,\,Z_{i})}\\
&=&{\frac{1}{n}}\, h \,\sum _{i=1}^{n}\Big[
\frac{1}{\sigma(\theta_{0}+n^{-\frac{1}{2}}v^{(n)}\, ,\,Z_{i})}
-\frac{1}{\sigma({\theta}_{0},\,Z_{i})}\Big]\,(v^{(n)})^\top \,
\partial\,
\sigma({\theta}_{0},\,Z_{i})^\top M_{f}({\epsilon}_{i})\,\frac{G(Z_{i})}{\sigma(\theta_{0}\,,\,Z_{i})}\\
&&+{\frac{1}{n}}\, h \,\sum _{i=1}^{n}\frac{(v^{(n)})^\top \,
\partial\,
\sigma({\theta}_{0},\,Z_{i})^\top}{\sigma(\theta_{0},\,Z_{i})}\,M_{f}({\epsilon}_{i})\,\frac{G(Z_{i})}{\sigma(\theta_{0}\,,\,Z_{i})}\\
&=& -n^{-\frac{1}{2}}{\frac{1}{n}}\, h \,\sum
_{i=1}^{n}\frac{(v^{(n)})^\top \, \partial\,
\sigma({\tilde{\theta}}_{n},\,Z_{i})^\top}{\sigma(\theta_{0}+n^{-\frac{1}{2}}v^{(n)}\,
,\,Z_{i})}\,\frac{(v^{(n)})^\top \, \partial\,
\sigma({\theta}_{0},\,Z_{i})^\top}{\sigma(\theta_{0},\,Z_{i})}
M_{f}({\epsilon}_{i})\,\frac{G(Z_{i})}{\sigma(\theta_{0}\,,\,Z_{i})}\\
&&+{\frac{1}{n}}\, h \,\sum _{i=1}^{n}\frac{(v^{(n)})^\top \,
\partial\,
\sigma({\theta}_{0},\,Z_{i})^\top}{\sigma(\theta_{0},\,Z_{i})}\,M_{f}({\epsilon}_{i})\,\frac{G(Z_{i})}{\sigma(\theta_{0}\,,\,Z_{i})}\\
&=& I^{(1,1)}_{n,h,3} + I^{(1,2)}_{n,h,3}.
\end{eqnarray*}
\noindent By Cauchy-Schwartz's inequality followed by the use of
(\ref{c}), (\ref{asyconvex1}), (\ref{asyconvex2}) , $(A_{3.1})$,
$(A_{3.3})$, $(A_{4.1})$ and the ergodicity and the stationarity
of the model, we shall to prove that
\begin{eqnarray}
I^{(1,1)}_{n,h,3} &=&  -n^{-\frac{1}{2}}{\frac{1}{n}}\, h \,\sum
_{i=1}^{n}\frac{(v^{(n)})^\top \, \partial\,
\sigma({\tilde{\theta}}_{n},\,Z_{i})^\top}{\sigma(\theta_{0}+n^{-\frac{1}{2}}v^{(n)}\,
,\,Z_{i})}\,\frac{(v^{(n)})^\top \, \partial\,
\sigma({\theta}_{0},\,Z_{i})^\top}{\sigma(\theta_{0},\,Z_{i})}
M_{f}({\epsilon}_{i})\,\frac{G(Z_{i})}{\sigma(\theta_{0}\,,\,Z_{i})}\nonumber\\
&=&o_{P}(1).\label{terme311}
\end{eqnarray}
\noindent It remains to evaluate the term $I^{(1,2)}_{n,h,3}$,
where
\begin{eqnarray*}
I^{(1,2)}_{n,h,3}&=&{\frac{1}{n}}\, h \,\sum
_{i=1}^{n}\frac{(v^{(n)})^\top \,
\partial\,
\sigma({\theta}_{0},\,Z_{i})^\top}{\sigma(\theta_{0},\,Z_{i})}\,M_{f}({\epsilon}_{i})\,\frac{G(Z_{i})}{\sigma(\theta_{0}\,,\,Z_{i})}.
\end{eqnarray*}
\noindent Using the same  reasoning applied on the term
$I^{(2,2)}_{n,h,1}$ with changing the condition $(A_{3.5})$ by
$(A_{3.3})$ and using $(A_{2.1}),$ we shall prove that
\begin{eqnarray}
I^{(1,2)}_{n,h,3}&=& h\, (v^{(n)})^\top Q^\top   + o_{P}(1)\nonumber\\
&=& o_{P}(1),\label{terme312}
\end{eqnarray}
such that for all  $j \in\{1,\dots,p\},$ we have
\begin{eqnarray*}
  Q^\top  = (Q_{1},\cdot,\cdot,\cdot,Q_{p}),\\
  Q_{j} = \mathbb{E}\Big[\frac{
\frac{\partial\,\sigma(\theta_{0},\,Z_{0})}{\partial\,\theta_{j}
}}{\sigma(\theta_{0}\,,\,Z_{0})}\frac{M_{f}({\epsilon}_{0})\,G(Z_{0})}
{\sigma(\theta_{0}\,,\,Z_{0})}\Big] =0.
\end{eqnarray*}
\noindent \textbf{In summary}\\
From the equalities (\ref{32}), (\ref{terme311}) and
(\ref{terme312}), we deduce that
\begin{eqnarray}
I_{n,h,3} &=&  o_{P}(1)\label{EVALUATION TROIXIEMETERME}.
\end{eqnarray}
\subsubsection*{Evaluation of the term $I_{n,h,2}$}
 We have
\begin{eqnarray}
I_{n,h,2} &=& -n^{-\frac{1}{2}}\,  \frac{h}{2} \,\sum _{i=1}^{n}
(\tilde{{\epsilon}}_{i,n} -
{\epsilon}_{i})^2 \ddot{M}_{f}(\tilde{\tilde{{\epsilon}_{i}}})\, \frac{G(Z_{i})}{\,\sigma(\theta_{0}+n^{-\frac{1}{2}}v^{(n)}\,,\,Z_{i})}\nonumber\\
 &=&-n^{-\frac{1}{2}}\,  \frac{h}{2} \,\sum _{i=1}^{n}
(\tilde{{\epsilon}}_{i,n} - {\epsilon}_{i})^2
\mathcal{\mathcal{}}\ddot{M}_{f}(\tilde{\tilde{{\epsilon}_{i,n}}})\,
\frac{G(Z_{i})}{\,\sigma(\theta_{0}+n^{-\frac{1}{2}}v^{(n)}\,,\,Z_{i})}\nonumber\\
 &=&n^{-\frac{1}{2}}\,  \frac{h}{2} \,\sum _{i=1}^{n}
(\tilde{\epsilon}_{i,n} - {\epsilon}_{i})^2
\ddot{M}_{f}(\tilde{\tilde{{\epsilon}_{i,n}}})\,
\frac{G(Z_{i})}{\,\sigma(\theta_{0},\,Z_{i})}\nonumber\\
  &&-  n^{-\frac{1}{2}}\,  \frac{h}{2} \,\sum _{i=1}^{n}
(\tilde{\epsilon}_{i,n} - {\epsilon}_{i})^2
\ddot{M}_{f}(\tilde{\tilde{{\epsilon}_{i}}})\,
\frac{G(Z_{i})}{\,\sigma(\theta_{0},\,Z_{i})}\nonumber\\
&=&  -n^{-\frac{1}{2}}\,  \frac{h}{2} \,\sum _{i=1}^{n}
(\tilde{\epsilon}_{i,n} - {\epsilon}_{i})^2
\ddot{M}_{f}(\tilde{\tilde{{\epsilon}_{i,n}}})\,
\frac{G(Z_{i})}{\,\sigma(\theta_{0},\,Z_{i})} \nonumber \\
&&- \,n^{-\frac{1}{2}}\,  \frac{h}{2} \,\sum _{i=1}^{n}
(\tilde{\epsilon}_{i,n} - {\epsilon}_{i})^2
\ddot{M}_{f}(\tilde{\tilde{{\epsilon}_{i,n}}})\,
G(Z_{i})[\frac{1}{\sigma(\theta_{0}+n^{-\frac{1}{2}}v^{(n)},\,Z_{i})}
-\frac{1}{\sigma(\theta_{0}\,Z_{i})}]\nonumber\\
&=&I^{(1)}_{n,h,2} + I^{(2)}_{n,h,2}, \nonumber
\end{eqnarray}
\noindent where
\begin{eqnarray*}
I^{(1)}_{n,h,2}&=& -n^{-\frac{1}{2}}\,  \frac{h}{2} \,\sum
_{i=1}^{n} (\tilde{\epsilon}_{i,n} - {\epsilon}_{i})^2
\ddot{M}_{f}(\tilde{\tilde{{\epsilon_{i,n}}}})\,
\frac{G(Z_{i})}{\,\sigma(\theta_{0},\,Z_{i})},\\
I^{(2)}_{n,h,2} &=& -n^{-\frac{1}{2}}\,  \frac{h}{2} \,\sum
_{i=1}^{n} (\tilde{\epsilon}_{i,n} - {\epsilon}_{i})^2
\ddot{M}_{f}(\tilde{\tilde{{\epsilon}_{i,n}}})\,
G(Z_{i})[\frac{1}{\sigma(\theta_{0}+n^{-\frac{1}{2}}v^{(n)},\,Z_{i})}
-\frac{1}{\sigma(\theta_{0}\,Z_{i})}].
\end{eqnarray*}
\noindent From (\ref{epsilon}) and after majoration and the use of
the Cauchy Schwartz's inequality, it results that:
\begin{eqnarray}
(\hat{\epsilon}_{i,n} - {\epsilon}_{i})^2&\leq&\frac{2}{n}\Big\{
[\frac{(u^{(n)})^\top\,\partial\,m(\tilde{\rho}_{n},\,Z_{i})^\top}{\sigma(\theta_{0}+n^{-\frac{1}{2}}v^{(n)}\,,\,Z_{i})}]^2
+ [\frac{(v^{(n)})^\top\,\partial\,
{\sigma(\tilde{\theta}_{n},Z_{i})^\top}){\epsilon}_{i}}{\sigma(\theta_{0}+n^{-\frac{1}{2}}v^{(n)}\,,\,Z_{i})}]^2\Big\}\nonumber\\
&\leq& \frac{2}{n}
\|u^{(n)}\|^2_{\ell}\,\Big\{\frac{\|\partial\,m(\tilde{\rho}_{n},\,Z_{i})\|_{\ell}}{\sigma(\theta_{0}+n^{-\frac{1}{2}}v^{(n)}\,,\,Z_{i})}\Big\}^2
+\frac{2}{n}
\|v^{(n)}\|^2_{p}\,\Big\{\frac{\|\partial\,\sigma(\tilde{\theta}_{n},\,Z_{i})\|_{p}}{\sigma(\theta_{0}+n^{-\frac{1}{2}}v^{(n)}\,,\,Z_{i})}
\Big\}^2{\epsilon}^2_{i}.~~~~~~~~~~\label{majepsi}
\end{eqnarray}
Then
\begin{eqnarray}
 |I^{(1)}_{n,h,2}| &\leq&  n^{-\frac{1}{2}}\, |h| \,\frac{1}{n}\sum _{i=1}^{n}
\|u^{(n)}\|^2_{\ell}\,\Big\{\frac{\|\partial\,m(\tilde{\rho}_{n},\,Z_{i})\|_{\ell}}{\sigma(\theta_{0}+n^{-\frac{1}{2}}v^{(n)}\,,\,Z_{i})}\Big\}^2
|\ddot{M}_{f}(\tilde{\tilde{{\epsilon}_{i}}})\,
\frac{G(Z_{i})}{\,\sigma(\theta_{0},\,Z_{i})}|\nonumber\\
&&+ n^{-\frac{1}{2}}\,  |h|  \,\frac{1}{n}\sum _{i=1}^{n}
\|v^{(n)}\|^2_{p}\,\Big\{\frac{\|\partial\,\sigma(\tilde{\theta}_{n},\,Z_{i})\|_{p}}{\sigma(\theta_{0}+n^{-\frac{1}{2}}v^{(n)}\,,\,Z_{i})}
\Big\}^2{\epsilon}^2_{i}
|\ddot{M}_{f}(\tilde{\tilde{{\epsilon}_{i}}})\,
\frac{G(Z_{i})}{\,\sigma(\theta_{0},\,Z_{i})}|\nonumber\\
&\leq&  n^{-\frac{1}{2}}\,|h|
\,[\sup_{n}[({{\tau}^{(n)})}^\top({\tau}^{(n)})]\Big[\frac{1}{n}\sum
_{i=1}^{n}
\Big\{\frac{\|\partial\,m(\tilde{\rho}_{n},\,Z_{i})\|_{\ell}}{\sigma(\theta_{0}+n^{-\frac{1}{2}}v^{(n)}\,,\,Z_{i})}\Big\}^2
|\ddot{M}_{f}(\tilde{\tilde{{\epsilon}_{i}}})|\,
|\frac{G(Z_{i})}{\,\sigma(\theta_{0},\,Z_{i})}|\nonumber\\
&&+ \,\frac{1}{n}\sum _{i=1}^{n}
\Big\{\frac{\|\partial\,\sigma(\tilde{\theta}_{n},\,Z_{i})\|_{p}}{\sigma(\theta_{0}+n^{-\frac{1}{2}}v^{(n)}\,,\,Z_{i})}
\Big\}^2{\epsilon}^2_{i}
|\ddot{M}_{f}(\tilde{\tilde{{\epsilon}_{i}}})|\,
|\frac{G(Z_{i})}{\,\sigma(\theta_{0},\,Z_{i})}|\Big].\nonumber
\end{eqnarray}
\noindent Since the second derivative  $\ddot{M}_{f}$ is bounded,
then there exist a positive real $\vartheta$ such that
\\
$\forall x \in \mathbb{R},$  we have
\begin{eqnarray}
  |\ddot{M}_{f}(x)| &\leq& \vartheta .\label{sd}
\end{eqnarray}
\noindent It follows that
\begin{eqnarray}
 |I^{(1)}_{n,h,2}| &\leq& \vartheta\, n^{-\frac{1}{2}}\, |h|
\,[\sup_{n}[({{\tau}^{(n)})}^\top({\tau}^{(n)})]\Big[\frac{1}{n}\sum
_{i=1}^{n}
\Big\{\frac{\|\partial\,m(\tilde{\rho_{n}},\,Z_{i})\|_{\ell}}{\sigma(\theta_{0}+n^{-\frac{1}{2}}v^{(n)}\,,\,Z_{i})}\Big\}^2
\,
|\frac{G(Z_{i})}{\,\sigma(\theta_{0},\,Z_{i})}|\nonumber\\
&&+ \,\frac{1}{n}\sum _{i=1}^{n}
\Big\{\frac{\|\partial\,\sigma(\tilde{\theta_{n}},\,Z_{i})\|_{p}}{\sigma(\theta_{0}+n^{-\frac{1}{2}}v^{(n)}\,,\,Z_{i})}
\Big\}^2{\epsilon}^2_{i}\,
|\frac{G(Z_{i})}{\,\sigma(\theta_{0},\,Z_{i})}|\Big].\nonumber
\end{eqnarray}
\noindent From (\ref{c}), (\ref{asyconvex1}), (\ref{asyconvex2}),
$(A_{3.1})$, $(A_{3.7})$, $(A_{4.1})$ and the ergodic theorem, it
follows asymptotically that
\begin{eqnarray}
 I^{(1)}_{n,h,2}&=&o_{P}(1). \label{adern}
\end{eqnarray}
\noindent It remains to evaluate the term $I^{(2)}_{n,h,2}$.\\
\noindent By Taylor's expansion with order $1$ of the function
$\sigma(\theta, \cdot)$ around $\theta_{0}$, the expression
$I^{(2)}_{n,h,2} $ can also be written
\begin{eqnarray*}
 I^{(2)}_{n,h,2} &=& {\frac{1}{n}}\, \frac{h}{2} \,\sum
_{i=1}^{n}\frac{(v^{(n)})^\top \, \partial\,
\sigma({\tilde{\theta}}_{n},\,Z_{i})^\top}{\sigma(\theta_{0}+n^{-\frac{1}{2}}v^{(n)}\,
,\,Z_{i})}(\tilde{\epsilon}_{i,n} - {\epsilon}_{i})^2
\ddot{M}_{f}(\tilde{\tilde{{\epsilon}_{i}}})\,
\frac{G(Z_{i})}{\sigma(\theta_{0},\,Z_{i})}.
\end{eqnarray*}
\noindent From the inequalities (\ref{majepsi}) and (\ref{sd})
followed by Cauchy -Schwartz inequality and   a simple majoration,
we obtain
\begin{eqnarray}
|I^{(2)}_{n,h,2}|&\leq& \frac{1}{n} \vartheta\,|h|
\,[\sup_{n}[({{\tau}^{(n)})}^\top({\tau}^{(n)})]^{\frac{3}{2}}\Big[\frac{1}{n}\sum
_{i=1}^{n}
\Big\{\frac{\|\partial\,m(\tilde{\rho}_{n},\,Z_{i})\|_{\ell}}{\sigma(\theta_{0}+n^{-\frac{1}{2}}\,,\,Z_{i})}\Big\}^2
\, |\frac{G(Z_{i})}{\,\sigma(\theta_{0},\,Z_{i})}||
\frac{(v^{(n)})^\top
\partial\,\sigma(\tilde{\theta}_{n},\,Z_{i})^\top}{\sigma(\theta_{0}+n^{-\frac{1}{2}}v^{(n)}\,,\,Z_{i})}|\nonumber\\
&&+ \,\frac{1}{n}\sum _{i=1}^{n}
\Big\{\frac{\|\partial\,\sigma(\tilde{\theta}_{n},\,Z_{i})\|_{p}}{\sigma(\theta_{0}+n^{-\frac{1}{2}}v^{(n)}\,,\,Z_{i})}
\Big\}^2{\epsilon}^2_{i}\,
|\frac{G(Z_{i})}{\,\sigma(\theta_{0},\,Z_{i})}|| \frac{
\partial\,\sigma(\tilde{\theta}_{n},\,Z_{i})^\top}{\sigma(\theta_{0}+n^{-\frac{1}{2}}v^{(n)}\,,\,Z_{i})}|\Big]\nonumber\\
&\leq&\frac{1}{n} \vartheta\,|h|
\,[\sup_{n}[({{\tau}^{(n)})}^\top({\tau}^{(n)})]^{\frac{3}{2}}\Big[\frac{1}{n}\sum
_{i=1}^{n}
\Big\{\frac{\|\partial\,m(\tilde{\rho}_{n},\,Z_{i})\|_{\ell}}{\sigma(\theta_{0}+n^{-\frac{1}{2}}v^{(n)}\,,\,Z_{i})}\Big\}^3
\, |\frac{G(Z_{i})}{\,\sigma(\theta_{0},\,Z_{i})}|\nonumber\\
&&+ \,\frac{1}{n}\sum _{i=1}^{n}
\Big\{\frac{\|\partial\,\sigma(\tilde{\theta}_{n},\,Z_{i})\|_{p}}{\sigma(\theta_{0}+n^{-\frac{1}{2}}v^{(n)}\,,\,Z_{i})}
\Big\}^3{\epsilon}^2_{i}\,
|\frac{G(Z_{i})}{\,\sigma(\theta_{0},\,Z_{i})}|\Big].\nonumber
\end{eqnarray}
 \noindent From  (\ref{asyconvex1}), (\ref{asyconvex2}), (\ref{c}),
$(A_{3.1})$,$(A_{3.7})$, $(A_{4.2})$,
 and the ergodicity of the model, it follows
that
\begin{eqnarray}
 I^{(2)}_{n,h,2}&=&o_{P}(1).\label{derniertaux}
\end{eqnarray}
\noindent From the equalities (\ref{adern}) and
(\ref{derniertaux}), we deduce that
\begin{eqnarray}
 I_{n,h,2}&=&o_{P}(1).\label{dt3}
\end{eqnarray}
\textbf{In summary, we have the following equalities}
\begin{eqnarray*}
\tilde{ r}_{f,h,n} - r_{f,h,n} &=& I_{n,h,1} + I_{n,h,2} +
I_{n,h,3},\\
 I_{n,h,1} &=& h\, (u^{(n)})^\top K^\top  +h\,(v^{(n)})^\top J^\top +o_{P}(1),\\
 I_{n,h,2}&=&o_{P}(1),\\
 I_{n,h,3} &=&  o_{P}(1).
\end{eqnarray*}
We deduce that
\begin{eqnarray}
\tilde{ r}_{f,h,n} - r_{f,h,n} &=& h\, (u^{(n)})^\top K^\top  +h\,(v^{(n)})^\top J^\top +o_{P}(1).\label{soussuitecentrale1}\\
\nonumber
\end{eqnarray}
 \noindent In order to evaluate the term  $\tilde{q}_{f,h',n} -  q_{f,h',n} $, we consider the difference between
the equations (\ref{r4}) et  (\ref{r2}), then we obtain
\begin{eqnarray*}
  \hat{ q}_{f,h',n} - q_{f,h',n} &=& -n^{-\frac{1}{2}}\, h' \,\sum _{i=1}^{n} [\frac{N_{f}(\hat{\epsilon}_{i,n})}
  {\sigma(\theta_{0}+n^{-\frac{1}{2}}v^{(n)}\,,\,Z_{i})} -
  \frac{N_{f}({\epsilon}_{i})}{\sigma(\theta_{0},\,Z_{i})}]S(Z_{i}).
\end{eqnarray*}
Using the same reasoning that  (\ref{DIFFER}), it results that
 \begin{eqnarray}
  \frac{N_{f}(\hat{\epsilon}_{i,n})}{\sigma(\theta_{0}+n^{-\frac{1}{2}}v^{(n)}\,,\,Z_{i})} - \frac{N_{f}({\epsilon}_{i})}{\sigma(\theta_{0},\,Z_{i})} &=&
  \frac{N_{f}(\hat{\epsilon}_{i,n})  - N_{f}({\epsilon}_{i})}{\sigma(\theta_{0}+n^{-\frac{1}{2}}v^{(n)}\,,\,Z_{i})}-
  \frac{n^{-\frac{1}{2}} (v^{(n)})^\top \, \partial\, \sigma(\tilde{\theta}_{n},\,Z_{i})^\top}
  {\sigma(\theta_{0}+n^{-\frac{1}{2}}v^{(n)}\,,\,Z_{i})\sigma(\theta_{0}\,,\,Z_{i})}\,N_{f}({\epsilon}_{i})\nonumber\\
&=&\frac{(\hat{\epsilon}_{i,n} -
{\epsilon}_{i})\dot{N}_{f}({\epsilon}_{i})}{\sigma(\theta_{0}+n^{-\frac{1}{2}}v^{(n)}\,,\,Z_{i})}
+ \frac{(\hat{\epsilon}_{i,n} - {\epsilon}_{i})^2
\ddot{N}_{f}(\tilde{\tilde{{\epsilon}_{i}}})}{2\sigma(\theta_{0}+n^{-\frac{1}{2}}v^{(n)}\,,\,Z_{i})}
\nonumber\\
&&- \frac{n^{-\frac{1}{2}} (v^{(n)})^\top \, \partial\,
\sigma(\tilde{\theta}_{n},\,Z_{i})^\top}
  {\sigma(\theta_{0}+n^{-\frac{1}{2}}v^{(n)}\,,\,Z_{i})\sigma(\theta_{0}\,,\,Z_{i})}\,N_{f}({\epsilon}_{i}).\nonumber\\
  \label{DIFFER1}
 \end{eqnarray}
Hence
\begin{eqnarray*}
 \tilde{ q}_{f,h',n} - q_{f,h',n} &=& I'_{n,h',1} + I'_{n,h',2} +
 I'_{n,h',3},
\end{eqnarray*}
with
\begin{eqnarray}
  I'_{n,h',1} &=& -n^{-\frac{1}{2}}\, h' \,\sum_{i=1}^{n}\frac{(\hat{\epsilon}_{i,n} -{\epsilon}_{i})
  \dot{N}_{f}({\epsilon}_{i})}{\sigma(\theta_{0}+n^{-\frac{1}{2}}v^{(n)}\,,\,Z_{i})}\,S(Z_{i}),\\
I'_{n,h',2} &=& -n^{-\frac{1}{2}}\, h' \,\sum _{i=1}^{n}
\frac{(\hat{\epsilon}_{i,n} -
{\epsilon}_{i})^2}{2\,\sigma(\theta_{0}+n^{-\frac{1}{2}}v^{(n)}\,,\,Z_{i})}\ddot{N}_{f}(\tilde{\tilde{{\epsilon}_{i}}})\,S(Z_{i}),\\
I'_{n,h',3} &=& {\frac{1}{n}}\, h' \,\sum
_{i=1}^{n}\frac{(v^{(n)})^\top \, \partial\,
\sigma(\tilde{\theta}_{n},\,Z_{i})^\top}
{\sigma(\theta_{0}+n^{-\frac{1}{2}}v^{(n)}\,,\,Z_{i})\sigma(\theta_{0}\,,\,Z_{i})}\,N_{f}({\epsilon}_{i})\,S(Z_{i}).
\end{eqnarray}
\subsubsection*{Evaluation of the term  $I'_{n,h',1}$}
Firstly, from (\ref{derivéepremiereden}), we remark that
\begin{eqnarray}
|\dot{N}_{f}(x)| &\leq& |M_{f}(x)| +| x\,\dot{ M}_{f}(x)|.
\label{maj}
 \end{eqnarray}
\noindent It results from the application of the Lemma
(\ref{lemmainequality}) on the inequality (\ref{maj}) and the use
of the conditions $(A_{3.3})$ and $(A_{3.6})$, that
\begin{itemize}
    \item  $(A'_{3.5})$ :\\There exist $\lambda >0$ such that :
$\mathbf{E}|\dot{{N}}_{f}({{\epsilon}_{0})}|^{{\lambda}+2}<+\infty.$
\end{itemize}
We have from the equality (\ref{derivéepremiereden}), the
following equality
\begin{eqnarray*}
x \dot{N}_{f}(x) &=& xM_{f}(x) + x^2\,\dot{ M}_{f}(x).
\end{eqnarray*}
By applying on this last equality the Lemma
(\ref{lemmainequality}) combined with the conditions $(A_{3.4})$
and $(A_{3.8}),$ we deduce that\\ ($A'_{3.6}$): There existe
$\lambda>0$, such that :
~~$\mathbf{E}\Big|\epsilon_{0}{\dot{N}}_{f}(\epsilon_{0})\Big|^{{\lambda}
+ 2}<+\infty.$ \noindent By changing respectively $(A_{3.1})$,
$(A_{3.5})$ and $(A_{3.6})$ by $(A_{3.2})$, $(A'_{3.5})$ and
$(A'_{3.6})$ and with applying on the expression $I'_{n,h',1}$ the
same previous reasoning applied on the expression $I_{n,h,1}$, we
shall prove that
\begin{eqnarray*}
I'_{n,h,1} &=& h'\, (u^{(n)})^\top K'^\top  + h'(v^{(n)})^\top
J'^\top +o_{P}(1),\label{dt'1}
\end{eqnarray*}
such that
\begin{eqnarray*}
  K'^\top  &=&  (K'_{1},\cdot,\cdot,\cdot,K'_{\ell}),\\
  K'_{\ell}&=&  \mathbb{E}\Big[\frac{
\frac{\partial\,m(\rho_{0},\,Z_{0})}{\partial\,\rho_{j}
}}{\sigma(\theta_{0}\,,\,Z_{0})}\frac{\dot{N}_{f}({\epsilon}_{0})\,S(Z_{0})}
{\sigma(\theta_{0}\,,\,Z_{0})}\Big],\\
 J'^\top  &=&  (J'_{1},\cdot,\cdot,\cdot,J'_{p}),\\
 J'_{k}&=& \mathbb{E}\Big[\frac{
\frac{\partial\,\sigma(\theta_{0},\,Z_{0})}{\partial\,\theta_{k}
}}{\sigma(\theta_{0}\,,\,Z_{0})}\frac{{\epsilon}_{0}\dot{N}_{f}({\epsilon}_{0})\,S(Z_{0})}
{\sigma(\theta_{0}\,,\,Z_{0})}\Big].\\
\end{eqnarray*}
\subsubsection*{Evaluation of the term $I'_{n,h',2}$}
 In this case, the condition (\ref{sd}) is replaced by the following
 condition :
 \begin{eqnarray}
  |\ddot{N}_{f}(x)| &\leq& \vartheta' ,\label{sdn}
\end{eqnarray}
where $ \vartheta'$ is strictly positive real.\\
 \noindent By changing $(A_{3.1})$  by $(A_{3.2})$ and with applying on the
expression $I'_{n,h',2}$ the same previous reasoning applied on
the expression $I_{n,h,2}$, we shall prove that
\begin{eqnarray}
 I'_{n,h',2}&=&o_{P}(1).\label{d't2}
\end{eqnarray}
\subsubsection*{Evaluation of the $I'_{n,h',3}$}
From the definition of the function $N_{f},$ and using the
condition $(A_{2.2})$, we obtain the following condition
:$(A'_{2.1})$:~~~~$\mathbf{E}\left\{
N_{f}(\epsilon_{0})\right\}=0.$ \noindent By changing respectively
$(A_{2.1})$  by $(A'_{2.1})$
 and with applying on the expression $I'_{n,h',3}$
the same previous reasoning applied on the expression $I_{n,h,3}$,
we shall prove that
\begin{eqnarray}
 I'_{n,h',3}&=& o_{P}(1). \label{d't3}
\end{eqnarray}
\textbf{ In summary } It follows from the equalities (\ref{dt'1}),
(\ref{d't2}) and (\ref{d't3})
\begin{eqnarray}
   \tilde{q}_{f,h',n} - q_{f,h',n} &=& h'\, (u^{(n)})^\top K'^\top  +h'\,(v^{(n)})^\top J'^\top +o_{P}(1).\label{soussuitecentrale2}
\end{eqnarray}
Hence the proposition is established.

\subsection*{Proof of the  Proposition \ref{kreiss}}
 \noindent The proof of proposition \ref{kreiss}  is a
consequence of the works of \cite{L} and \cite{K}. The interested
reader  can refer to   in
\cite[Lemma ($4.4$)]{K} for more details.\\
\subsection*{Proof of the Theorem} \label{proofoptimality}
Consider again the equality\begin{eqnarray}
 \mathcal{\widehat{V}}_{n,h,h'}-\mathcal{V}_{n,h,h'}&=&\sqrt{n}(\hat{\rho}_{n} - \rho_{0})^\top ( h\,K^\top + h'\,K'^\top) +
  \sqrt{n}(\hat{\theta}_{n} - \theta_{0})^\top ( h\,J^\top + h'\,J'^\top)
  +o_{P}(1).~~~~~~\label{prop3'}
\end{eqnarray} and let

\begin{eqnarray*}
 D_{n,h,h'}&=&-\Big(\sqrt{n}(\hat{\rho}_{n} - \rho_{0})^\top ( h\,K^\top + h'\,K'^\top) +
  \sqrt{n}(\hat{\theta}_{n} - \theta_{0})^\top ( h\,J^\top +
  h'\,J'^\top)\Big),\label{bounded ecart}
\end{eqnarray*} clearly, $|D_{n,h,h'}|=O_{P}(1)$, in fact by
applying the Cauchy Schwartz inequality combined with the triangle
inequality, it follows that:\begin{eqnarray*}
|D_{n,h,h'}|\leq\sqrt{n} \|\hat{\rho}_{n} - \rho_{0}\|_{\ell}
\|h\,K^\top + h'\,K'^\top\|_{\ell} + \sqrt{n}\|\hat{\theta}_{n}
-\theta_{0}\|_{p} \|h\,J^\top + h'\,J'^\top\|_{p}.
\end{eqnarray*}
Since the estimates $\rho_n$ and $\theta_n$ are consistent, it
follows that $D_{n,h,h'}=O_{P}(1),$ therefore the equality
(\ref{prop3'}) can also rewritten \begin{eqnarray}
 \mathcal{\widehat{V}}_{n,h,h'}-\mathcal{V}_{n,h,h'}&=- D_{n,h,h'}  +o_{P}(1).~~~~~~\label{prop3''}
\end{eqnarray}
From the assumption $(P.0)$, there exists another estimate
$\bar{\Omega}_n=\Omega_{n}^{(1,j_{n}})$ of the unknown parameter
 $\Omega$ such that
\begin{eqnarray}
 {\mathcal{V}}_{n,h,h'}(\bar{\Omega}_n)&=\mathcal{V}_{n,h,h'} +o_{P}(1).~~~~~~\label{prop3'''}
\end{eqnarray}
Under a additional assumptions $(P.1)$, $\bar{\Omega}_n$  is
$\sqrt{n}$-root  consistent , see \cite[Subsection 1.2]{TL2012}
 The equality (\ref{prop3'''}), enables us to deduce that, with $o_{P}(1)$ close, the replacing in
the expression (\ref{test}) of the test of the central sequence
$\mathcal{V}_{n,h,h'}(\Omega)$ by the estimate central sequence
$\mathcal{V}_{n,h,h'}(\Omega_{n}^{(1,j_{n})})$ has no effect.\\
From the continuity of the function ${\tau}^2(\cdot,\cdot)$ and
the convergence in probability of the random sequence
$\Omega_{n}^{(1,j_{n})}$ to the unknown parameter $\Omega$, it
follows that under the hypothesis $ H_{0}$ and under contiguous
alternatives, we get
\begin{eqnarray*}
  I{\Big\{{\frac{\mathcal{V}_{n,h,h'}(\Omega_{n}^{(1,j_{n})})}{{\tau}_{h,h'}({{\bar{\rho}}_{n}} ,{\hat{\theta}_{n}})
}\geq Z(u)}\Big\}} &=&
I{\Big\{{\frac{\mathcal{V}_{n,h,h'}(\Omega)}{{\tau}_{h,h'}(\rho,\theta)}\geq
Z(u)}\Big\}}  +o_{P}(1).
\end{eqnarray*}
\noindent The  two sequences of tests
$\hat{T}_{n}=I{\Big\{{\frac{\mathcal{V}_{n,h,h'}(\Omega_{n}^{(1,j_{n})})}{{\tau}_{h,h'}({{\bar{\rho}}_{n}}
,{\hat{\theta}_{n}}) }\geq Z(u)}\Big\}} $ and
$T_{n}=I{\Big\{{\frac{\mathcal{V}_{n,h,h'}(\Omega)}{{\tau}_{h,h'}(\rho,\theta)}\geq
Z(u)}\Big\}}$ are locally and asymptotically equivalent, hence the
optimality of the test. The asymptotic power of this test is equal
to   $1 -\Phi(Z(\alpha)-{\tau^2( \bar{\rho}_{n})} )$, see
\cite[Theorem 3]{HB}.

\begin{Remark}
We can also  get the optimality of the test  when we replace the
estimate $(\Omega_{n}^{(1,j_{n})})$ by the
$(\Omega_{n}^{(2,k_{n})})$ in this previous proof.
\end{Remark}
\begin{Conclusion}
On a basis of the discrete estimates and for each step $n$, we
have modified one component of our estimate in order to absorb the
error, this new estimate was constructed  on the tangent space of
the discrete estimate in each step $n$, so the introduction of
this kind of estimate has enabled us to get the optimality of the
test which is based on the Neyman-Pearson statistic when we
replace in the expression of this statistic the unknown parameter
by the M.D.E.\\
In practise, we shall obtain a good   M.D.E. when the errors $\|
\hat{\rho}_{n} - \rho_{0}\|_{\ell}  $ and  $\| \hat{\theta}_{n} -
\theta_{0}\|_{p}$ are best estimated, in this case , we shall used
the bootstrap methods.
\end{Conclusion}

\subsection*{Proof of the  Lemma (\ref{about sufficuent condition}]}

\noindent For the AR(m) model, the expression of the central
sequence is given by :
\begin{eqnarray*}
\mathcal{V}_{n}(\rho_0)=-\frac{1}{\sqrt{n}}\sum_{i=1}^{n}M_{f}(\epsilon_i)G(Y(i-1))
-\frac{1}{\sqrt{n}}\sum_{i=1}^{n}N_{f}(\epsilon_i)S(Y(i-1)),\quad
\mbox{where}\quad N_{f}(\epsilon_i)=1 + \epsilon_i
M_{f}(\epsilon_i).
\end{eqnarray*}
In order to evaluate the  difference between the two partial
derivatives central sequences, we calculate the derivative with
respect to the component $ \rho_j,$  then  we obtain
:\begin{eqnarray*} \mbox{For each integer i}  \in  \quad
\{1,\dots,m\}\quad \mbox{we have}:  \frac{\partial\epsilon_i
}{\partial \rho_j}= -Y_{i - j},\quad
\dot{M}_f(\epsilon_i)=-1,\quad \mbox{and} \quad
\dot{N}_f(\epsilon_i)=-2\epsilon_i,
\end{eqnarray*}
With a simple calculation, we shall prove that:
\begin{eqnarray}
\frac{1}{\sqrt{n}}\frac{\partial\mathcal{V}_n(\rho)}{\partial
\rho_j}&=&\frac{-1}{n}\sum_{i=1}^{n}\frac{\partial
(\epsilon_{i})}{\partial \rho_j} \dot{M}_f(\epsilon_i)G(Z_i) -
\frac{1}{n}\sum_{i=1}^{n}\frac{\partial (\epsilon_{i})}{\partial
\rho_j} \dot{N}_f(\epsilon_i)S(Z_i),\label{first derivative
general }\\ \mbox{~~and,}\\
\frac{1}{\sqrt{n}}\frac{\partial\mathcal{V}_n(\hat{\rho}_n)}{\partial
\rho_j}&=&\frac{-1}{n}\sum_{i=1}^{n}\frac{\partial
(\hat{\epsilon}_{i,n})}{\partial \rho_j}
\dot{M}_f(\hat{\epsilon}_{i,n})G(Z_i) -
\frac{1}{n}\sum_{i=1}^{n}\frac{\partial
(\hat{\epsilon}_{i,n})}{\partial \rho_j}
\dot{N}_f(\hat{\epsilon}_{i,n})S(Z_i)\label{first estimate
derivative general }.
\end{eqnarray}
From the difference between the equalities (\ref{first estimate
derivative general }) and (\ref{first derivative general }), it
follows that:
\begin{eqnarray}
\frac{1}{\sqrt{n}}\frac{\partial\mathcal{V}_n(\rho)}{\partial
\rho_j}-\frac{1}{\sqrt{n}}\frac{\partial\mathcal{V}_n(\hat{\rho}_n)}{\partial
\rho_j}&=&\frac{-1}{n}\sum_{i=1}^{n}\Big(\frac{\partial
(\hat{\epsilon}_{i,n})}{\partial \rho_j}
\dot{M}_f(\hat{\epsilon}_{i,n}) - \frac{\partial
(\epsilon_{i})}{\partial \rho_j} \dot{M}_f(\epsilon_i)\Big)
G(Z_i)\nonumber \\
- \frac{1}{n}\sum_{i=1}^{n}\Big(\frac{\partial
(\hat{\epsilon}_{i,n})}{\partial \rho_j} \dot{
N}_f(\hat{\epsilon}_{i,n}) - \frac{\partial
(\epsilon_{i})}{\partial \rho_j}
\dot{N}_f(\epsilon_i)\Big)S(Z_i),\nonumber\\
&=& \frac{-2}{n}\sum_{i=1}^{n} (\hat{\epsilon}_{i,n} -
\epsilon_i) Y_{i-j}S(Z_i).\nonumber\\
\end{eqnarray}
From the equalities the previous equalities, it follows
that:\begin{eqnarray}
\frac{1}{\sqrt{n}}\frac{\partial\mathcal{V}_n(\rho)}{\partial
\rho_j}-\frac{1}{\sqrt{n}}\frac{\partial\mathcal{V}_n(\hat{\rho}_n)}{\partial
\rho_j}=  (\hat{\rho}_{1,n} -\rho_1)\times \frac{-2}{n}
\sum_{i=1}^{n}Y_{i-1}Y_{i-j}S(Z_i)\\ + \dots + (\hat{\rho}_{m,n}
-\rho_m)\times \frac{-2}{n}\sum_{i=1}^{n}Y_{i-m}Y_{i-j}S(Z_i).\nonumber\\
\end{eqnarray}
 For all integers $i$ and $j$, we have the following equalities:
 \begin{eqnarray}
|Y_{i-m}Y_{i-j}S(Z_i)|\leq \frac{1}{2}\Big[ |Y_{i-m}Y_{i-j}|^2 +
|S(Z_i)|^2\Big]\leq \frac{1}{4}\Big[ |Y_{i-m}|^4  +
|Y_{i-j}|^4\Big] +\frac{1}{2}|S(Z_i)|^2 .
\end{eqnarray}
By applying this last equalities on the next previous equality, it
results that:\begin{eqnarray}
|\frac{1}{\sqrt{n}}\frac{\partial\mathcal{V}_n(\rho)}{\partial
\rho_j}-\frac{1}{\sqrt{n}}\frac{\partial\mathcal{V}_n(\hat{\rho}_n)}{\partial
\rho_j}|&\leq&  (\hat{\rho}_{1,n} -\rho_1)\times\Big[\frac{1}{2n}
\sum_{i=1}^{n}Y_{i-1}^4 + \frac{1}{2n} \sum_{i=1}^{n}Y_{i-j}^4 +
\frac{1}{n} \sum_{i=1}^{n}S^2(Z_i)\Big] \nonumber\\
  &+&  \dots + (\hat{\rho}_{m,n}
-\rho_m)\times \Big[\frac{1}{2n}\sum_{i=1}^{n}Y_{i-m}^4 +
\frac{1}{2n} \sum_{i=1}^{n}Y_{i-j}^4 +
\frac{1}{n} \sum_{i=1}^{n}S^2(Z_i)\Big].\nonumber\\
\end{eqnarray}
Recall that the estimator
$\hat{\rho}_{n}=\Big(\hat{\rho}_{n,1},\dots,\hat{\rho}_{n,m})^\prime$
is consistent, it follows that, for each integer\\
$k\in\{1,\dots,m\}$,  the quantity $\hat{\rho}_{k,n} -\rho_k
\stackrel{P}{\longrightarrow}0$ as $n \longrightarrow\infty,$
remark that this convergence in probability is one consequence of
the continuous mapping theorem, see for instance \cite{W}. Since
the model is ergodic with finite second and fourth moments, we
obtain under $H_0$:\begin{eqnarray*}
\frac{1}{\sqrt{n}}\frac{\partial\mathcal{V}_n({\hat{\rho}}_n)}{\partial
\rho_j}=\frac{1}{\sqrt{n}}\frac{\partial\mathcal{V}_n(\rho_0)}{\partial
\rho_j} + o_{p}(1).
\end{eqnarray*}

%
\section*{Appendix}
\noindent We prove the results which are stated in the remark
\ref{remarquederivéébornée}, more precisely when $f$ is density of
a student distribution with a degree of freedom $l$ greater than
$3,$ the functions  $x\longmapsto\dot{M}_{f}(x)$, $x \longmapsto
\ddot{M}_{f}(x)$ and $x \longmapsto x\ddot{M}_{f}(x)$ are
bounded.\\
 We have
\begin{eqnarray*}
  f(x) &=&C_{l}(1+ \frac{x^2}{l})^{-\frac{l+1}{2}},
\end{eqnarray*}
where $
C_{l}=\frac{\Gamma(\frac{l+1}{2})}{\sqrt{{\Pi}l}\,\Gamma(\frac{l}{2})},$
and $\Gamma$ is the gamma function. Then we have
\begin{eqnarray*}
M_f(x)&=&- \frac{l+1}{l}\,\frac{x}{(1+ \frac{x^2}{l})}.\\
\dot{ M}_f(x)&=&-\frac{l+1}{l}\Big[\frac{\frac{2x^2}{l}}{{(1+
\frac{x^2}{l})}^2} -\frac{1}{1+\frac{x^2}{l}}\Big].\\
\ddot{M}_f(x)&=&-\frac{l+1}{l}\Big[
\frac{\frac{8x^3}{l^2}}{{(1+\frac{x^2}{l})}^3} -
\frac{\frac{4x}{l}}{{(1+\frac{x^2}{l})}^2} -
\frac{\frac{2x}{l}}{{(1+\frac{x^2}{l})}^2}
   \Big].
\end{eqnarray*}
We have
\begin{eqnarray*}
|\dot{ M}_f(x)|&\leq&\frac{l+1}{l}\Big[\frac{\frac{2x^2}{l}}{{(1+
\frac{x^2}{l})}^2} + \frac{1}{1+\frac{x^2}{l}}\Big].\\
\end{eqnarray*}
We can remark that
\begin{eqnarray*}
\frac{\frac{2x^2}{l}}{{(1+
\frac{x^2}{l})}^2}&=&\frac{1}{2}\frac{\frac{2x}{\sqrt{l}}}{(1+
\frac{x^2}{l})}.\frac{\frac{2x}{\sqrt{l}}}{(1+ \frac{x^2}{l})}.
\end{eqnarray*}
Since $\frac{2x}{\sqrt{l}}$ $\leq(1+ \frac{x^2}{l})$ and $1\leq(1+
\frac{x^2}{l}),$ it results that
\begin{eqnarray}
|\dot{
M}_f(x)|&\leq&\frac{3(l+1)}{2l}\label{boundedfirstderivativeM}.\\
\nonumber
\end{eqnarray}
We have
\begin{eqnarray*}
|\ddot{M}_f(x)|&\leq&\frac{l+1}{l}\Big[
|\frac{\frac{8x^3}{l^2}}{{(1+\frac{x^2}{l})}^3} |+
|\frac{\frac{4x}{l}}{{(1+\frac{x^2}{l})}^2}| +
|\frac{\frac{2x}{l}}{{(1+\frac{x^2}{l})}^2}|
   \Big].
\end{eqnarray*}
We can remark that
\begin{eqnarray*}
\frac{\frac{8x^3}{l^2}}{{(1+\frac{x^2}{l})}^3}&=&\frac{\sqrt{l}}{l}\frac{\frac{2x}{\sqrt{l}}}{{(1+\frac{x^2}{l})}}.
\frac{\frac{2x}{\sqrt{l}}}{{(1+\frac{x^2}{l})}}\frac{\frac{2x}{\sqrt{l}}}{{(1+\frac{x^2}{l})}}.\\
\frac{\frac{4x}{l}}{{(1+\frac{x^2}{l})}^2}&=& \frac{2}{\sqrt{l}}\frac{\frac{2x}{\sqrt{l}}}{{(1+\frac{x^2}{l})}}\frac{1}{{(1+\frac{x^2}{l})}}.\\
\frac{\frac{2x}{l}}{{(1+\frac{x^2}{l})}^2}&=&\frac{\sqrt{l}}{l}\frac{\frac{2x}{\sqrt{l}}}{{(1+\frac{x^2}{l})}}\frac{1}{{(1+\frac{x^2}{l})}}.
\end{eqnarray*}
It results that
\begin{eqnarray}
|\ddot{M}_f(x)|&\leq&
\frac{(l+1)(4\sqrt{l})}{l^2}.\label{boundedsecondderivativeM}
\end{eqnarray}
It remains to show that the function $x \longmapsto
x\ddot{M}_{f}(x)$ is bounded. In fact, we have
\begin{eqnarray*}
x\ddot{M}_f(x)&=&-\frac{l+1}{l}\Big[
\frac{\frac{8x^4}{l^2}}{{(1+\frac{x^2}{l})}^3} -
\frac{\frac{4x^2}{l}}{{(1+\frac{x^2}{l})}^2} -
\frac{\frac{2x^2}{l}}{{(1+\frac{x^2}{l})}^2}
   \Big].
\end{eqnarray*}
We have
\begin{eqnarray*}
\frac{\frac{8x^4}{l^2}}{{(1+\frac{x^2}{l})}^3}&=&8\frac{\frac{x^2}{l}}{{(1+\frac{x^2}{l})}^2}\frac{\frac{x^2}{l}}{(1+\frac{x^2}{l})}.\\
\frac{\frac{4x^2}{l}}{{(1+\frac{x^2}{l})}^2}&=&4\frac{\frac{x^2}{l}}{(1+\frac{x^2}{l})}\frac{1}{(1+\frac{x^2}{l})}.\\
\frac{\frac{2x^2}{l}}{{(1+\frac{x^2}{l})}^2}&=&2\frac{\frac{x^2}{l}}{(1+\frac{x^2}{l})}\frac{1}{1+\frac{x^2}{l}}.
\end{eqnarray*}
Since $\frac{x^2}{l}\leq (1+ \frac{x^2}{l})\leq {(1+
\frac{x^2}{l})}^2,$ it result that
\begin{eqnarray}
|x\ddot{M}_f(x)|&\leq&\frac{14(l+1)}{l}.\label{boundedsecondderivativexM}
\end{eqnarray}
\noindent Using the equality (\ref{dsss}) and from the equalities
(\ref{boundedfirstderivativeM}) and
(\ref{boundedsecondderivativexM}), it results that the second
derivative $\ddot{N}_f$ is bounded. Obviously, this previous
results remain satisfied when the value of the degree of freedom
is smaller than $3.$ {document}
\bibliographystyle{natbib}
\bibliography{biblio}
\end{document}